\documentclass[pre,aps,eqsecnum,preprint]{revtex4}
\usepackage{graphicx}
\usepackage[dvips]{epsfig}
\usepackage{color}
\definecolor{red}{rgb}{1,0,0}
\definecolor{green}{rgb}{0.13,0.55,0.13}
\definecolor{blue}{rgb}{0,0,1}

\usepackage{amsmath}
\usepackage{amsfonts}
\usepackage{amssymb}
\begin{document}

\title{Critical Behavior in Rectangles with Mixed Boundaries}

\author{ E. Eisenriegler }

\affiliation{
	Theoretical Soft Matter and Biophysics, Institute of Complex
	Systems,\\
	Forschungszentrum J\"ulich, D-52425 J\"ulich, Germany}

\date{\today}

\begin{abstract}

Density profiles are investigated arising in a critical Ising model in two dimensions which is confined to a rectangular domain with uniform or mixed boundary conditions and arbitrary aspect ratio. For the cases in which the two vertical sides of the rectangle have up-spin boundary conditions + and the two horizontal sides with either down-spin boundary conditions $-$ or with free-spin boundary conditions $f$, exact results are presented for the density profiles of the energy and the order parameter which display a surprisingly rich behavior. The new results follow by means of conformal transformations from known results in the half plane with $+-+-+$ and $+f+f+$ boundary conditions. The corners with mixed boundary conditions lead to interesting behavior, even in the limit of a half-infinite strip. The behavior near these corners can be described by a ``Corner-Operator-Expansion'', which is discussed in the second part of the paper. The analytic predictions agree very well with simulations, with no adjustable parameters.

\end{abstract}

\maketitle

\section{INTRODUCTION}

Due to the macroscopic correlation length in a critical system, the effects of boundaries penetrate deeply into the bulk. Two or more boundaries, even when separated by a macroscopic distance, induce critical density profiles which are, in general, not simple superpositions of the single boundary-profiles. Both the density profiles and the free energy of interaction of macroscopic range between the boundaries depend in a nontrivial way on the configuration of the boundaries. 

An important feature is the detail-independence or ``universality'' on large length scales of bulk and boundary critical phenomena \cite{BinderDombLebo, Diehl, CardBoundaryCritPhen}. This paper concentrates on two-dimensional systems in the Ising universality class. The boundaries belong to the ``ordinary''  and ``normal'' boundary universality classes which we denote by $f$ and $+$ or $-$ since these two universality classes are realized in the Ising lattice model by boundary spins that are {\it f}ree of outside bonds or fixed in the + or $-$ direction, respectively. The simplest geometry in two dimensions with a boundary is the upper half plane bounded by the horizontal coordinate axis. Besides {\it uniform} boundaries with one of the three classes or boundary conditions $f, \, +$, and $-$ extending along the entire horizontal axis \cite{nocrossov},  the interesting case of {\it mixed} boundaries has also been studied \cite{Cardytab, BX, TWBG1, TWBG2, BE21}. In the simplest case, the boundary condition along the negative horizontal axis is different from that along the positive axis, i.e. the boundary condition switches at the origin. In Refs. \cite{TWBG1, TWBG2, BE21} explicit expressions for the density profiles of the order parameter, the energy, and the stress tensor in the half plane \cite{unlike} were obtained for multiple switching points between + and $-$, between + and $f$, and for $-f+$ at arbitrary switch points with a macroscopic distance between them. 

Sec. \ref{recmix} of this paper is devoted to evaluate the critical density profiles in a {\it rectangle} with mixed boundaries and their dependence on the aspect ratio. Of main interest is the case in which the common universality class of the horizontal boundaries is different from the common universality class of the vertical boundaries. While extensive studies exist for (the universal part of) the free energy of this system, see e.g. Refs. \cite{CardPe,KV, Hucht, Imamura, Wu, BondesSaleur}, density profiles in rectangles with mixed boundary conditions have been investigated to a much lesser extent \cite{domainwall}. Assuming the system is at its bulk-critical point, the exact density profiles in the rectangle are derived from those in the upper half plane by means of conformal transformations.
In Sec. \ref{seminfstrip} density profiles are studied in the simpler geometry of a semi-infinite rectangle or semi-infinite strip. In addition to the densities of the order parameter $\sigma$ and the energy $\epsilon$, attention is payed to the density of the stress tensor $T$ since, apart from its importance in  the conformal theory and for the free energy, it plays a key role in the discussion of the near-boundary behavior. At internal points of the rectangle the density profiles of $\sigma$, $\epsilon$, and $T$ provide local information about the amount of preference for one of the two Ising directions, the degree of disorder, and the orientation-dependence of short distance correlations, respectively.

The {\it corners} of a rectangle where two different universality classes meet have a profound effect on the density profiles, in particular on the energy density \cite{engydens}, and there is an interesting dependence on the aspect-ratio. Another motivation for considering the rectangular geometry is that the  theoretical results can be conveniently compared with simulations \cite{OAV}.  

In Sec. \ref{coordaxes} of the paper, operator expansions are introduced that hold in the vicinity of corners with arbitrary angles, both with sides of the same and of different universality classes. The boundary operators in the expansions are located right at the tips of the corners. These ``Corner Operator Expansions (COE)'' are interesting in their own right. They are used to evaluate the near-corner behavior of the density profiles and study the dependence on the size and aspect ratio \cite{FdG} of the rectangle. The expansions apply not only to the Ising model but also to other conformally invariant models in two spatial dimensions. They are similar in spirit to the short distance expansion of an operator product in the bulk or the boundary operator expansions (BOE) for a flat boundary, see, e.g., Appendix \ref{BOEmid} and Ref. \cite{BE21}.

\section{DENSITY PROFILES IN SEMI-INFINITE STRIPS WITH A MIXED \\ BOUNDARY} \label{seminfstrip}

For the later study of rectangles it is instructive to begin with the simple system of a semi-infinite strip $0 < y < {\cal W}, \, 0 < x < +\infty$ in the $z=x+iy$ plane. Exact results for the density profiles are obtained by means of the conformal mapping 
\begin{eqnarray} \label{semiinfstrip}
H(z)=\cosh \tilde{z} \, , \quad \tilde{z} \equiv \pi z / {\cal W} = \tilde{x} +i \tilde{y}  \, 
\end{eqnarray}
of the semi-infinite strip onto the upper half $H$ plane, $H=G+iJ$ with $J>0$ \cite{unlike}. The corners at $z=0$ and $z=i {\cal W}$ of the strip are mapped onto $H=1$ and $H=-1$, respectively, and the midline $z=x+i{\cal W}/2$ of the strip is mapped onto the imaginary axis $H=iJ$, with $J= \sinh \tilde{x}$. Generally, $G=\cosh \tilde{x} \times \cos \tilde{y}$ and $J=\sinh \tilde{x} \times \sin \tilde{y}$, so that a point and its mirror image about the midline of the strip are mapped to a point and its mirror image about the imaginary axis of the upper half $H$ plane.

Densities of primary operators $\phi$ such as \cite{primaryfields} $\epsilon$ and $\sigma$ in the strip follow from those in the half plane via
\begin{eqnarray} \label{stripfromplane}
\langle \phi (x,y) \rangle=|dH/dz|^{x_{\phi}} \times \langle \phi (G,J) \rangle \, ,
\end{eqnarray}
where $x_{\phi}$ is the scaling dimension of $\phi$.

For the case of {\it uniform} boundary condition $a$, where $\langle \phi (G,J) \rangle = {\cal A}_a^{(\phi)} J^{-x_{\phi}}$, Eq. (\ref{stripfromplane}) yields \cite{primaryfields,unisemstrip}
\begin{eqnarray} \label{uniformstrip}
\langle \phi(x,y) \rangle &=& |dH/dz|^{x_{\phi}} \times {\cal A}_a^{(\phi)} J^{-x_{\phi}} = {\cal A}_a^{(\phi)} \, \Psi^{x_{\phi}} \, , \nonumber\\
\Psi &\equiv& {\pi \over {\cal W}} {|\sinh \tilde{z}| \over {\rm Im} \cosh \tilde{z} } =  {\pi \over {\cal W}}  \, \Biggl({1 \over \sinh^2 \tilde{x}} + {1 \over \sin^2 \tilde{y}}  \Biggr)^{1/2} \, .
\end{eqnarray}
This allows to rewrite Eq. (\ref{stripfromplane}) in the form
\begin{eqnarray} \label{generalstrip}
\langle \phi (x,y) \rangle = \Psi^{x_{\phi}} 
\times \bigl[ J^{x_{\phi}} \, \langle \phi (G,J) \rangle  \bigr]_{G=\cosh \tilde{x} \times \cos \tilde{y}, \, J=\sinh \tilde{x} \times \sin \tilde{y}}  \, ,
\end{eqnarray}
which turns out to be convenient. The expressions (\ref{uniformstrip}), (\ref{generalstrip}) and (\ref{tressstrip})-(\ref{tressstripaba}) below are quite general and not limited to the Ising model.	 

In the Ising model three types of {\it mixed}  boundary conditions are of particular interest: These are $ABC=-+-, \, f+f$, and $-f+ $, where in counterclockwise order $A$ refers to the upper horizontal edge, $B$ to the vertical edge, and $C$ to the lower horizontal edge. The corresponding boundary conditions in the upper half $H$
plane are $A$ for $-\infty < G < -1$, $B$ for $-1 <G<1$, and $C$ for $1<G<+ \infty$.

\subsection{Stress tensor density in the strip geometry} \label{stresssistrip}

The stress tensor density in the semiinfinite strip follows via the general transformation formula in \cite{Ttrafo} from the stress tensor in the upper half $H$ plane. The latter can be obtained from Eq. (1.3) in \cite{BE21}, and one finds
\begin{eqnarray} \label{tressstrip}
\langle T(z) \rangle_{ABC} = \Bigl( {\pi \over {\cal W}} \Bigr)^2 \Bigl\{ {\hat{c} \over 12} \Bigl(1-{3 \over 2} {\rm coth}^2 \tilde{z}  \Bigr) + t_{AC} - 2 \, \Bigl[  {t_{AB} \over 1+ {\rm cosh}\tilde{z}} + {t_{BC} \over 1- {\rm cosh}\tilde{z}} \Bigr]  \Bigr\} \, . 
\end{eqnarray}
Here $t_{AB}$ is the amplitude in the expression $\langle T(H) \rangle_{AB} = t_{AB}/H^2$ of the stress tensor in the upper half plane, with a single switch from $A$ to $B$ on the boundary at $H=0$, see Ref. \cite{BX}, which vanishes for $A=B$. 

We mention a few properties of Eq. (\ref{tressstrip}). 

(i) For $\tilde{x} \gg 1$ it takes the well-known form
\begin{eqnarray} \label{tressstripprimeprime}
\langle T(z) \rangle_{ABC}  \to \Bigl( {\pi \over {\cal W}} \Bigr)^2  \Bigl[- {\hat{c} \over 24} +t_{AC} \Bigr]  \, , \quad x \gg {\cal W} 
\end{eqnarray}
of the stress tensor in an infinite $A$C strip, which is independent of $z$ and the distant vertical boundary $B$. 

(ii) Along the midline $y={\cal W}/2$ it takes the form
\begin{eqnarray} \label{tressstripprime}
\langle T(z= x+ i{\cal W}/2) \rangle_{ABC} &=& \Bigl( {\pi \over {\cal W}} \Bigr)^2  \Bigl\{ {\hat{c} \over 12} \Bigl(1-{3 \over 2} {\rm tanh}^2 \tilde{x}  \Bigr) + t_{AC} - \nonumber \\
&& - {2 \over 1+{\rm sinh}^2 \tilde{x}}   \Bigl[ t_{AB} + t_{BC} +i (t_{AB} - t_{BC}) \, {\rm sinh} \tilde{x} \Bigr]\Bigr\}  
\end{eqnarray}
implying that
\begin{eqnarray} \label{tressstripprime}
\langle T(z= i{\cal W}/2) \rangle_{ABC} = \Bigl( {\pi \over {\cal W}} \Bigr)^2  \Bigl\{ {\hat{c} \over 12} + t_{AC} - 2 \bigl[ t_{AB} + t_{BC} \bigr] \Bigr\} \, . 
\end{eqnarray}
in the center $z=i{\cal W}/2$ of the vertical boundary.

(iii) $\langle T(z) \rangle_{ABC}$ is finite for all $z$ including the boundaries, except at the corners  $z=0$ and $z=i {\cal W}$, where it diverges. For $B=C$ the leading and next-to-leading terms near the corner at $z=0$, where the vertical boundary of class $B$ meets the lower horizontal boundary of class $C$, are given by
\begin{eqnarray} \label{tressstripcorner}
\langle T(z) \rangle_{ABB} \to -{\hat{c} \over 8} {1\over z^2} + \Bigl( {\pi \over {\cal W}} \Bigr)^4  \, {1 \over 4} \Bigl(- {\hat{c} \over 30} +t_{AB} \Bigr) z^2 \, .
\end{eqnarray}
For $B \neq C$ 
\begin{eqnarray} \label{tressstripcornerprime}
\langle T(z) \rangle_{ABC} \to \Bigl(-{\hat{c} \over 8} + 4 t_{BC} \Bigr) {1\over z^2} + \Bigl( {\pi \over {\cal W}} \Bigr)^2  \,  \Bigl(t_{AC} - t_{AB} - {1 \over 3} t_{BC} \Bigr) \, . 
\end{eqnarray}
Thus the next-to-leading orders are $\propto z^2$ and $\propto z^0$ in the cases $B=C$ and $B \neq C$, respectively.

(iv) For $ABC=aba$ and $bab$ the stress tensors are equal, like their counterparts in the upper half plane, and are given by 
\begin{eqnarray} \label{tressstripaba}
&&\langle T(z) \rangle_{aba} = \langle T(z) \rangle_{bab} =  \\ 
&&= \Bigl( {\pi \over {\cal W}} \Bigr)^2 \Bigl\{ {\hat{c} \over 12} \Bigl(1-{3 \over 2} {\rm coth}^2 \tilde{z}  \Bigr) - 2 \, t_{ab} \Bigl[ {1  \over 1+ {\rm cosh}\tilde{z}} + {1 \over 1- {\rm cosh}\tilde{z}} \Bigr]  \Bigr\} \, . \nonumber
\end{eqnarray}

As shown below, the expressions (\ref{tressstripprime}) and (\ref{tressstripcorner}), (\ref{tressstripcornerprime}) for the {\it stress tensor} determine the {\it order parameter} and {\it energy density} profiles in the strip near the end point $z=i {\cal W}/2$ of its midline and near its lower corner $z=0$ according to boundary and corner operator expansions, respectively. See Eqs. (\ref{product})-(\ref{productprimeprimeprime}) in Appendix \ref{BOEmid} and Sec. \ref{Yforstrip} ff., respectively.

The expressions (\ref{uniformstrip})-(\ref{tressstripcornerprime}) below are quite general and not limited to the Ising model.	In the Ising model
\begin{eqnarray} \label{IsingA}
{\cal A}_a^{(\phi)} &=& 2^{1/8}, - 2^{1/8}, 0, -1/2, -1/2, 1/2 
\end{eqnarray}
for the pairs $(\phi,a)=(\sigma,+),\,(\sigma,-),\, (\sigma,f), \,(\epsilon,+),\,(\epsilon,-),\,(\epsilon,f)$, compare Eqs. (2.20), (2.21) in Ref. \cite{BE21}, and
\begin{eqnarray} \label{Isingt}
\hat{c}=1/2 \, , \quad t_{+-}= 1/2 \, , \quad t_{+f}=t_{-f} = 1/16 \, ,  
\end{eqnarray}
see above Eqs. (4.1) in Ref. \cite{BX}.

\subsection{Energy density in the strip geometry} \label{epssistrip}

The energy density profiles in the upper half $H$ plane for the three sets of boundary conditions mentioned above are for the Ising model given by
\begin{eqnarray} \label{epssistripABC}
&&\langle \epsilon (G,J) \rangle_{-+-}=-{1 \over 2J} \Bigl( 4 {\cal C}^2 -3 \Bigr) \, ,  \nonumber \\
&&\langle \epsilon (G,J) \rangle_{f+f}= - \langle \epsilon (G,J) \rangle_{+f+}= {1 \over 2J} \, {\cal C}\, , \nonumber  \\
&&\langle \epsilon (G,J) \rangle_{-f+} = {1 \over 2J} \, {-G^2 +3J^2+1 \over Q} \, .
\end{eqnarray}
where 
\begin{eqnarray} \label{calC}
{\cal C} \equiv {G^2 +J^2-1 \over Q} \, , \quad Q \equiv\sqrt{[G^2 +J^2]^2 +1 + 2 (J^2 - G^2)} \, .
\end{eqnarray}
The first two profiles have a boundary with only two different boundary conditions and, as discussed in \cite{BXprof}, are obtained from the profiles with a single switch between them, which are given in Eq. (4.1) in Ref. \cite{BX} and in Eqs. (\ref{scalephiH}) and (\ref{fIsing}) below. The third profile with a $-f+$ boundary of three different boundary conditions follows from Eq. (2.63) of Ref. \cite{BE21}. All three energy profiles in (\ref{epssistripABC}) are even in $G$, so that their counterparts in the semi-infinite strip are symmetric about the midline, as expected. Three simple features help understand the functional form of the energy densities in the strips:

(i) The ``zero-lines'' of $\langle \epsilon \rangle$ in the $x,y$ plane along which $\langle \epsilon (x,y) \rangle$ vanishes and which separate regions with positive and negative $\langle \epsilon \rangle$, i.e. regions with a short range order that is weaker and stronger, respectively, than in the bulk \cite{primaryfields,domainwallprime}. These lines are quite different in the three cases and follow immediately from their counterparts in the $G,J$ plane. Like the switching points $H=1$ and $H=-1$, the two corners of the strip are end points of the lines.

(ii) For $x\gg {\cal W}$ the energy densities reduce to the $x$-independent profiles in the infinite strip, $\langle \epsilon (x \to \infty, y) \rangle_{ABC} \to \langle \epsilon (y) \rangle_{AC}$. The limiting densities
\begin{eqnarray} \label{epsinfstrip}
\langle \epsilon (y) \rangle_{--}=-\langle \epsilon (y) \rangle_{ff} = - {\pi \over 2 {\cal W} \sin \tilde{y}} \, , \quad \langle \epsilon (y) \rangle_{-+} = {\pi \over 2 {\cal W}}\Biggl( 4 \sin \tilde{y} -{1 \over \sin \tilde{y}} \Biggr) \,   
\end{eqnarray}
follow from Eqs. (4.1) in Ref. \cite{BX} and are all symmetric about the midline $\tilde{y} =\pi /2$, as mentioned above. While the $--$ and $ff$ profiles are negative and positive for all $y$, respectively, the $-+$ profile changes sign at $y={\cal W}/6$ and $5 {\cal W}/6$ and is positive in between.

(iii) The behavior along the midline of the strips for which Eq. (\ref{stripfromplane}) with $G=0$ yields
\begin{eqnarray} \label{epssistripmidABC}
&&\langle \epsilon (x,y={\cal W}/2) \rangle_{-+-}=-{\pi \over {\cal W}\,  \sinh (2 \tilde{x})} \, \Biggl[ \cosh^2 \tilde{x} -16 +{16 \over \cosh^2 \tilde{x}} \Biggr]  \, , \nonumber \\
&&\langle \epsilon (x,y={\cal W}/2) \rangle_{f+f}= - {\pi \over {\cal W}\,  \sinh (2 \tilde{x})} [1-\sinh^2 \tilde{x}] \, , \nonumber  \\
&&\langle \epsilon (x,y={\cal W}/2) \rangle_{-f+}= {\pi \over {\cal W}\,  \sinh (2 \tilde{x})} [1+3 \sinh^2 \tilde{x}] \, .
\end{eqnarray}
For $0<x \ll {\cal W}$ and $x \gg {\cal W}$ the three $ABC$ expressions reduce, respectively, to the behavior of $\langle \epsilon \rangle$ near an infinite vertical $B$ wall and to the value of $\langle \epsilon \rangle$ on the midline of the infinite $AC$ strip addressed above. The {\it next-to-leading} behavior for $0<x \ll {\cal W}$ is related to a boundary-operator-expansion as we discuss in Appendix \ref{BOEmid}.

The energy density is now discussed for each of the three strips in more detail. 

\subsubsection{Semi-infinite $-+-$ strip} \label{semi-+-}

In this case the conflicting tendencies of the vertical and the two horizontal boundaries to align the Ising spins up and down, respectively, lead to a remarkable distribution of order ($\langle \epsilon \rangle <0$) and disorder ($\langle \epsilon \rangle >0$) inside the strip, which is discussed in some detail.

Due to the symmetry of $\langle \epsilon \rangle$ it is sufficient to determine its zero lines in the lower half $0<y<{\cal W}/2$ of the strip corresponding to the upper right quarter of the $H$ plane. There are two zero lines, which for the strip read 
\begin{eqnarray} \label{zero-+-}
z=z_{0 \pm} (J) &\equiv& {{\cal W} \over \pi} \times {\rm arccosh}\Bigl(G_{\pm}(J) +iJ \Bigr) \, , \quad 0<J<2 \pm \sqrt{3} \; , \nonumber \\
G_{\pm}(J) &\equiv& \sqrt{1 \pm 2 J \sqrt{3}-J^2}.
\end{eqnarray}
The parametric representation (\ref{zero-+-}) arises from the corresponding zero lines $G=G_{\pm} (J)  \, , \; 0<J<2 \pm \sqrt{3}$ in the $H$ plane, seen in the first of Eqs. (\ref{epssistripABC}), which are the two circular segments addressed in Ref. \cite{BXprof}. The two lines $z=z_{0+}(J)$ and $z=z_{0-}(J)$ are upward bending curves starting at $J=0$ at the lower corner $z=z_{0 \pm}(J=0) =0$ with  tangent unit vectors $\exp (i \pi /12)$ and $\exp (5i \pi /12)$, respectively. Including, for later comparison with the COE, the next order correction near the corner, their form follows from expanding \cite{tangent} the rhs of Eq. (\ref{zero-+-}) to orders $J^{1/2}$ and $J^{3/2}$ and can be expressed as
\begin{eqnarray} \label{zero-+-prime}
{\rm arg} \bigl(z_{0+} \, ,  \, z_{0-} \bigr) \, = \, \Bigl({\pi \over 12} \, , \, {5 \pi \over 12} \Bigr) +{1 \over 24} \, \Bigl({\pi \over {\cal W}} \Bigr)^2 \bigl( |z_{0+}|^2 \, ,  \, |z_{0-}|^2 \bigr)  \, .
\end{eqnarray}

For the upper limits of $J$, where $G_{\pm}$ vanishes, the zero lines arrive at the midline of the strip at $z=z_{0 \pm}(2 \pm \sqrt{3} ) = i({\cal W}/2)+x_{\pm}$ with $x_+ =0.6454 \, {\cal W}$ and $x_- =0.0843 \, {\cal W}$. When extended to the entire strip, the region in between $z_{0 -}(J)$ and $z_{0 +}(J)$ has the shape of a waxing moon with its tips located at the two corners of the semi-infinite strip, similar to the left moon in FIG. \ref{Henkel} (a).
Inside and outside the moon, $\langle \epsilon \rangle >0$ and $\langle \epsilon \rangle <0$, respectively.

On the midline $y={\cal W}/2$ the energy density $\langle \epsilon \rangle$ is given by the first of Eqs. (\ref{epssistripmidABC}), with the expected limiting behavior $-1/(2x)$ and $-\pi /(2 {\cal W})$ for $x \ll {\cal W}$ and $x \gg {\cal W}$, respectively. In the positive region between $x_-$ and $x_+$, $\langle \epsilon \rangle$ displays a striking maximum at $x=x_{\rm m} \equiv ({\cal W}/ \pi)\times {\rm arccosh} \Bigl(2 \sqrt{(10+\sqrt{7})/31} \Bigr) \approx 0.232 \times {\cal W}$,  where $\langle \epsilon \rangle \approx 7.059 / {\cal W}$. This maximum reappears in the corresponding rectangle of Sec. \ref{+-+-} when its horizontal extension ${\cal H}$ is sufficiently large compared to its vertical width ${\cal W}$, see curve (a) in FIG. \ref{-+completehorizontalOleg} for ${\cal H}/{\cal W}=2.2$. 

From the perspective of the entire strip, this midline-maximum is actually a saddle point of the energy density. Moving at fixed $x=x_{\rm m}$ away from it in $y$ direction, $\langle \epsilon \rangle$ {\it increases} and, in the lower half of the strip, for example, reaches a maximum at the point $(x=x_{\rm m} \, , y \approx 0.22 \, {\cal W})$ where $\langle \epsilon \rangle \approx 8.881 / {\cal W}$.
This point belongs to a line inside the moon-shaped region which is the projection of the top-line of a ridge in the $\langle \epsilon \rangle$ landscape that at this point is rather broad. The ridge ascends and sharpens on decreasing $x$, and finally the line approaches the lower corner $x=y=0$ with a tangent unit vector $\exp (i \pi /4)$. There $\langle \epsilon \rangle \to 3/|z|$ diverges, reaching its maximal value, i.e., maximal disorder, in the lower half of the strip. Due to mirror symmetry about the midline, there is corresponding behavior in the upper half of the strip. In contrast, approaching the horizontal or vertical boundaries of the strip one is {\it outside} the moonlike region, and $\langle \epsilon \rangle$ tends to $-\infty$, corresponding to maximal order.

\subsubsection{Semi-infinite $f+f$ strip} \label{epsf+fstrip}

In this case the midline behavior of $\langle \epsilon \rangle$ is given by the second equation in (\ref{epssistripmidABC}) and has the limiting behavior  $-1/(2x)$ and $\pi /(2 {\cal W})$ of $\langle \epsilon \rangle$ for $x \to 0$ and $x \to \infty$. Here $\langle \epsilon (x, y={\cal W}/2) \rangle$ increases monotonically with $x$, and there is a single change of sign at $x=x_0 =({\cal W}/ \pi) \ln (1+ \sqrt{2}) \approx 0.280 \times {\cal W}$, corresponding to the single zero $J=J_0 =1$ of $\langle \epsilon (G=0, J) \rangle$. This zero is the intersection point of the midline with the zero-line  along which $\langle \epsilon (x,y) \rangle$ vanishes. The zero line is the image of the upper unit circle $H=H_0 = \sqrt{1-J^2} +iJ$ with parametric representation 
\begin{eqnarray} \label{zerof+f}
z=z_{0} (J) &\equiv& {{\cal W} \over \pi} \times {\rm arccosh} \bigl(  \sqrt{1-J^2} +iJ \bigr) \, ,\quad 0<J<1 \, ,
\end{eqnarray}
with a positive square root for the lower half of the strip. The line is an upward bending curve starting at $J=0$ from the lower corner $z=0$ with a tangent vector $\exp(i \pi /4)$. The leading and next-to-leading behavior
\begin{eqnarray} \label{zerof+fprime}
{\rm arg}z_0 \to {\pi \over 4}  + {1 \over 12} \Bigl( {\pi|z_0| \over {\cal W}} \Bigr)^2
\end{eqnarray}
of the zero line near the corner follows from the expansion \cite{tangent} of the rhs of Eq. (\ref{zerof+f}) for small $J$ to orders $J^{1/2}$ and $J^{3/2}$.

\subsubsection{Semi-infinite $-f+$ strip}

\paragraph{Midline}

The $x$-dependence of $\langle \epsilon \rangle$ along the midline is given by the third equation in (\ref{epssistripmidABC}). $\langle \epsilon \rangle$ is always positive there with limiting behavior $1/(2x)$ and $3 \pi/(2 {\cal W}) \approx 4.712 /{\cal W}$ for $x \ll {\cal W}$ and $x \gg {\cal W}$, respectively, and displays a shallow minimum $\langle \epsilon \rangle \approx 4.443 /{\cal W} $ at $x = ({\cal W}/ \pi) \ln (1+ \sqrt{2}) \approx 0.280 \times {\cal W}$. The location of the minimum in the $-f+$ strip happens to be the same as the location of the zero in the $f+f$ strip.

\paragraph{BOE at the left end of the midline}The boundary operator expansion (BOE)  \cite{Diehl},\cite{CardBoundaryCritPhen},\cite{BE21} is a useful tool to evaluate the behavior of density profiles close to a locally flat and uniform boundary, such as the vertical boundary of the semi-infinite strips. Here the BOE follows from Eq. (\ref{vertBOE}) as described in the paragraph above Eq. (\ref{product}) and involves the stress tensor $\langle T(i {\cal W}/2) \rangle$ given in Eq. (\ref{tressstripprime}). With $t_{AC}=1/2$, $t_{AB}=t_{BC}=1/16$ and $\hat{c}=1/2$ the BOE predicts that $\langle \epsilon (x,{\cal W}/2) \rangle \to (2x)^{-1} [1+(7/3)\tilde{x}^2]$ which is in agreement with the expression in the third of Eqs. (\ref{epssistripABC}).  

\paragraph{Zero lines}

There are two zero-lines. One,
\begin{eqnarray} \label{zero-f+}
z=z_{\rm lower}(J) \equiv {{\cal W} \over \pi} \times {\rm arccosh}\bigl( \sqrt{3J^2 +1} +iJ \bigr) \, , \quad 0<J<\infty \, ,
\end{eqnarray}
starts from the corner $z=0$ at $J=0$ with a tangent vector $\exp(i \pi /4)$ and, on increasing  monotonically, reaches the horizontal line $z=i {\cal W}/6$ for $J \to \infty$, which is the lower zero-line in the {\it infinite} $-+$ strip. The other one, $z=z_{\rm upper}(J)$, is its mirror image wrt the midline. In between and outside the two zero-lines, $\langle \epsilon \rangle$ is positive and negative, respectively.

\subsection{Order parameter densities in the strip}

The counterparts of Eqs. (\ref{epssistripABC}) for the order parameter are 
\begin{eqnarray} \label{sigsistripABC}
&&\langle \sigma (G,J) \rangle_{-+-}=-\Biggl({2 \over J}\Biggr)^{1/8}\,  {\cal C} \, , \nonumber \\
&&\langle \sigma (G,J) \rangle_{f+f}=\Biggl({2 \over J}\Biggr)^{1/8} \, 2^{-1/4} \, \Bigl( 1- {\cal C} \Bigr)^{1/4} \, , \nonumber \\
&&\langle \sigma (G,J) \rangle_{-f+}=\Biggl({2 \over J}\Biggr)^{1/8} \, 2^{-1/4} \, \Bigl[ \bigl(1+ {\cal C}\bigr)^{1/2} - J \bigl(1- {\cal C}\bigr)^{1/2} \Bigr]^{1/2}\, .
\end{eqnarray}
Here ${\cal C}$ is the quantity considered in Ref. \cite{BXprof} and given in Eq. (\ref{calC}). Obviously the line in the semi-infinite strip along which $\langle \sigma (x,y) \rangle_{-+-}$ vanishes equals the line of $\langle \epsilon (x,y) \rangle_{f+f}$ discussed in Eq. (\ref{zerof+f}) above. The zero-lines of $\langle \sigma \rangle_{f+f}$ in the upper half $H$ plane and the semiinfinite strip are the two $f$ boundaries which correspond to $\cos \theta =1$ of the single switch profile $\langle \sigma \rangle_{+f}$ in the $z_{\rm BX}$ plane, see Refs. \cite{BX,BXprof}.

\section{DENSITY PROFILES IN RECTANGLES WITH A MIXED BOUNDARY} \label{recmix}

Next we consider a critical system defined on the rectangular domain $-{\cal H}/2<x_{\rm M}<{\cal H}/2, \, -{\cal W}/2<y_{\rm M}<{\cal W}/2$ centered about the origin of the $z_{\rm M}=x_{\rm M}+i y_{\rm M}$ plane and discuss the profiles $\langle \phi(x_{\rm M}, y_{\rm M}) \rangle$ of the energy density $\phi=\epsilon$ and the order parameter $\phi=\sigma$ for the boundary conditions $+,-$ or $f$ on the four sides of the rectangle. 

Of primary interest is the dependence on the aspect ratio ${\cal H}/{\cal W}$. While for ${\cal H} \gg {\cal W}$ or ${\cal H} \ll {\cal W}$ the behavior corresponds to adjacent semi-infinite strips, new features arise for ${\cal H}/{\cal W}$ of order one. This is true, in particular, for the topology of the zero-lines, the behavior along the midlines $y_{\rm M}=0$ and $x_{\rm M}=0$ of the rectangle, including its center, and near the corners. 

In the following denote, in counterclockwise order, the NE, NW, SW, and SE corners of the rectangle at $({\cal H}+i{\cal W})/2, \, (-{\cal H}+i{\cal W})/2, \, (-{\cal H}-i{\cal W})/2$, and $({\cal H}-i{\cal W})/2$ by I, II, III, and IV. Likewise the four corresponding quarters of the rectangle inside which the coordinates  $(x_{\rm M}, \, y_{\rm M})$ are (positive, positive), (negative, positive), (negative, negative), and (positive, negative) are denoted by i, ii, iii, and iv.  

Using conformal invariance at the critical point \cite{CardBoundaryCritPhen} the profiles $\langle \phi(x_{\rm M},y_{\rm M}) \rangle$ in the rectangle can be evaluated from the known profiles $\langle \phi(G,J) \rangle$ in the upper half $H=G+iJ$ plane with appropriate boundary conditions \cite{BE21}. The corresponding conformal transformation proceeds through the intermediate geometry of the circular unit-disk \cite{facildisk}, as presented in Appendix \ref{RectMcircH}. It maps the boundary of the rectangle to the real axis  $H=G$, the centers of the N, W, S, and E sides of the rectangle at $ i{\cal W}/2, \, -{\cal H}/2, \, -i{\cal W}/2$, and ${\cal H}/2,$ to the points 
\begin{eqnarray} \label{sidecenter}
H= -1, \, 0, \, 1 \, , \infty 
\end{eqnarray}
and the corners I, II, III, IV to the points  
\begin{eqnarray} \label{GK}
G=G_{\rm I} \equiv - t^{-1}, \, G_{\rm II} \equiv - t, \, G_{\rm III} \equiv  t, \, G_{\rm IV} \equiv  t^{-1} \, , \quad 0<t<1  \, , \quad t \equiv \tan (\alpha /2) \, ,
\end{eqnarray}
which are denoted by (I), (II), (III), and (IV). The angle $\alpha$ is related to the aspect ratio of the rectangle by
\begin{eqnarray} \label{aspratios}
{{\cal H} \over {\cal W}} ={{\bf K}(\cos \alpha) \over {\bf K}(\sin \alpha)} \, , \quad 0< \alpha < \pi /2\, ,
\end{eqnarray}
where ${\bf K}$ is the complete elliptic integral of the first kind.  Thus $\alpha = \pi /4$  for the square ${\cal H}={\cal W}$, and $\alpha \to 0$ and $\alpha \to \pi /2$ for ${\cal H} \gg {\cal W}$ and ${\cal H} \ll {\cal W}$, respectively. 

Sometimes it is convenient to use the alternative notations \cite{confuse}
\begin{eqnarray} \label{alternative}
q \equiv \cos \alpha \, , \qquad q' \equiv \sin \alpha \, , \quad s \equiv \sin \alpha \, , \quad S \equiv \sin^2 \alpha \, 
\end{eqnarray}
to characterize the aspect ratio, see, e.g.,  Eqs. (\ref{aspq}), (\ref{Lgeo}), and (\ref{qqprime}), as well as the present section.	Eq. (\ref{aspratios}) implies in particular that $s \to  4 \exp [- (\pi /2) {\cal H}/{\cal W}]$ for ${\cal H} \gg {\cal W}$.

The horizontal and vertical midlines $y_{\rm M}=0$ and $x_{\rm M}=0$ of the rectangle are mapped to the imaginary axis $H=iJ$ with $0<J< \infty$ and to the upper half unit circle $H=i\exp(i \psi)$ with $- \pi /2 < \psi < \pi /2$, respectively, of the upper half $H$ plane, cf. the discussion above Eq. (\ref{invhalfrect}). Thus the four quarters i, ii, iii, and iv of the rectangle are mapped onto the four half-plane regions ``left of the imaginary axis and outside the unit circle'',  ``left of the imaginary axis and inside the unit circle'', ``right of the imaginary axis and inside the unit circle'', and ``right of the imaginary axis and outside the unit circle'', respectively. These regions are grouped in a counterclockwise manner around the point $H=i$, the image of the rectangle's center $z_{\rm M}=0$, and we denote them by (i), (ii), (iii), and (iv), respectively. Obviously the points (I), (II), (III), and (IV) lie on the boundaries of the regions (i), (ii), (iii), and (iv), respectively.

The simple relation
\begin{eqnarray} \label{relatecenter}
\langle \phi(x_{\rm M}=0,\, y_{\rm M}=0) \rangle = \Biggl( {2 \over \Lambda} \Biggr)^{x_{\phi}} \langle \phi(G=0, \, J=1) \rangle \, 
\end{eqnarray}
between the profile values at the {\it center} $z_{\rm M}=0$ of the rectangle and at $H=i$ in the upper half plane follows from the transformations (\ref{unitcircrect}), (\ref{ZMtozM}), and (\ref{Moebprime}). Here   the length $\Lambda$ depends on the size and aspect ratio of the rectangle, as defined in (\ref{Lgeo}), which in the notation (\ref{alternative}) reads 
\begin{eqnarray} \label{Lambdaprime}
{1 \over \Lambda} \equiv \Biggl( {{\bf K}(\cos \alpha) \over {\cal H}} \, {{\bf K}(\sin \alpha) \over {\cal W}} \Biggr)^{1/2}  \equiv {{\bf K}(\cos \alpha) \over {\cal H}} \equiv {{\bf K}(\sin \alpha) \over {\cal W}}    \, .
\end{eqnarray}
For later use note that
\begin{eqnarray} \label{uniformmid}
\langle \phi(x_{\rm M}=0,\, y_{\rm M}=0) \rangle_a = \Biggl( {2 \over \Lambda} \Biggr)^{x_{\phi}} \times {\cal A}_a^{(\phi)}
\end{eqnarray}
for rectangles with a {\it uniform} boundary condition $a$. This follows from the form $\langle \phi(G, \, J) \rangle_a = {\cal A}_a^{(\phi)} \, J^{- x_{\phi}} $ of the profiles in the half plane with a uniform boundary condition $a$. For the values of ${\cal A}_a^{(\phi)}$ in the Ising model see Eq. (\ref{IsingA}).

Assume uniform boundary conditions for each of the  N, W, S, and E sides, i.e., of the top, left, bottom, and right sides of the rectangle which are denoted by $A$, $B$, $C$, and $D$, respectively, which is in line with the notation below Eq. (\ref{generalstrip}) for semi-infinite strips. Thus the corresponding boundary conditions in the upper half $H$ plane are $D,A,B,C,D$ for $- \infty < G < G_{\rm I}, \, G_{\rm I} < G < G_{\rm II}, \, G_{\rm II} < G < G_{\rm III}, \, G_{\rm III} < G < G_{\rm IV},\, G_{\rm IV} <G < \infty$, respectively. In the following rectangles are considered where the two vertical boundaries have the same boundary condition $a$ and the two horizontal boundaries have the same boundary condition $b$, so that $ABCD=baba$ \cite{moregenbc} and in the upper half plane the boundary conditions are $ababa$. 

In the two limits $\alpha = 0$ ($t=0$) and $\alpha = \pi /2$ ($t=1$) of infinite horizontal and vertical strips, this implies uniform boundary conditions $b$ and $a$, respectively, in the upper half plane, so that the rectangle profiles reduce to profiles in the infinite strip with $b$ and $a$ boundary conditions, respectively. In particular, $\langle \phi(G=0, \, J=1) \rangle$ reduces to ${\cal A}_b^{(\phi)}$ and ${\cal A}_a^{(\phi)}$, respectively, which is consistent with the profile value (\ref{relatecenter}) at the center reducing to the corresponding profile value (\ref{uniformmid}) on the midline of the infinite strip. The leading correction to the infinite strip value follows from viewing the rectangle as two semi-infinite strips stiched together. For finite ${\cal W}$ and ${\cal H} \to \infty$, where $\alpha \to 0$, it corresponds to two semi-infinite $bab$ strips, and the correction is given by 
\begin{eqnarray} \label{stiched}
&&\langle \phi(x_{\rm M}=0,\, y_{\rm M}=0) \rangle - \Biggl( {\pi \over {\cal W}} \Biggr)^{x_{\phi}} {\cal A}_b^{(\phi)} \to \nonumber \\
&& \to 2 \Biggl[\langle \phi(x = {\cal H}/2, \, y= {\cal W}/2 ) \rangle_{bab}  - \Biggl( {\pi \over {\cal W}} \Biggr)^{x_{\phi}} {\cal A}_b^{(\phi)} \Biggr] \, , \quad {\cal H} \gg {\cal W} \, .
\end{eqnarray}
For increasing ${\cal H}/{\cal W}$ the expression on the rhs of (\ref{stiched}) decays exponentially to zero. For example, for $a=b$ it is given by $\bigl( \pi / {\cal W} \bigr)^{x_{\phi}} {\cal A}_b^{(\phi)} \times 4 x_{\phi} \exp [- \pi {\cal H}/{\cal W}]$, see (\ref{uniformstrip}), and for $\phi=\sigma, \, a=+, \, b=f$, where ${\cal A}_b^{(\phi)} \equiv {\cal A}_f^{(\sigma)}$ vanishes, by $\bigl( \pi / {\cal W} \bigr)^{1/8} 2^{13/8} \exp [- (\pi /4) {\cal H}/{\cal W}]$ which follows from Eq. (\ref{sigsistripABC}) together with $1-{\cal C}=2/(\sinh^2 \tilde{x}+1)$. These results follow directly from the expressions for the rectangle on the lhs of (\ref{stiched}), the first one via Eq.  (\ref{uniformmid}) and the second one via Eq. (\ref{sig+f+f+center}) below, taking into account the exponential decay of $s$ given below Eq. (\ref{alternative}) and the corresponding behavior of $1/\Lambda$ determined by Eq. (\ref{Lambdaprime}).   
 
The density profiles $\langle \phi(x_{\rm M}, \, y_{\rm M}) \rangle$ of the rectangle and the profiles $\langle \phi(u, \, v) \rangle$ in the intermediate geometry of the unit disk introduced in Appendix \ref{RectMcircH} are mirror-symmetric about the coordinate axes \cite{facildisk}, since the above-mentioned boundary conditions have this symmetry. In the upper half $H$ plane, by the same argument the profiles are mirror symmetric about the vertical coordinate axis. Moreover, since the conformal transformation $H \to -1/H \equiv -\bar{H}/|H|^2$ maps the upper half plane with the boundary conditions onto itself, the profiles must be reproduced by this transformation. Thus
\begin{eqnarray} \label{invH}
\langle \phi (G,J) \rangle = \langle \phi (-G,J) \rangle = {1 \over (G^2 +J^2)^{x_{\phi}}} \times \Bigg\langle \phi \Biggl(G \to {G \over G^2 +J^2}, \, J \to {J \over G^2 +J^2}\Biggr) \Bigg\rangle \, .
\end{eqnarray}

\subsection{Rectangle with horizontal $-$ and vertical $+$ boundaries} \label{+-+-}

Here consider a rectangle with boundary condition $a=+$ on the two vertical edges and $b=-$ on the two horizontal edges, i.e. $ABCD=-+-+$ in the above notation. This corresponds to an upper half plane with boundary conditions $+-+-+$ in the intervals (\ref{GK}). 

\subsubsection{Energy density} \label{eps+-+-}

The energy density $\langle \epsilon (G,J) \rangle$ follows from Eqs. (16) in Ref. \cite{TWBG2} and can be written as 
\begin{eqnarray} \label{epsrec+-+-+}
\langle \epsilon (G,J) \rangle = - {1 \over 2J} \Biggl\{ 1-{4J^2 \over P} \Biggl[ {16 \, G^2 \over M(t) \, M(-t) \, M(t^{-1}) \, M(-t^{-1})} + \nonumber \\ 
+ \Biggl( {t^2 \over M(t) \, M(-t)} + {t^{-2} \over M(t^{-1}) \, M(-t^{-1})} \Biggr) \Biggr] \Biggr\}
\end{eqnarray}
where  
\begin{eqnarray} \label{epsrec+-+-+prime}
M(\tau)=(G-\tau)^2 + J^2 \, , \;  P= {4 \over (t^2 -t^{-2})^2}+{1\over 4} = {1 \over 4} \Bigl[{1 \over 1-S} - S \Bigr]
\end{eqnarray}
where $P$ is the corresponding form of the Pfaffian ${\rm Pf}^{(4)} \zeta_{ij}^{-1}$ in \cite{TWBG2}. For $t = 0$ and $t=1$, corresponding to uniform $-$ and $+$ boundaries, the rhs of Eq. (\ref{epsrec+-+-+}) reduces to $-1/(2J)$, as expected. On approaching a switching point, say $(G,J) \to (t,0)$, Eq. (\ref{epsrec+-+-+}), (\ref{epsrec+-+-+prime}) yields $\langle \epsilon (G,J) \rangle \to -  \{ 1-4J^2 /M(t) \} /( 2J) = -  \{ 1-4 \sin^2 \theta \} /( 2J)$, consistent with expression (4.1) of Ref. \cite{BX} for a single $+-$ switch at $G=t$. Here $\theta$ is the angle $G-t+iJ$ forms with the positive real axis. 

\paragraph{Horizontal and vertical midline of the circular unit-disk}

According to Appendix \ref{RectMcircH} the horizontal midline $y_{\rm M}=0$ corresponds to $G=0$, for which Eq. (\ref{epsrec+-+-+}) implies
\begin{eqnarray} \label{epsrecmid+-+-+}
\langle \epsilon (G=0,J) \rangle = -{1 \over 2J} \Bigl\{ 1- {4 \over P} \, {\Theta (J^2 +J^{-2}) +4 \over (\Theta +J^2 +J^{-2})^2} \Bigr\} \, , \quad \Theta = t^2 +t^{-2} \equiv (4/S)-2 \, .
\end{eqnarray}
On using the transformation formula $\langle \epsilon (u,v=0) \rangle = 2 (1-u)^{-2} \, \langle \epsilon (G=0,J) \rangle$, $J=(1+u)/(1-u)$ from (\ref{Moebprime}) that maps $(G=0,J)$ to the real axis $w=u$ inside the circular unit-disk in the $w=u+iv$ plane, Eq. (\ref{epsrecmid+-+-+}) then yields
\begin{eqnarray} \label{epscirc+-+-+mid}
&&\langle \epsilon (u,0) \rangle = \, :e(u,S) \, ,  \\
&&e(u,S)=  - {1 \over 1-u^2} \Bigl\{ 1-16 {S(1-S) \over 1-S(1-S)}  \, (1-u^2)^2 \, {(2-S) (1+6u^2 +u^4) +S (1-u^2)^2 \over \big[ S(1+6u^2 +u^4)+(2-S) (1-u^2)^2 \big]^2} \Bigr\} \nonumber
\end{eqnarray}

In addition to the general mirror symmetry about the coordinate axes for fixed $S$ considered in between Eqs. (\ref{uniformmid}) and (\ref{invH}) above,  $\langle \epsilon (u,v) \rangle$ is invariant on replacing  
\begin{eqnarray} \label{epsaddsymm}
(u,v;S) \to (v,u;1-S) \, ,
\end{eqnarray}
which for $u=0$ implies that 
\begin{eqnarray} \label{epsaddsymmprime}
\langle \epsilon (0,v) \rangle = e(v,1-S) \, ,
\end{eqnarray}
so that Eq. (\ref{epscirc+-+-+mid}) not only determines the energy density along the {\it horizontal} but also along the {\it vertical} midline. For the corner images in the disk given below Eq. (\ref{qqprime}) the disks characterized by $S$ and $1-S$, i.e. by $\alpha$ and $(\pi /2) - \alpha$, follow from each other simply by a rotation of 90 degrees together with an exchange of boundary conditions + and $-$, which has {\it no} effect on the profile of the energy density. 

In the case of a disk with $+f+f+$ boundary condition that is considered below in subsection \ref{+f+f}, exchanging  $+$ and $f$ has a {\it nontrivial} effect on $\langle \epsilon \rangle$ --- see the discussion in \cite{fweak} --- and a simple relation such as (\ref{epsaddsymm}) does not exist. As a consequence there is no simple relation between the corresponding two midline expressions (\ref{epshoriu}) and (\ref{epsvertv}). 

\paragraph{Center values, special aspect ratios, and zero lines} \label{aspandzerolines}

In an expansion about the center of the unit disk a second order term $\propto uv$ is absent due to the mirror symmetries, and Eqs. (\ref{epscirc+-+-+mid}) and (\ref{epsaddsymmprime}) imply
\begin{eqnarray} \label{epsaddcircsymm1}
\langle \epsilon (u,v) \rangle \to e_{0}(S)+u^2 \, e_{2}(S) + v^2 \, e_{2}(1-S) + O (|u+iv|^4) \, 
\end{eqnarray}
where
\begin{eqnarray} \label{epsaddcircsymm1a}
e_{0}(S) &=&  -  {1-9 S(1-S) \over 1- S(1-S)}  \equiv  -  {1-9 \sin^2 \alpha \, \cos^2 \alpha \over 1- \sin^2 \alpha \, \cos^2 \alpha } \, ,  \nonumber \\
e_{2}(S) &=& - \Bigl[  1-8 {S(1-S) \over 1-S(1-S)} (9-12S)  \Bigr]
\end{eqnarray}
are the coefficients in the expansion $e(u,S) \to e_0 (S) +u^2 e_2 (S) + ...$ of $e(u,S)$ in (\ref{epscirc+-+-+mid}) for small $u$. 

The corresponding expression for the rectangle in the $z_{\rm M}$ plane follows from the transformation (\ref{primarytrafo}), (\ref{unitcircrect}), (\ref{ZMtozM}) and reads  
\begin{eqnarray} \label{epsaddcircsymm3}
&& \Lambda \langle \epsilon (x_{\rm M},y_{\rm M}) \rangle \to e_{0}(S)+(x_{\rm M}/\Lambda)^2 \, \bigl[e_{2}(S) + (2S-1)e_{0}(S) \bigr] \,+ \nonumber \\
&& \qquad \qquad +(y_{\rm M}/\Lambda)^2 \,\bigl[e_{2}(1-S) + (1-2S)e_{0}(S)  \bigr]  + O\bigl( (|x_{\rm M}+iy_{\rm M}|/\Lambda)^4 \bigr) \, ,
\end{eqnarray}
where $\Lambda$ is from Eq. (\ref{Lambdaprime}) and the second terms in the square brackets arise from the rescaling factor $|dw/dZ_{\rm M}|$. 

As expected the expression for the energy density at the center of the rectangle,
\begin{eqnarray} \label{epsaddreccent}
\langle \epsilon (x_{\rm M}=0, y_{\rm M}=0) \rangle = e_0 (S) /\Lambda \, ,
\end{eqnarray}
is invariant under $\alpha \to (\pi /2)- \alpha$, i.e., under $S \to 1-S$ which exchanges the values of $\cal H$ and $\cal W$. It vanishes for the two values $S= S_> \equiv (1/2) + (\sqrt{5}/6)$ and $S = S_< \equiv (1/2) - (\sqrt{5}/6)$, which correspond to the aspect ratios ${\cal W}=1.5172 \, {\cal H}$ and ${\cal H}= 1.5172 \, {\cal W}$, respectively \cite{cancellation}. For the $S$-interval in between, $\langle \epsilon (x_{\rm M}=0, y_{\rm M}=0) \rangle$ is positive while it is negative outside the interval. In particular, for the square ${\cal H}={\cal W}$, where $S=1/2$, it equals
\begin{eqnarray} \label{epsaddreccenter}
\langle \epsilon (x_{\rm M}=0, y_{\rm M}=0) \rangle = (5/3) {\bf K}(1/\sqrt{2}) /{\cal H} = 3.0902 /{\cal H} \, , \quad  {\cal H}={\cal W} \, ,
\end{eqnarray}
and for $S =0$ and $S=1$, where ${\cal H}/{\cal W} \to \infty$ and $\to 0$, it reproduces the  values $-\pi /(2{\cal W})$ and $-\pi /(2{\cal H})$, respectively, for the infinite strip. The different signs arise from the difference how the proximities of the center to the ordering sides compare with those to the disordering corners of the rectangle.

\begin{figure}
	\begin{center}
		\includegraphics[width=0.4\textwidth]{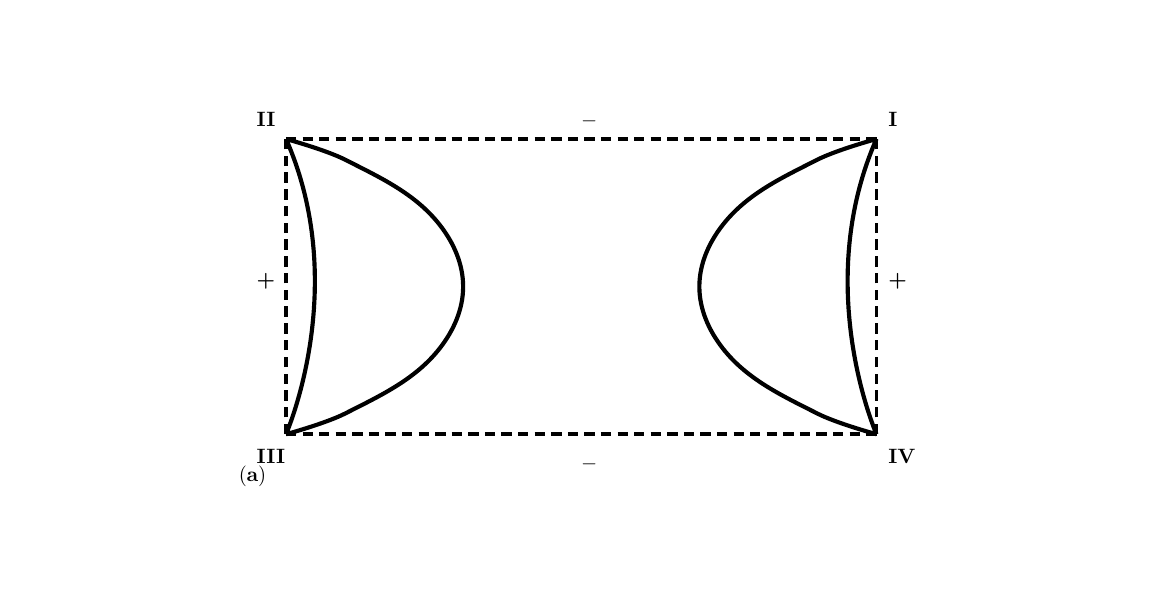}
		\includegraphics[width=0.24\textwidth]{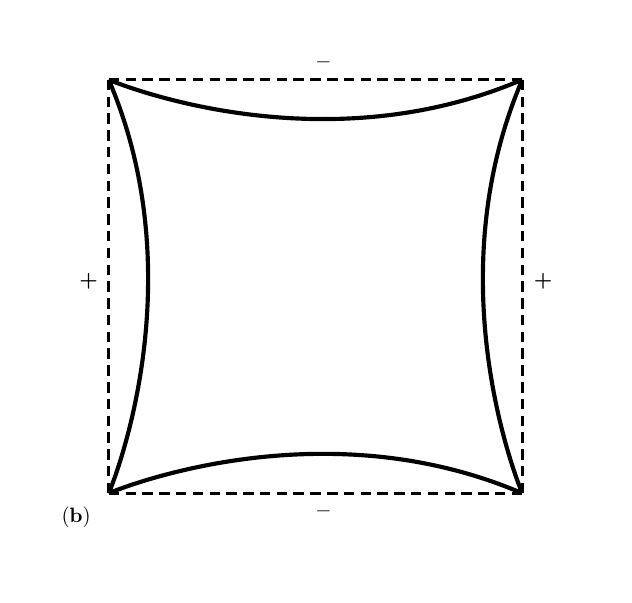}\qquad
		\includegraphics[width=0.24\textwidth]{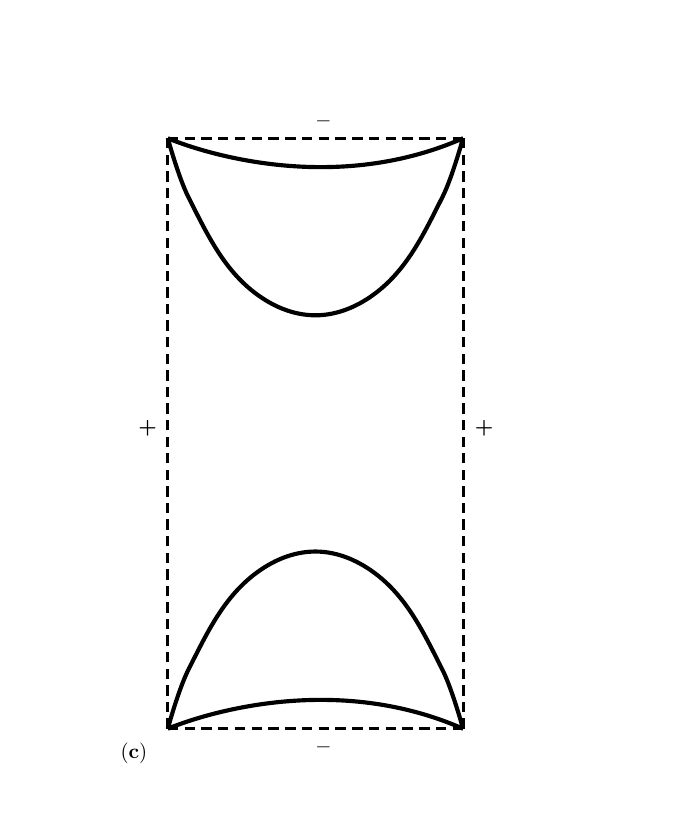}\\
	\end{center}
	\caption{Zero lines of $\langle \epsilon \rangle$ in the rectangle with boundary conditions $-$ and  $+$ on the horizontal and vertical sides with lengths ${\cal H}$ and ${\cal W}$, respectively. The lines separate regions of positive and negative $\langle \epsilon \rangle$ induced by the disordering $-+$ corners and the ordering $-$ or $+$ sides, respectively. Results for the aspect ratios ${\cal H}/{\cal W} = 2.2, \, 1.0, \, 1/2.2$ are shown in (a), (b), (c). In (a) and (c) the energy density is positive inside the two moon-like regions while in the three regions outside the moons, one of which includes the rectangle's center, it is negative. For the square in (b), however, the energy density is positive in the region that includes the center. 
	As described in points (1)-(3) of Sec. \ref{aspandzerolines}, the topology of the lines changes at the two aspect ratios ${\cal H}/{\cal W}=1.5172$ and $1/1.5172$ for which two of the zero lines form an intersection at the center where $\langle \epsilon \rangle$ vanishes. On moving from (b) to (a) on increasing ${\cal H}/{\cal W}$ from 1.0 to 2.2, the upper and lower regions of negative $\langle \epsilon \rangle$ in (b) approach each other and coagulate at ${\cal H}/{\cal W}=1.5172$ to finally form in (a) the central region of negative $\langle \epsilon \rangle$. At the same time the central region of positive $\langle \epsilon \rangle$ in (b) splits into a left and right region that form in (a) the two moon-like regions of positive $\langle \epsilon \rangle$. The situation near the coagulation point ${\cal H}/{\cal W}=1.5172$ is detailed in the paragraph containing Eq. (\ref{epsaddcircsymm4}). In particular, right at ${\cal H}/{\cal W}=1.5172$ the opening angle of the upper and lower regions of negative $\langle \epsilon \rangle$ at the intersection equals $0.352 \, \pi$. The present coagulation process from (b) to (a) should be compared with its counterpart in FIG. \ref{Ted} where there are only two zero lines, where on moving from (a) to (e) the right and left regions of negative $\langle \epsilon \rangle$ coagulate to form the central region, and where their opening angle at the intersection in (c) equals $\pi /3$.}\label{Henkel}
\end{figure}

Next consider the {\it zero lines}, separating regions of positive and negative $\langle \epsilon (x_{\rm M}, y_{\rm M}) \rangle$, the sources and sinks of which are the corners of the rectangle. {\it Two} zero lines (with an asymptotically enclosed angle of $\pi /3$) leave from each of our $+-$ corners (see Sec. \ref{CEOzeroline} and FIG. \ref{Henkel}) and arrive at one or two of the other corners, so there are four zero lines in the rectangle. Their topology must change at the borders of the abovementioned interval $S_> > S > S_<$, i.e., $1/1.5172 < {\cal H}/{\cal W} < 1.5172$, since for aspect ratios inside the interval the center of the rectangle is the center of a region of positive $\langle \epsilon \rangle$ while outside the interval it is the center of a region of negative $\langle \epsilon \rangle$. We now describe the corresponding topologies of the lines: 

(1) For aspect ratios inside the interval the four zero lines with obvious notation $L_{\rm I,II}$, $L_{\rm II,III}$, $L_{\rm III,IV}$, and $L_{\rm IV,I}$ connect, without intersecting, corners I with II, II with III, III with IV, and IV with I, enclosing a region where $\langle \epsilon \rangle$ is positive and which includes the rectangle's center. For the square with $S=1/2$, in particular, this region has the symmetry of the square, and its smallest diameter is along the coordinate axes with a size of $0.3963 \times 2 \times ({\cal H}={\cal W})$, as argued in the paragraph above Eq. (\ref{zerosem}) and shown in FIG. \ref{Henkel} (b). This should be compared with the central region of the square with $+f$ corners discussed in (1) of Sec. \ref{competitionf+} which has a different shape, cf. FIG. \ref{Ted} (e), and where $\langle \epsilon \rangle$ is negative, and the present center value (\ref{epsaddreccenter}) is to be compared with (\ref{ep+f+f+00SQ}).

(2) On decreasing $S$ from $S_>$ down to $0$, i.e., increasing ${\cal H}/{\cal W}$ from $1/1.5172$ up to $\infty$, the lines $L_{\rm II,III}$ and $L_{\rm IV,I}$ persist all the way, with their crossing points with the horizontal midline remaining at a finite distance from the center. However, the lines $L_{\rm I,II}$ and $L_{\rm III,IV}$ do not persist beyond the value $S = S_<$, i.e. beyond ${\cal H}/{\cal W} = 1.517$. Approaching this value, their crossing points with the vertical midline move to the center so that right at $S = S_<$ the two lines combine to form two lines $L_{\rm I,III}$ and $L_{\rm II,IV}$ that connect corners I with III and II with IV and intersect with a finite angle at the center. The value of this angle is determined by
\begin{eqnarray} \label{epsaddcircsymm4}
&& \Lambda \langle \epsilon (x_{\rm M},y_{\rm M}) \rangle \to (x_{\rm M}/\Lambda)^2 \, e_{2}\Bigl( (1/2)-(\sqrt{5}/6) \Bigr)   \,+ (y_{\rm M}/\Lambda)^2 \,e_{2} \Bigl( (1/2)+(\sqrt{5}/6) \Bigr) \, = \nonumber \\
&& \qquad \qquad \qquad =2 \Bigl[ (x_{\rm M}/\Lambda)^2 \, \bigl(\sqrt{5}+1 \bigr)   \,-  (y_{\rm M}/\Lambda)^2\, \bigl(\sqrt{5}-1 \bigr) \Bigr] \, ,
\end{eqnarray}
which follows from Eqs. (\ref{epsaddcircsymm1a}), (\ref{epsaddcircsymm3}) for our case $e_0 =0$. Thus the intersection separates upper and lower regions with negative $\langle \epsilon \rangle$ from left and right regions with positive $\langle \epsilon \rangle$, and the opening angle $2 {\rm arctan}\sqrt{(\sqrt{5}-1)/(\sqrt{5}+1)} = 0.352 \, \pi$ of the upper and lower regions is {\it smaller} than that of the left and right regions. This qualitative difference from the {\it larger} opening angle of the upper and lower regions at the crossing in FIG. \ref{Ted} (c), corresponding to paragraph (2) of Sec. \ref{competitionf+}, is not surprising, since here the zero lines leave the corners on enclosing the small angle $\pi /12$ with the horizontal sides of the rectangle while in case of FIG. \ref{Ted} it is $\pi /4$ (and there are only two zero lines). 

(3) On decreasing $S$ beyond $S_<$, i.e., increasing ${\cal H}/{\cal W}$ beyond $1.5172$ the intersecting lines $L_{\rm I,III}$ and $L_{\rm II,IV}$ split into two lines $L_{\rm II,III}'$ and $L_{\rm IV,I}'$ which together with the lines $L_{\rm II,III}$ and $L_{\rm IV,I}$ form the shape of a crescent moon and its reflected partner with tips at the corner-pairs II,III and IV,I, respectively, see FIG. \ref{Henkel} (a). For $S \to 0$, i.e., ${\cal H}/{\cal W} \to \infty$, the former one reduces to the crescent moon in the $-+-$ semiinfinite horizontal strip addressed below Eq. (\ref{zero-+-}). Inside the two moons the energy density is positive, while in the region between $L_{\rm II,III}'$ and $L_{\rm IV,I}'$, which includes the center and the entire vertical midline of the rectangle, it is negative.

(4) The development of the zero lines on increasing $S$ within the interval $S_< < S < 1$ follows easily from their development on decreasing $S$ within the interval $S_> > S > 0$ described above since the consequence of $S  \to 1-S$ for the shapes of the rectangle and its zero lines is a mere rotation by 90 degrees. Thus in the new interval it is the lines $L_{\rm I,II}$ and $L_{\rm III,IV}$ that persist all the way and in its sub-interval $S_> < S < 1$ the lines  $L_{\rm II,III}$ and $L_{\rm IV,I}$ are changed to lines $L_{\rm I,II}'$ and $L_{\rm III,IV}'$ enclosing a region of negative $\langle \epsilon \rangle$ that contains the center and the horizontal midline of the rectangle, see FIG. \ref{Henkel} (b) and (c). 

Finally consider the behavior of the zero lines near the corners as predicted by Section \ref{CEOzeroline} on the basis of the corner-operator expansion. Due to symmetry it is sufficient to discuss the behavior near the SW corner III where the tangent vectors of the two zero lines form the angles $\pi /12$ and $5 \pi /12$ with the lower horizontal boundary. Since for $1/2 > S > 0$  the average $\langle {\cal Y} \rangle$ of the corner operator in Eq. (\ref{opdiffWEDGE}) is $< 0$, cf. Eq. (\ref{rect-+-+}) for $q>q'$, the relation (\ref{zero_-+}) implies that the two zero lines bend away from their tangents in counterclock direction 
for all ${\cal H} > {\cal W}$. In particular, for $1/2 > S > S_<$, i.e. $1<{\cal H}/{\cal W} < 1.5172$, this implies that the lines $L_{\rm I,II}$ and $L_{\rm III,IV}$ in (1) have two inflection points, i.e., resemble a saucer and a bell, respectively. 

For the square ${\cal H}/{\cal W} =1$, i.e. $S=1/2$, there is no violation of the symmetry about its diagonal since  $\langle {\cal Y} \rangle$ vanishes and the bending away from the tangents is of higher order, not described by the COE, cf. point (ii) in Sec \ref{SQmixed}.   

\paragraph{Complete behavior along the midlines of the rectangle} \label{comp-+} 

Due to the invariance (\ref{epsaddsymm}), (\ref{epsaddsymmprime}), the value of $\langle \epsilon (x_{\rm M}=0 , y_{\rm M}= \xi ) \rangle$ on the vertical midline equals $\langle \epsilon (x_{\rm M}=\xi , y_{\rm M}= 0 ) \rangle$ on the horizontal midline on interchanging the magnitudes of ${\cal H}$ and ${\cal W}$ so that $S \to 1-S$. 

The complete dependence of $\langle \epsilon \rangle$ along the horizontal midline of the rectangle is conveniently represented in terms of the scale-invariant function 
\begin{eqnarray} \label{scinvhoreps}
{\cal E}({\cal X},S) \equiv  \bigl({\cal H} \times {\cal W}\bigr)^{1/2} \, \langle \epsilon (x_{\rm M}= {\cal X} \, {\cal H}  \, , \, y_{\rm M}=0) \rangle   \, , \quad -1/2 < {\cal X} < 1/2  \, ,
\end{eqnarray}
which depends on $S$, i.e., on the aspect ratio ${\cal H}/{\cal W}$ of the rectangle, but not on its size. Applying the above invariance to the rectangle's center yields ${\cal E}(0,S)={\cal E}(0,1-S)$. The form of $\cal E$ follows from substituting Eq. (\ref{epscirc+-+-+mid}) in the transformation formula (\ref{primarytrafo}) and using the relations (\ref{unitcircrect}), (\ref{ZMtozM}), as well as (\ref{alternative}) and (\ref{Lambdaprime}), and reads 
\begin{eqnarray} \label{scinvhorepsprime}
&&{\cal E}({\cal X},S) =  \Bigl( {\bf K}(\sqrt{1-S}) \,  {\bf K}(\sqrt{S}) \Bigr)^{1/2} \, |du/dX_{\rm M}| \, e(u,S) \, , \nonumber \\
&& u=u \bigl( X_{\rm M}, q(S) \bigr) \, , \quad X_{\rm M} \equiv x_{\rm M}/ \Lambda ={\cal X} \times {\bf K}(\sqrt{1-S}) \, .
\end{eqnarray}
In FIG. \ref{-+completehorizontalOleg} (a) and (c) explicit results for the ${\cal X}$-dependence of ${\cal E}$ are plotted for $S=63/64=1-1/64$ and $S=1/64$, corresponding to ${\cal W}=2.2040 \, {\cal H}$ and ${\cal H}=2.2040 \, {\cal W}$, respectively, that coincide for ${\cal X}=0$, while in (b) we show the dependence for the square ${\cal H}= {\cal W}$ where $S=1/2$. 

To understand the gross behavior, consider the limiting cases. For $S$ close to 1, where ${\cal H} \ll {\cal W}$, the corners of the rectangle are far away from all points of the horizontal midline, and $\langle \epsilon (x_{\rm M},  y_{\rm M}=0) \rangle$ equals the profile $- \pi [2 {\cal H} \cos (\pi x_{\rm M} / {\cal H})]^{-1}$ in an {\it infinite} vertical strip of width ${\cal H}$. However, for $S$ close to 0 where ${\cal H} \gg {\cal W}$, $\langle \epsilon (x_{\rm M},  y_{\rm M}=0) \rangle$ takes the (midline) value $-\pi /(2 {\cal W})$ of an {\it infinite} horizontal strip of width ${\cal W}$ only outside the two regions near the ends of the midline, where the distance from the corners is of order ${\cal W}$ or less, while inside these regions it displays the midline behavior in a {\it semi-infinite} horizontal strip with the maximum discussed in Sec. \ref{semi-+-}. Features of the limits ${\cal H} \ll {\cal W}$ and ${\cal H} \gg {\cal W}$ are still visible in the curves (c) and (a), respectively, of FIG. \ref{-+completehorizontalOleg}. See, in particular, the maximum in (a). 

In agreement with the discussion in (1)-(4) of Sec. \ref{aspandzerolines} and FIG. \ref{Henkel} (a) , $\langle \epsilon (x_{\rm M},0) \rangle$ has four zeros for $0<S<S_<$, two inner ones, $\pm x_{\rm M-}$, and two outer ones, $\pm x_{\rm M+}$  where $|x_{\rm M-}| < |x_{\rm M+}|$. For $S_<<S<S_>$ only the two outer ones survive, and for $S_> <S<1$ there are no zeros on the horizontal midline, cf. FIG. \ref{Henkel} (b) and (c), respectively. The locations of the zeros follow via (\ref{invhalfrect}) from the values of $J$ for which the curly bracket in (\ref{epsrecmid+-+-+}) vanishes and are given by
\begin{eqnarray} \label{zeros'}
{|x_{\rm M\pm}| \over {\cal H}}={1 \over 2 {\bf K}(\sqrt{1-S})} F\Biggl(\Bigl( {k_{\pm}(S)-1 \over k_{\pm}(S)+1} \Bigr)^{1/2} ,(1-S)^{1/2} \Biggr)\, 
\end{eqnarray}
where
\begin{eqnarray} \label{zerosprime}
k_{\pm}(S)={(2-S) (7-7S-S^2) \pm 8 \sqrt{3} (1-S)^{3/2} \over S(1-S+S^2)}\, .
\end{eqnarray}
The functions $k_+ (S)$ and $k_- (S)$ decrease monotonically with increasing $S$ and approach the value 1 at $S=S_>$ and $S=S_<$, respectively, for which, consistent with the discussion in (1)-(4) above, the corresponding zero $x_{\rm M+}$ and $x_{\rm M-}$ is at the midpoint of the rectangle and beyond which it disappears. 

Applying Eq. (\ref{zeros'}) to the square, where $S=1/2$, one finds $|x_{\rm M+}| / ({\cal H}= {\cal W}) = 0.3963$ and for the rectangle with $S=1/64$, i.e. ${\cal H}= 2.2 \times {\cal W}$, the values $\bigl[|x_{\rm M-}|, |x_{\rm M+}| \bigr] / {\cal H} = [0.2068, 0.4617]$, in agreement with FIGS. \ref{Henkel} and \ref{-+completehorizontalOleg}.

Also note the form
\begin{eqnarray} \label{zerosem}
[({\cal H}/2) -|x_{\rm M \pm}|] /{\cal W} = \Biggl[{\bf K}\Bigl( (1-S)^{1/2} \Bigr)-F\Biggl(\Bigl( {k_{\pm}(S)-1 \over k_{\pm}(S)+1} \Bigr)^{1/2} ,(1-S)^{1/2} \Biggr) \Biggr] \Big/ \Bigl[2 {\bf K}(S^{1/2})  \Bigr] \nonumber \\
\end{eqnarray}
of the normalized distances between the two left zeros and the left boundary of the rectangle. For $S \searrow 0$ this reproduces the two corresponding distances $x_{\mp} / {\cal W}=\Bigl[0.0843, \, 0.6454  \Bigr]$ in the semi-infinite strip mentioned below Eq. (\ref{zero-+-}).

\subsubsection{Order parameter} \label{sig+-+-}

The density $\langle \sigma (G,J) \rangle$ of the order parameter follows from Eqs. (17) and (18a) in Ref. \cite{TWBG2} and reads
\begin{eqnarray} \label{sigGJ+-+-}
\langle \sigma (G,J) \rangle = (2/J)^{1/8} \times {\cal M}_{\sigma} (G,J;S) \, ,
\end{eqnarray}
where 
\begin{eqnarray} \label{MsigGJ+-+-}
{\cal M}_{\sigma} (G,J;S) &=& \Bigl( M(t) \, M(-t) \, M(t^{-1}) \, M(-t^{-1}) \Bigr)^{-1/2} \times \nonumber \\
&& \times \Biggl[ (G^2 +J^2)^2 +1 - 2{2-S \over S} \, \Biggl(G^2 + {1-S-S^2 \over 1-S+S^2} J^2  \Biggr) \Biggr] \, ,
\end{eqnarray}
with $M(\tau)$ from Eq. (\ref{epsrec+-+-+prime}). The corresponding profile in the unit circle is 
\begin{eqnarray} \label{Msiguv+-+-}
\langle \sigma (u,v) \rangle &=& \Biggl({4 \over 1-u^2 -v^2} \Biggr)^{1/8} \times {\cal N}_{\sigma} (u,v;S) \, , \nonumber \\
{\cal N}_{\sigma} (u,v;S) &\equiv& {\cal M}_{\sigma} \Biggl( {-2v \over 1-2u+u^2 +v^2} \, , \, {1-u^2 -v^2 \over 1-2u+u^2 +v^2} \, ; \, S \Biggr) \, ,
\end{eqnarray}
which checks with the symmetry relation
\begin{eqnarray} \label{Nsiguv+-+-}
{\cal N}_{\sigma} (u,v;S) = - {\cal N}_{\sigma} (v,u;1-S) \, . 
\end{eqnarray}

\paragraph{Horizontal midline}

The horizontal midline corresponds to $G=0$, for which Eq. (\ref{MsigGJ+-+-}) yields
\begin{eqnarray} \label{midMsigGJ+-+-}
{\cal M}_{\sigma} (G=0,J;S) &=& {J^2 +J^{-2}-2k(S) \over J^2 +J^{-2}+2(2-S)/S} \, , \quad k(S) \equiv {2-S \over S} \times {1-S-S^2 \over 1-S+S^2} \, .
\end{eqnarray}
Here it was used that the first factor on the rhs of (\ref{MsigGJ+-+-}) equals $J^4 +1+\Theta J^2$, with $\Theta \equiv t^2 +t^{-2} =(4/S)-2$.

As a first application note that at the center of the rectangle
\begin{eqnarray} \label{midpointsig+-+-}
\langle \sigma (x_{\rm M}=0, y_{\rm M}=0) \rangle &=& \langle \sigma (G=0,J=1) \rangle \times 2^{1/8} \, \Biggl( {{\bf K}(\sqrt{1-S}) \over {\cal H}} \, {{\bf K}(\sqrt{S}) \over {\cal W}} \Biggr)^{1/16} \nonumber \\
\langle \sigma (G=0,J=1) \rangle &=& 2^{1/8} \, {2[S-(1/2)] \over (3/4)+[S-(1/2)]^2} \, .	
\end{eqnarray}
As expected, the order parameter at the center of the $+-+-$ rectangle vanishes for the square ${\cal H}={\cal W}$ where $S=1/2$.

For $0<S\leq 1/2$ Eq. (\ref{midMsigGJ+-+-}) implies that  $\langle \sigma (G=0,J) \rangle$ vanishes for $J^2 =k(S)\pm \sqrt{k(S)^2 -1}$. For $1/2 <S<1$ it does not vanish on the imaginary axis $H=iJ$, since in this case $2k(S)<2$ while $J^2 +J^{-2}$ cannot be smaller than 2. To determine for the rectangle via Eq. (\ref{invhalfrect}) the zeros $x_{\rm M}=\pm x_{\rm M0}$ where $\langle \sigma (x_{\rm M}, \, y_{\rm M}=0) \rangle$ vanishes from the vanishing of the numerator of ${\cal M}_{\sigma}$,  one uses the identity $(J^2-1)/(J^2+1)=[(J^2 +J^{-2}-2)/(J^2 +J^{-2} +2)]^{1/2}$ and obtain  
\begin{eqnarray} \label{sigzeros'}
{|x_{\rm M0}| \over {\cal W}}={1 \over 2 {\bf K}(S^{1/2})} F\Biggl(\Bigl( {k(S)-1 \over k(S)+1} \Bigr)^{1/2} ,(1-S)^{1/2} \Biggr)\, , \quad 0<S<1/2
\end{eqnarray}
with $k(S)$ given by Eq. (\ref{midMsigGJ+-+-}). This checks with the expected result for the square ${\cal H}/{\cal W}=1$, in which, by symmetry, the order parameter vanishes along the two diagonals so that $|x_{\rm M0}|=0$. This is consistent with (\ref{sigzeros'}) since $S=1/2$ and $k(S)=1$ for the square. 

As in Eq. (\ref{zerosem}) consider the difference $[{\cal H}/2 - |x_{\rm M0} (S)|]/{\cal W}$. This is again given by the expression on the rhs of (\ref{zerosem}), except that $k_{\pm}(S)$ is replaced by $k(S)$. On increasing $S$ from 0 to 1/2 the difference monotonically increases from the value 0.28055 of the semi-infinite $-+-$ strip mentioned above to the value 0.5000 for the square.   

\paragraph{Zero lines in the rectangle} \label{zerosig+-}

Dropping the restriction to the midline and considering the entire rectangle, one finds that $\langle \sigma \rangle$ vanishes along two non-intersecting parabolic-like lines with mirror symmetry with respect to the vertical and horizontal midlines. For ${\cal H}>{\cal W}$ ($S<1/2$) one line, $L_{\rm II,III}$, connects the corners II and III and intersects the horizontal midline at $z_{\rm M}=-|x_{\rm M0}|$. The other, $L_{\rm I,IV}$, connects the corners I and IV. For ${\cal H}<{\cal W}$ ($S>1/2$) one, $L_{\rm I,II}$, connects I with II and the other, $L_{\rm III,IV}$, III with IV. For ${\cal H}={\cal W}$ the lines reduce to the two diagonals. 

A parametric representation, similar to (\ref{zero-+-}), of the zero lines in the rectangle follows from 
that of the corresponding lines ${\cal L}_{\rm II,III}$ etc. in the upper half $H$ plane, for which the square bracket in (\ref{MsigGJ+-+-}) vanishes and which can be written as 
\begin{eqnarray} \label{sigzerosentire}
G=\pm |G(J,S)|_{\pm} \, , \quad \bigl(|G(J,S)|_{\pm}\bigr)^2 \equiv {2-S \over S}-J^2 \pm 2 \Biggl( {1 \over S^2} -{1 \over S} -J^2 {(2-S)S \over 1-S+S^2} \Biggr)^{1/2}  .
\end{eqnarray}
For $G>0$ the curves ($G=|G(J,S)|_-,J$) and ($G=|G(J,S)|_+ ,J$) describe for $S<1/2$ the right halves of ${\cal L}_{\rm II,III}$ and ${\cal L}_{\rm I,IV}$, respectively, while for $S>1/2$ they describe two segments composing ${\cal L}_{\rm III,IV}$. This implies that for $J=0$ , $|G(J;S)|_{-} = \tan(\alpha /2) \equiv G_{\rm III}$ and $|G(J,S)|_{+} = \cot(\alpha /2) \equiv G_{\rm IV}$, which is easily checked. As expected, for $S<1/2$, $|G(J,S)|_{-}$ and $|G(J,S)|_{+}$ vanish, respectively at the smaller and larger $J$ value for the vanishing of ${\cal M}_{\sigma} (G=0,J;S)$ in (\ref{midMsigGJ+-+-}) discussed above.

To obtain the quantitative shapes of the zero lines $L$ in the rectangle, it is sufficient to discuss for $0<S<1/2$ the lower half of $L_{\rm II,III}$, originating from corner III, that corresponds to the right half of ${\cal L}_{\rm II,III}$, originating from switching point (III). From the symmetries of the  $+-+-$ rectangle all the zero lines $L$ follow. The corresponding parametric representation reads 
\begin{eqnarray} \label{transsigzero}
{z_{\rm M}(J,S) \over {\cal W}}&=&{Z_{\rm M}(w(J,S),S) \over {\bf K} (S^{1/2})} \, , \quad w(J,S)={H(J,S)-i \over H(J,S)+i} \, , \nonumber \\
H(J,S)&=&|G(J,S)|_- +iJ \, , \quad 0<J<\sqrt{k(S)-\sqrt{(k(S))^2 -1}} 
\end{eqnarray}
as follows from the transformations $Z_{\rm M}(w_{\rm M},S)$ given in (\ref{invcircrect}) and $w(H)=(H-i)/(H+i)$, which is the inverse of the Moebius transformation in (\ref{Moebprime}). 

\subsection{Rectangle with horizontal $f$ and vertical $+$ boundaries} \label{+f+f}

Here the boundary conditions of the corresponding upper half $H=G+iJ$ plane are the same as in the previous Subsection except that $-$ is replaced by $f$. 

Let us start with the expressions for $\langle \epsilon \rangle$ and $\langle \sigma \rangle$ in the half plane given in Eqs. (2.41) and (2.40) of Ref. \cite{BE21}, which can be written as
\begin{eqnarray} \label{arbeps}
\langle \epsilon (G,J) \rangle = -{1 \over 2J} \, \Bigl[ C_{\rm II,I} \, C_{\rm IV,III} + \Xi \, S_{\rm II,I} \, S_{\rm IV,III} \Bigr] \, ,
\end{eqnarray}
\begin{eqnarray} \label{arbsig}
\langle \sigma (G,J) \rangle = \Biggl({2 \over J} \Biggr)^{1/8} \, 2^{-1/2} \,  \Bigl[ \sqrt{(1+ C_{\rm II,I})(1+ C_{\rm IV,III})} + \Xi \sqrt{(1- C_{\rm II,I})(1- C_{\rm IV,III})} \Bigr]^{1/2} \, , \nonumber \\ \, , 
\end{eqnarray}
where
\begin{eqnarray} \label{C1C3}
&&\Bigl[ C_{\rm II,I} \, , \, S_{\rm II,I} \Bigr]= \Bigl[ N_{\rm II,I} \, , \, J (G_{\rm II} - G_{\rm I}) \Bigr] \Big/ \sqrt{N_{\rm II,I}^2 +J^2 (G_{\rm II} - G_{\rm I})^2 } \, ,  \\
&&N_{\rm II,I} \equiv (G-G_{\rm II}) (G-G_{\rm I}) + J^2 \, ; \quad \Bigl[C_{\rm IV,III}, \, S_{\rm IV,III}  \Bigr] = \Bigl[C_{\rm II,I}, \, S_{\rm II,I}  \Bigr]\Big|_{G_{\rm II}\to G_{\rm IV} ,\, G_{\rm I}\to G_{\rm III}} \, , \nonumber
\end{eqnarray}
and
\begin{eqnarray} \label{Xi}
\Xi ={1-\chi^2 \over 1+\chi^2} \, , \quad \chi^2 = \Biggl[ {(G_{\rm III}-G_{\rm I}) (G_{\rm IV}-G_{\rm II}) \over (G_{\rm III}-G_{\rm II}) (G_{\rm IV}-G_{\rm I})} \Biggr]^{1/2} \, ,
\end{eqnarray}
and which are valid for an {\it arbitrary} configuration $G_{\rm I} < G_{\rm II} < G_{\rm III} < G_{\rm IV}$ of the switching points of the $+f+f+$ boundary conditions. For the configuration (\ref{GK}) considered here, 
\begin{eqnarray} \label{Xiprime}
\Xi =- {1-s \over 1+s} \, , \quad s \equiv \sin \alpha \, . 
\end{eqnarray}
For the shifted rectangle considered in Appendices \ref{firstquadrant} and \ref{COE+f+frec}, the configuration is different and the corresponding $\Xi$ is given in Eq. (\ref{CS}). 

\subsubsection{Energy density}

The energy density profile $\langle \epsilon (x_{\rm M} , \, y_{\rm M}) \rangle$ in the rectangle considered here allows to study interesting features of the competition between order and disorder induced by the two types of boundaries. Of particular interest is the dependence on the aspect ratio ${\cal H}/{\cal W}$. 

For the configuration (\ref{GK}) of switching points, the general expression for the energy density in (\ref{arbeps})-(\ref{Xi}) reduces to 
%
\begin{eqnarray} \label{ep+f+f+}
\langle \epsilon (G,J) \rangle = -{1 \over 2J} \,  \times B \, , \quad
B={1 \over R_{+} R_{-}} \Bigl[{\cal C}_+ {\cal C}_- + \Xi \, {\Sigma} \Bigr]\, ,
\end{eqnarray}
\begin{eqnarray} \label{ep+f+f+prime}
\langle \epsilon (G,J) \rangle = -{1 \over 2J} \, {1\over 1+ s^{-1}} \times A \, , \nonumber  \\ 
A={1 \over R_{+} R_{-}} \Bigl[\bigl( 1+s^{-1} \bigr) \, {\cal C}_+ {\cal C}_- + \bigl( 1-s^{-1} \bigr) \, {\Sigma} \Bigr]
\end{eqnarray}
where $\Xi$ is from (\ref{Xiprime}) and
\begin{eqnarray} \label{ep+f+f+prime}
{\cal C}_{\pm} &=& |H|^2 +1 \pm 2G s^{-1} \, , \quad {\Sigma}=4J^2 \Bigl(s^{-2} -1 \Bigr) \, , \nonumber \\
R_{\pm}&=&\Bigl( {\cal C}_{\pm}^2 +  {\Sigma} \Bigr)^{1/2} \, , \quad s \equiv S^{1/2} = \sin \alpha \, .
\end{eqnarray}
One may check that this result obeys the symmetry properties (\ref{invH}). 

According to Eq. (\ref{ep+f+f+}) the zero lines in the $H$ plane along which $\langle \epsilon (G,J) \rangle$ vanishes are determined by
\begin{eqnarray} \label{ep+f+f+primeprime}
(G^2 +J^2 +1)^2 = 4 \Bigl[ G^2 s^{-2} +J^2 (s^{-1}-1)^2 \Bigr] \, .
\end{eqnarray}
The zero lines in the rectangle follow from (\ref{ep+f+f+primeprime}) via the inverse of the Möbius transformation (\ref{Moebprime}) and Eqs. (\ref{ZMtozM}), (\ref{invcircrect}). Results are shown in FIG. \ref{Ted} and discussed in paragraph \ref{competitionf+} below. 

An important special case is the behavior of $\langle \epsilon \rangle$ along the horizontal and vertical midlines $y_{\rm M}=0$ and $x_{\rm M}=0$ of the rectangle. As explained below Eq. (\ref{aspratios}), this corresponds to the imaginary axis and the half unit circle, respectively, in the upper half $H$ plane. In the first case Eqs. (\ref{ep+f+f+}) and (\ref{ep+f+f+prime}) imply
\begin{eqnarray} \label{horizontal}
B= {1 \over 1+s}  \Biggl[s+ {(J^2+1)^2 - 4J^2 (s^{-2} -1) \over (J^2-1)^2 +4J^2 s^{-2}} \Biggr] \, ,
\end{eqnarray}
\begin{eqnarray} \label{horizontalprime}
A=1+{1 \over s} \, {(J^2+1)^2 - 4J^2 (s^{-2} -1) \over (J^2-1)^2 +4J^2 s^{-2}} \, ,
\end{eqnarray}
and in the second case, in which $G=- \sin \psi, \, J=\cos \psi$, and $R_{\pm} =2(s^{-1} \mp \sin \psi)$,
\begin{eqnarray} \label{vertical}
B= -1+{2s \cos^2 \psi \over 1-s^2 \sin^2 \psi} \, ,
\end{eqnarray}
\begin{eqnarray} \label{verticalprime}
R_+ R_- \times A= -4 \, s^{-3} \, (1+s) \, \Bigl[1-2s + \sin^2 \psi \times s(2-s)  \Bigr] \, ,
\end{eqnarray}
which is noted for later use.

The imaginary axis and the unit circle intersect at $H=i$, which corresponds to the center $z_{\rm M}=0$ of the rectangle. In this case Eq. (\ref{ep+f+f+}), together with (\ref{ep+f+f+prime}) or with (\ref{horizontal}), (\ref{vertical}), yield 
\begin{eqnarray} \label{ep+f+f+01}
\langle \epsilon (G=0,J=1) \rangle = {1 \over 2} - \sin \alpha \, , 
\end{eqnarray}
so that from Eq. (\ref{relatecenter}) 
\begin{eqnarray} \label{ep+f+f+00}
\langle \epsilon (x_{\rm M}=0, \, y_{\rm M}=0) \rangle = {1- 2 \sin \alpha \over \Lambda} \,,
\end{eqnarray}
with $1/\Lambda$ given by Eq. (\ref{Lambdaprime}). In the limits $\alpha \to 0$ and $\alpha \to \pi /2$ where the rectangle reduces to an infinite horizontal strip with $f$ boundaries of width ${\cal W}$ and an infinite vertical strip with $+$ boundaries of width ${\cal H}$, respectively, Eq. (\ref{ep+f+f+00}) reproduces the corresponding midline values $(2/ \Lambda){\cal A}_f^{(\epsilon)} =\pi / (2 {\cal W})$ and $(2/ \Lambda){\cal A}_+^{(\epsilon)} =- \pi / (2 {\cal H})$ of $\langle \epsilon \rangle$, respectively.

\paragraph{Order-disorder competition, zero lines, and three special aspect ratios} \label{competitionf+}

It is interesting that in the competition between the disorder induced by the two horizontal $f$ boundaries and the order induced by the two vertical $+$ boundaries of the rectangle, the latter is stronger. 

(1) For the square ${\cal H}/{\cal W}=1$ where $\alpha = \pi /4$, $1- 2 \sin \alpha = 1-\sqrt{2}$, so that the energy density at the midpoint,
\begin{eqnarray} \label{ep+f+f+00SQ}
\langle \epsilon (x_{\rm M}=0, \, y_{\rm M}=0) \rangle = -\big(\sqrt{2}-1 \big) {\bf K} \big( 1/ \sqrt{2} \big) /{\cal H} = - 0.768 /{\cal H }\, , \quad {\cal W}={\cal H} \, ,
\end{eqnarray}
is negative. One of the two zero lines, $L_{\rm I,II}$, connects the NE with the NW corner and the other one, $L_{\rm III,IV}$, the SW with the SE corner, see FIG. \ref{Ted} (e). In the region in between them, which contains the entire horizontal axis, the energy density is negative. Their crossing points $(x_{\rm M},y_{\rm M})=(0, \pm y_{0+})$ with the vertical midline have a considerable mutual separation $2y_{0+}$ which is only smaller by a factor 0.41 than the entire side-length ${\cal W}={\cal H}$ of the square, see Eq. (\ref{Y0Sq}) below, FIG. \ref{Ted} (e), and curve (ii) in FIG. \ref{f+completehorizontalOleg}. The behavior of $\langle \epsilon \rangle$ along the diagonals of the square, which also belong to the negative region, is presented in Eq. (\ref{epshordiaprime+f}) below and shown in curve (iii) in FIG. \ref{f+completehorizontalOleg}. The corresponding behavior (\ref{epshordia+f}) in the upper half plane is another instructive example in which + dominates $f$. 

(2) In order that $1- 2 \sin \alpha$ and the energy density at the midpoint vanish, the length ${\cal H}$ of the $f$ boundaries must be longer than the length ${\cal W}$ of the $+$ boundary. This happens for $\alpha = \pi /6$, corresponding via Eq. (\ref{aspratios}) to ${\cal H}/{\cal W} = {\bf K}(\sqrt{3}/2)/{\bf K}(1/2) = 1.279$. 

For ${\cal H}/{\cal W}<1.279$ the zero lines qualitatively behave as in (1), see e.g. FIG. \ref{Ted} (d). For ${\cal H}/{\cal W}=1.279$ they combine to two lines, $L_{\rm I,III}$ and $L_{\rm II,IV}$, that connect the NE with the SW corner and the NW with the SE corner, respectively, and cross at the center, see FIG. \ref{Ted} (c). These lines separate the rectangle in upper and lower regions with positive $\langle \epsilon \rangle$, each with an opening angle of $2 \pi /3$ near the center and in left and right regions with negative $\langle \epsilon \rangle$, each with an opening angle of $\pi /3$ near the center, see below Eq. (\ref{epsnearcenterprime}) and Ref. \cite{crossangle}. That the former angle is {\it larger} than the latter is plausible, since ${\cal H}>{\cal W}$ and since the zero lines leave the corners at an angle of 45 degrees between their boundaries, see Sec. \ref{CEOzeroline}. For ${\cal H}/{\cal W}>1.279$, however, the zero lines connect the NE with the SE corner ($L_{\rm I,IV}$) and the NW with the SW corner ($L_{\rm II,III}$). They cross the horizontal axis at points $(x_{\rm M},y_{\rm M})=(\pm x_{0+}, 0)$, and in the region between them, which contains the entire vertical axis, the energy density is positive, see FIG. \ref{Ted} (a) and (b).

(3) The zero lines flow into the corners with a tangent equal to the symmetry line that encloses 45 degrees with the bounding $+$ and $f$ sides of the corner. The leading deviation from this asymptotic behavior changes at the aspect ratio ${\cal H}/{\cal W} = {\bf K} \Bigl(\sqrt{(3 \sqrt{5}-5)/2} \Bigr) \big/ {\bf K} \bigl((3- \sqrt{5})/2 \bigr)= 1.471$, corresponding to $\sin \alpha =(3-\sqrt{5})/2$. For ${\cal H}/{\cal W} < 1.471$ the lines bend away from the tangent towards the region where $\langle \epsilon \rangle$ is positive, i.e., towards the $f$ side of the corner, and for ${\cal H}/{\cal W} > 1.471$ towards the + side of the corner. As explained in Sec \ref{CEOzeroline} this follows from the change in sign of the average corner operator $\langle {\cal Y} \rangle$ implied by Eq. (\ref{rectf+f+}).

\paragraph{Zeros on the midlines}

As in Eqs. (\ref{zeros'})-(\ref{zerosem}) in paragraph \ref{comp-+}, compact analytic expressions are presented now for the location of the zeros of the energy density on the {\it midlines} of the rectangle that describe their dependence on the aspect ratio.

According to (\ref{primarytrafo}) the vanishing of $\langle \epsilon (x_{\rm M}, y_{\rm M}) \rangle$ follows from the vanishing of $\langle \epsilon (G,J) \rangle$. Due to Eqs. (\ref{horizontal}) and (\ref{vertical}) the latter vanishes for
\begin{eqnarray} \label{Jplusminus}
G=0 \, , \quad J^2 = J_{\pm}^2 \equiv s^{-2} \times \Bigl[2(1-2s)+s^2 \pm 2(1-s) \sqrt{1-2s}\Bigr] \, , \quad 0<\alpha < \pi /6 \, ,   
\end{eqnarray}
and for
\begin{eqnarray} \label{sinpsi}
H=i e^{i \psi} \, , \quad \sin \psi =  \sin \psi_{\pm} \equiv \pm \Biggl( {2s-1 \over s(2-s)} \Biggr)^{1/2}\, , \quad \pi /6 < \alpha < \pi /2 \, , 
\end{eqnarray}
which for ${\cal H}/{\cal W}>1.279$ and ${\cal H}/{\cal W}<1.279$ determines the values of the zeros $x_{0+}$ and $y_{0+}$ addressed above. The explicit result for the latter,
\begin{eqnarray} \label{Y0}
y_{0+} = {{\cal W} \over{\bf K}(s)} Y_{0+} \, , \quad   Y_{0+} = {1 \over 2} F\Biggl( \Biggl[ {2s-1\over s(2-s)} \Biggr]^{1/2}\, , \, s  \Biggr) \, ,\quad {1 \over 2} < s \equiv \sin \alpha< 1\, ,
\end{eqnarray}
follows from  (\ref{sinpsi}), (\ref{ZMtozM}), and (\ref{invhalfrectprime}). Here $F$ is the elliptic integral of the first kind as defined below Eq. (\ref{invcircrect''}). For the square with $s=1/\sqrt{2}$ and ${\cal H}/{\cal W}=1$, this yields
\begin{eqnarray} \label{Y0Sq}
{2 y_{0+} \over {\cal W}} \equiv {2 Y_{0+} \over {\bf K}(s) } = 0.41646 \, , \quad s= {1 \over \sqrt{2}} \, .
\end{eqnarray}

As expected, the two zeros are equidistant from the rectangle's midpoint in accordance with the symmetry of the boundary conditions. This is obvious for the zeros on the vertical midline, and, since $J_{-}^2 = J_{+}^{-2}$, together with (\ref{invhalfrect}), it applies as well to the zeros at $x_{\rm M}= \pm x_{0+}$ on the horizontal axis. Eqs. (\ref{Jplusminus}) and (\ref{invhalfrect}) imply
\begin{eqnarray} \label{X0}
x_{0+} = {{\cal H} \over {\bf K}(\sqrt{1-s^2})} X_{0+} \, , \quad   X_{0+} = {1 \over 2} F\Biggl( { \sqrt{1-2s} \over 1-s} \, , \, \sqrt{1-s^2}  \Biggr) \, ,\quad 0 < s \equiv \sin \alpha< {1 \over 2} \, .
\end{eqnarray}
For ${\cal H}/{\cal W}=2$ for example, where $s=\tan^2 (\pi/8)$ and ${\cal H} \langle \epsilon (x_{\rm M}=0, \, y_{\rm M}=0) \rangle =2.079$ by (\ref{ep+f+f+00}),  Eq. (\ref{X0}) yields $x_{0+}/{\cal H}=0.33716$. 

It is instructive to consider the dimensionless distance $d_{\rm horiz} \equiv ({\cal H}/2 - x_{0+})/{\cal W} $ of the right zero from its closer vertical boundary in a rectangle with ${\cal H}=L$, ${\cal W}=\ell$ and compare it with its counterpart $d_{\rm vert} \equiv ({\cal W}/2 - y_{0+})/{\cal H}$ in a rectangle with ${\cal H}= \ell$, ${\cal W}= L$. Thus $d_{\rm horiz} = (L/2 -x_{0+})/{\ell}$ is to be compared with $d_{\rm vert} = (L/2 -y_{0+})/{\ell}$. In order that both zeros $x_{0+}$ and $y_{0+}$ exist, we require $L/ \ell \geq 1.279$, i.e., $\alpha = \tilde{\alpha}$ and $\alpha =(\pi /2)-\tilde{\alpha}$ in the two cases with $\tilde{\alpha} \leq \pi /6$, cf. FIG. \ref{Ted}. Thus, apart from a 90 degree rotation, the two rectangles have the same shape but the boundary conditions $f$ and $+$ are interchanged. As expected, $d_{\rm horiz} >  d_{\rm vert}$, in general, reflecting that $+$ dominates $f$. For $L/ \ell = 1.279$, i.e. $\alpha=\pi /6$ and $\alpha = \pi /2-\pi /6 = \pi /3$, for example, this follows from $x_{0+}=0$ and $y_{0+}>0$, compare the remark below Eq. (\ref{sinpsi}). Using Eqs. (\ref{Jplusminus}), (\ref{sinpsi}), and (\ref{invhalfrect}), (\ref{invhalfrectprime}) or (\ref{Y0}) one finds, e.g., $d_{\rm horiz}=0.6396,\, 0.3564$ for $\alpha = \pi/6,\, \pi/12$ and $d_{\rm vert}= 0.2823, \, 0.2806$ for $\alpha = \pi/3, \, 5 \pi/12$.  However, for $L/{\ell} \to \infty$,  $d_{\rm horiz} \searrow  d_{\rm vert}$, and both approach the same finite value $\pi^{-1} \ln (1+ \sqrt{2}) = 0.2805499$. The reason is that in this limit the two rectangles are like semi-infinite strips, one of width ${\cal W}=\ell$ with boundary condition $f$ in the infinitely long edges and $+$ in the finite edge and the other with $f$ and $+$ exchanged. In this case {\it duality} \cite{fweak} implies the same $\epsilon$ profiles apart from the sign, so that the distance of the zeros on the midlines from the finite edge are the same and given by the above value as shown in Section \ref{epsf+fstrip}.

\paragraph{Behavior of $\epsilon$ along the midlines}

The expression 
\begin{eqnarray} \label{epsuv} 
\langle \epsilon(u,v) \rangle= -{1 \over 1-u^2 -v^2} \times \bigl[ B \bigr]_{H=i(1+w)/(1-w)}
\end{eqnarray}
for the energy density in the unit disk follows from Eqs. (\ref{ep+f+f+}), (\ref{uniprimaryprim}), and (\ref{Moebprime}). 

Along the horizontal midline $v=0$ of the disk, Eq. (\ref{horizontal}) leads to
\begin{eqnarray} \label{epshoriu} 
\langle \epsilon(u,v=0) \rangle= {1 \over u^2 -1} \times {(u^4 +1) (2 S^{1/2} -1) +2 u^2 (2S-2S^{1/2} +1) \over u^4 +1+2u^2 (2S-1)} \, .
\end{eqnarray}
For $u \to \pm 1$, $S>0$, where one of the two + boundaries is approached, the rhs of (\ref{epshoriu}) diverges as $-1/(1-u^2)$ towards $- \infty$, while for $S=0$ and $S=1$, in which case the boundary becomes uniformly $f$ and $+$, the rhs has over the entire interval $-1<u<1$ the positive and negative dependence $1/(1-u^2)$ and $-1(1-u^2)$, respectively.

Along the vertical midline $u=0$, Eq. (\ref{vertical}) together with $(\sin \psi ,\, \cos \psi)= (2v, \, 1-v^2)/(1+v^2) $ yields
\begin{eqnarray} \label{epsvertv} 
\langle \epsilon(u=0,v) \rangle= {1 \over 1-v^2}-2S^{1/2}{1-v^2 \over (1+v^2)^2 -4Sv^2} \, .
\end{eqnarray}
As above, this checks with the known limits for $v \to \pm 1$, $S=0$, and $S=1$. For the square, where $S=1/2$, the corresponding behavior along the horizontal and vertical midlines is shown in curves (i) and (ii), respectively, of FIG. \ref{f+completehorizontalOleg}.

The expressions
\begin{eqnarray} \label{epsnearcenter} 
&&\langle \epsilon(u,v) \rangle \to E_0 + u^2 E_2^{(\rm horiz)} + v^2 E_2^{(\rm vertic)} \, ; \quad E_0 =1-2S^{1/2} \, ,  \nonumber \\
&&E_2^{(\rm horiz)}=1-2S^{1/2}-8S+8S^{3/2} \, , \quad E_2^{(\rm vertic)}=1+6S^{1/2}-8S^{3/2}
\end{eqnarray}
near the center of the circle follow from expanding Eqs. (\ref{epshoriu}), (\ref{epsvertv}), and the absence of a term $\propto uv$ due to symmetry. The corresponding expression for the rectangle in the $z_{\rm M}$ plane follows from the transformation (\ref{primarytrafo}), (\ref{unitcircrect}), (\ref{ZMtozM}) and reads  
\begin{eqnarray} \label{epsnearcenterprime}
&& \Lambda \langle \epsilon (x_{\rm M},y_{\rm M}) \rangle \to E_{0}+(x_{\rm M}/\Lambda)^2 \, \bigl[E_{2}^{(\rm horiz)} + (2S-1)E_{0} \bigr] \,+ \nonumber \\
&& \qquad \qquad +(y_{\rm M}/\Lambda)^2 \,\bigl[E_{2}^{(\rm vertic)} + (1-2S)E_{0}  \bigr]  + O\bigl( (|x_{\rm M}+iy_{\rm M}|/\Lambda)^4 \bigr) \, ,
\end{eqnarray}
where the second terms in the square brackets arise from the rescaling factor $|dw/dZ_{\rm M}|$. For $S=1/4$ where $E_0 =0$, $E_2^{(\rm horiz)}=-1$, and $E_2^{(\rm vertic)}=3$, Eq. (\ref{epsnearcenterprime}) determines the form $x_{\rm M}=\pm \sqrt{3} \, y_{\rm M}$ of the intersecting zero lines near the center, so that the right and left sectors with negative  $\langle \epsilon \rangle$ enclose an angle of $\pi /3$ (60 degrees). Due to the angle-invariance of conformal mappings, this value can be determined directly in the half-plane geometry, see Ref. \cite{crossangle} where it is denote by $2 \beta$.   

\subsubsection{Order parameter density} 

The profile $\langle \sigma (G,J) \rangle$ follows from inserting the expressions (\ref{GK}) for the switching points into Eqs. (\ref{arbsig})-(\ref{Xi}). For $G=0$, which determines the behavior  along the horizontal midline of the rectangle, this yields
\begin{eqnarray} \label{sig+f+f+mid}
\langle \sigma (G=0,J) \rangle = \Biggl( {2 \over J} \Biggr)^{1/8} \sqrt{{s \over 1+s}} \Biggl( 1+ {1+J^2 \over \sqrt{(1-J^2)^2 s^2 +4J^2}} \Biggr)^{1/2}  \, , \quad s \equiv S^{1/2} \, .
\end{eqnarray}
Together with Eq. (\ref{relatecenter}), this implies the dependence 
\begin{eqnarray} \label{sig+f+f+center}
\langle \sigma (x_{\rm M}=0, x_{\rm M}=0) \rangle = \Biggl( {1 \over \Lambda} \Biggr)^{1/8} 2^{3/4} \sqrt{{s \over 1+s}}
\end{eqnarray}
of the center value of $\langle \sigma \rangle$	on the aspect ratio. Here Eqs. (\ref{Lambdaprime}) and (\ref{aspratios}) must be taken into account. 

\section{CORNER BOUNDARY EXPANSIONS} \label{coordaxes}

A useful tool for investigating the behavior near the boundary of a critical system is the boundary-operator-expansion (BOE), in which a bulk operator $\phi$ is expanded in terms of ``boundary operators'' located on the boundary, see the discussions belonging to Eqs. (3.167) and (23) in the first and second review, respectively, of Diehl \cite{Diehl}. These expansions were first developed for uniform boundary conditions. Recently, in two spatial dimensions, expansions about points where the boundary condition switches between two different boundary universality classes have been considered in \cite{BE21}. Here these expansions, in which the boundary is a straight line, are generalized to corners where the boundary abruptly changes directions, expanding in terms of operators located at the apex \cite{brief}. This is clearly of interest to critical systems bounded by a wedge or a polygon, in particular, by a rectangle. 

As other operator expansions, the corner operator expansion (COE) describes how distant perturbations affect the critical behavior in the vicinity of the expansion point, here the point where the boundaries intersect. The leading term in the expansion is the apex-operator of lowest scaling dimension multiplied by an amplitude ${\cal F}^{(\phi)}$ which depends on the bulk-operator $\phi$ in question. While the distant perturbations in a given case affect the corresponding average of the apex-operator, the amplitude depends, apart from $\phi$, solely on the  enclosed angle of the corner  and the two boundary conditions meeting at the apex. Due to this local nature the amplitude may be calculated for the corner of a wedge with the most convenient type of perturbations, which are switches between boundary conditions at distant points on the sides of the wedge. The reason is that these perturbations do not change the shape, and both in their presence and absence the mapping to the corresponding half-plane situations is the {\it same}.

As in the upper half plane, at the corner there are qualitative differences, notably in the scaling dimension of the apex operator and its amplitude, depending on whether the two boundary conditions $a$ and $b$ meeting at the apex are equal or not. In the first and second of the following subsections  wedges with $a=b$ and $a \neq b$, respectively, are discussed and the corresponding amplitudes are determined. It is checked in the third subsection that the amplitudes are local properties with the same form near corners in the quite different geometry of a semi-infinite strip. Lateron this check is extended to rectangles.

The corner considered in the following encloses an angle $\gamma \equiv \pi / g$ and its apex is at the origin $z=0$ of the $z=x+iy$ plane. One of its edges is directed along the positive real axis $z=|z|$ with boundary condition $b$, and the other along the direction $z=|z| \exp (i \gamma)$ with boundary condition $a$. When unperturbed, it is the corner of a wedge where the two edges extend with {\it uniform} boundary conditions from $|z|=0$ to $|z|= \infty$, and we denote it by $a|b$.  

\subsection{Corner with equal boundary conditions meeting at the apex} \label{equalbc}

To derive the expansion in this case, begin with the two-dimensional version of the well known BOE near the uniform boundary of semi-infinite critical systems, see Refs. \cite{Diehl,CardBoundaryCritPhen,EES,EEKD}. The BOE applies as well on approaching a flat part of the boundary where the boundary condition is uniform and outside which the boundary might be non-uniform and have a more complicated shape, see Sec. IIIA of Ref. \cite{BE21} and Appendix A of the present paper. On approaching a boundary interval with boundary condition $a$ of the upper half $H=G+iJ$ plane, it reads
\begin{eqnarray} \label{uniformBOE}
\phi(G, J)-\langle \phi \rangle_a \to \mu_a^{(\phi)}\, J^{2-x_{\phi}} T(G)  \, , \quad  J \to 0 \, . \label{BOE}
\end{eqnarray}
Here $\langle \phi \rangle_a \equiv {\cal A}_a^{(\phi)} / J^{x_{\phi}}$ is the average for a boundary condition $a$ extending {\it uniformly} along the entire real axis, ``unperturbed'' by any switching point. Eq. (\ref{BOE}) applies to the pairs $(\phi, a)$ for which the leading boundary operator is the stress tensor $T(G)$, and the prefactors are given by $\mu_a^{(\phi)} = -(4 x_{\phi} /\hat{c}) {\cal A}_a^{(\phi)}$. For the Ising model with $\hat{c}=1/2$ the pairs are $(\phi,a)=(\sigma,+),\,(\sigma,-),\,(\epsilon,+),\,(\epsilon,-),\,(\epsilon,f)$  \cite{fsigma} and the values of ${\cal A}_a^{(\phi)}$ are given in Eq. (\ref{IsingA}).  

We focus on approaching the boundary of the upper half plane at the origin $H=G=0$, assuming that it is an internal point of the interval with boundary condition $a$. According to the previous discussion a simple example is a boundary with a single switching point at $G= \chi >0$ with boundary condition $a$ for $- \infty < G < \chi$ and $c$ for $\chi < G < + \infty$.  For the average $\langle \phi \rangle$ of $\phi$ in this system, Eq. (\ref{BOE}) yields
\begin{eqnarray} \label{avUHP}
\langle\phi(G, J)\rangle-\langle \phi \rangle_a &\to& \mu_a^{(\phi)}  \, J^{2-x_{\phi}}  \,\langle T(H=0)\rangle   \nonumber \\
\langle T(H)\rangle &\to& \langle T(H=0)\rangle \, .
\end{eqnarray}
Here included is the ``expansion'' for the stress tensor average which is regular at the boundary, away from switching points. For the above example $\langle T(H=0)\rangle =t_{ac} / \chi^2$, cf. the paragraph below Eq. (\ref{tressstrip}).

The conformal transformation
\begin{eqnarray} \label{halfvswedge}
H(z)=z^{g}
\end{eqnarray}
relates the upper half $H$ plane to the wedge in the $z=x+iy$ plane with opening angle ${\gamma} \equiv \pi /g$ mentioned above. In the example it has boundary condition $a$ except for the interval $\chi^{1/g}  \equiv \zeta < x < + \infty$ on the real axis, i.e., there is a switching point from $a$ to $c$ at $z= \zeta$. The two boundaries meeting at the apex both have boundary condition $a$, so that in the notation introduced above we are dealing with an $a|a$ corner. Applying the transformation laws given in (\ref{stripfromplane}) and in Ref. \cite{Ttrafo}, 
\begin{eqnarray} \label{transops}
\langle\phi(x, y)\rangle-\langle \phi(x,y) \rangle_{a|a} &=& |dH/dz|^{x_{\phi}} \,  \Bigl(\langle\phi(G, J)\rangle-\langle \phi \rangle_a  \Bigr) \, , \nonumber \\
\langle T(z)\rangle - \langle T(z)\rangle_{a|a} &=& (dH/dz)^2 \, \langle T(H)\rangle \, ,  
\end{eqnarray}
to the relations (\ref{avUHP}) yields their counterparts
\begin{eqnarray} \label{avWEDGE}
\langle\phi(x, y)\rangle-\langle \phi(x,y) \rangle_{a|a} &\to& g^{x_{\phi}} |z|^{(g -1)x_{\phi}} \mu_a^{(\phi)}  \, \Bigl( {\rm Im} z^{g} \Bigr)^{2-x_{\phi}}  \times \langle {\cal T}\rangle \equiv  \nonumber \\
&\equiv& \mu_a^{(\phi)} g^{x_{\phi}} |z|^{2g -x_{\phi}} \Bigl( \sin (\vartheta g) \Bigr)^{2-x_{\phi}} \times \langle {\cal T}\rangle  \equiv \nonumber \\ 
&\equiv& {\cal F}_a^{(\phi)} (x,y) \times \langle {\cal T}\rangle \, , \nonumber \\
\langle T(z)\rangle - \langle T(z)\rangle_{a|a} &\to& g^2 z^{2 g -2} \times \langle {\cal T} \rangle \equiv {\cal F}_a^{(T)} (z) \times \langle {\cal T}\rangle
\end{eqnarray}
for the wedge with the corresponding boundary conditions. Here $z=|z| \exp (i \vartheta)$, and the apex operator ${\cal T}$ is normalized by imposing the condition \cite{swvsshape}
\begin{eqnarray} \label{identapex}
\langle {\cal T}\rangle =  \langle T(H=0)\rangle \, .
\end{eqnarray}
In the simple example $\langle {\cal T}\rangle =  t_{ac} /\chi^2 \equiv t_{ac} /\zeta^{2 g} $. Generally $\langle {\cal T}\rangle$ scales as $ (1/{\rm length})^{2 g} \equiv (1/{\rm length})^{2 \pi / \gamma}$, i.e., ${\cal T}$ has the scaling dimension $2g$. For later reference note that \cite{fsigma}
\begin{eqnarray} \label{opWEDGE}
\phi(x, y)-\langle \phi(x,y) \rangle_{a|a} &\to& {\cal F}_a^{(\phi)} (x,y) \times  {\cal T} \, ; \quad (\phi,a)=(\sigma,+),\,(\sigma,-),\,(\epsilon,+),\,(\epsilon,-),\,(\epsilon,f) \, , \nonumber \\
T(z) - \langle T(z)\rangle_{a|a} &\to& {\cal F}_a^{(T)} (z) \times  {\cal T}
\end{eqnarray}
which follows from Eq. (\ref{avWEDGE}).

The above forms of the prefactors ${\cal F}_a^{(\phi)} (x,y)$ and ${\cal F}_a^{(T)} (z)$ of the apex operator should be compared with the forms $\langle \phi(x,y) \rangle_{a|a} = {\cal A}_a^{(\phi)} \Bigl[g \Big/ ( |z| \sin (\vartheta g) )  \Bigr]^{x_{\phi}}$ and $\langle T(z) \rangle_{a|a} =(\hat{c}/12) \, S(H(z)) =(1-g^2) \, \hat{c} /(24 z^2)$ of the profiles of $\phi$ and $T$, respectively, in the unperturbed $a|a$ wedge. Here $S$ is the Schwarzian derivative \cite{Ttrafo} of the transformation $H(z)$ in (\ref{halfvswedge}). Both ${\cal F}_a^{(T)}$ and $\langle T(z) \rangle_{a|a}$ are independent of $a$.

Both $ {\cal F}_a^{(\phi)} (x,y)$ and $\langle \phi(x,y) \rangle_{a|a}$ are invariant under mirror-imaging about the centerline of the wedge, i.e., under $\vartheta \to \gamma - \vartheta \equiv (\pi /g) - \vartheta$.

\subsection{Corner with different boundary conditions meeting at the apex} \label{diffbc}

Deriving the expansion for a corner with different boundary conditions $a|b$ follows pretty much the track presented in Subsec \ref{equalbc} for equal boundary conditions. Here one starts from the BOE about the $a|b$ switching point at $H=0$ in the nonuniform boundary of the upper half $H$ plane. The simple example is a nonuniform boundary with conditions $a, b$, and $c$ for $-\infty < G < 0$, $0 < G < \chi$, and $\chi < G < + \infty$, respectively, where $a \neq b$ and $b \neq c$. This has been discussed in detail in Eqs. (3.6) ff. of Ref. \cite{BE21}, and the counterpart of (\ref{avUHP}) is 
\begin{eqnarray} \label{avdiffUHP}
\langle\phi(G, J)\rangle-\langle \phi \rangle_{a|b} &\to& F_{ab}^{(\phi)} (G,J) \times\langle \Upsilon (H=0)\rangle \, , \quad \phi = \sigma, \, \epsilon \, ,  \nonumber \\
\langle T(H)\rangle - \langle T(H)\rangle_{a|b} &\to& {1 \over H}  \langle \Upsilon(H=0)\rangle \,\equiv F_{ab}^{(T)} \times \langle \Upsilon(H=0)\rangle \, ,
\end{eqnarray}
where $\Upsilon (H=0)$ is the boundary operator at the $ab$ switching point of lowest scaling dimension. For the simple example mentioned right above Eq. (\ref{avdiffUHP}) $\langle \Upsilon (H=0)\rangle = (t_{ab}+t_{bc}-t_{ac})/ \chi.$ Unlike (\ref{BOE}), (\ref{avUHP}), and (\ref{opWEDGE}), here and in Eq. (\ref{opdiffWEDGE}) below there is no restriction on combining $\phi$ with the pair $ab$. The average $\langle \phi \rangle_{a|b}$ is for a boundary with conditions $a$ and $b$ for $-\infty < G <0$ and $0 < G < + \infty$, ``unperturbed'' by further switchings. It has the scaling form
\begin{eqnarray} \label{scalephiH}
\langle \phi(G,J) \rangle_{a|b} = J^{-x_{\phi}} {\cal B}_{a|b}^{(\phi)} (\theta)		
\end{eqnarray}
where $\theta$ is the angle that $H=G+iJ$ encloses with the positive real axis such that ${\cal B}_{a|b}^{(\phi)} (\theta)$ equals ${\cal A}_a^{(\phi)}$ and ${\cal A}_b^{(\phi)}$ for $\theta=\pi$ and $\theta=0$, respectively. As shown in Eq. (3.29) of Ref. \cite{BE21} 
\begin{eqnarray} \label{scaleFH}
F_{a|b}^{(\phi)} (G,J) = {1 \over 2t_{ab}} \, J^{-x_{\phi}} |H| \times (\sin\theta) \, {d \over d \theta} \, {\cal B}_{a|b}^{(\phi)} (\theta) \, .
\end{eqnarray}
The relations in (\ref{scalephiH}) and (\ref{scaleFH}) as well as $\langle T(H) \rangle_{a|b} = t_{ab} /H^2$ for the stress tensor are of general validity, not limited to the Ising model. For the Ising model the form of the scaling functions ${\cal B}_{a|b}^{(\phi)} (\theta)$ follows from Eq. (4.1) in Ref. \cite{BX} and reads
\begin{eqnarray} \label{fIsing}
&&{\cal B}_{+|-}^{(\sigma)}=-{\cal B}_{-|+}^{(\sigma)}=-2^{1/8} \cos \theta \, , \quad {\cal B}_{+|-}^{(\epsilon)}={\cal B}_{-|+}^{(\epsilon)}=-{1 \over 2}(1-4 \sin^2 \theta)  \\
&&{\cal B}_{+|f}^{(\sigma)}=2^{1/8}\Biggl( \sin {\theta \over 2} \Biggr)^{1/2} \, , \quad {\cal B}_{f|+}^{(\sigma)}=2^{1/8}\Biggl( \cos {\theta \over 2} \Biggr)^{1/2} \, , \quad {\cal B}_{+|f}^{(\epsilon)} =-{\cal B}_{f|+}^{(\epsilon)} = {1 \over 2}\cos \theta \, . \nonumber
\end{eqnarray}

The conformal mapping leads to a wedge with $a|b$ switch at the apex, and the transformations (\ref{transops}) adapted to the present case with $a|a$ replaced by $a|b$ and $\langle \phi \rangle_{a}$ by $\langle \phi \rangle_{a|b}$ together with (\ref{avdiffUHP}) yields
\begin{eqnarray} \label{avdiffWEDGE}
\langle\phi(x, y)\rangle-\langle \phi(x,y) \rangle_{a|b} &\to& g^{x_{\phi}} |z|^{(g -1)x_{\phi}} F_{ab}^{(\phi)} (G,J) \times \langle {\cal Y}\rangle \nonumber \\ 
&\equiv& {\cal F}_{a|b}^{(\phi)} (x,y) \times \langle {\cal Y}\rangle \, , \nonumber \\
\langle T(z)\rangle - \langle T(z)\rangle_{a|b} &\to& g^2 z^{g -2} \times \langle {\cal Y} \rangle \equiv {\cal F}_{a|b}^{(T)} (z) \times \langle {\cal Y}\rangle \, .
\end{eqnarray}
Here we have normalized the apex operator ${\cal Y}$ by imposing the condition \cite{swvsshape} 
\begin{eqnarray} \label{identdiffapex}
\langle {\cal Y}\rangle = \langle {\Upsilon (H=0)}\rangle \, .
\end{eqnarray}
In the simple example its explicit form is $\langle {\cal Y}\rangle = (t_{ab}+t_{bc}-t_{ac})/ \chi \equiv (t_{ab}+t_{bc}-t_{ac})/ \zeta^g$. The scaling dimension $g$ of the present apex operator ${\cal Y}$ for $a \neq b$ should be compared with the scaling dimension $2g$ of the apex operator ${\cal T}$ for $a=b$. 

Combinig the scaling expressions in (\ref{scalephiH}) and (\ref{scaleFH}) with the transformation (\ref{halfvswedge}) and the relation (\ref{avdiffWEDGE}) between ${\cal F}_{a|b}^{(\phi)}$ and $F_{a|b}^{(\phi)}$ yields  
\begin{eqnarray} \label{scalephi}
\langle \phi(x,y) \rangle_{a|b} &=& \Biggl({g \over |z| \sin g \vartheta}  \Biggr)^{x_{\phi}}	 \, {\cal B}_{a|b}^{(\phi)} (g \vartheta)  \, , \nonumber \\
{\cal F}_{a|b}^{(\phi)} (x,y) 
&=& {1 \over 2t_{ab}} \, 
\Biggl( {g \over |z| \sin g \vartheta } \Biggr)^{x_{\phi}} \, |z|^g \sin g \vartheta  \times 
{d \, {\cal B}_{a|b}^{(\phi)}(g \vartheta) \over d \, g \vartheta} \, ,
\end{eqnarray}
where, like above, $\vartheta$ is the argument of $z= x+iy=|z| \exp (i \vartheta)$. For $g=1$, $\langle \phi(x,y) \rangle_{a|b}$ and ${\cal F}_{a|b}^{(\phi)}$ reduce to $\langle \phi(G,J) \rangle_{a|b}$ and $F_{a|b}^{(\phi)}$, respectively. For the Ising model the explicit expressions
\begin{eqnarray} \label{calF+f}
{\cal F}_{+|f}^{(\epsilon)}(x , y) &=& -4 g|z|^{g -1} \sin (\vartheta \, g) \, , \nonumber \\
{\cal F}_{+|f}^{(\sigma)}(x , y) &=& 4 g^{1/8} |z|^{g - 1/8} \Biggl( \sin {\vartheta g \over 2} \Biggr)^{3/8} \Biggl( \cos {\vartheta g \over 2} \Biggr)^{15/8} \, ,
\end{eqnarray}
and
\begin{eqnarray} \label{calF+-}
{\cal F}_{+|-}^{(\epsilon)}(x , y) &=& 2 g|z|^{g -1} \sin (2 \vartheta g) \, , \nonumber \\
{\cal F}_{+|-}^{(\sigma)}(x , y) &=& (2g)^{1/8}  |z|^{g - 1/8} \sin^{2-(1/8)}( \vartheta g) \, ,
\end{eqnarray}
as well as
\begin{eqnarray} \label{epssig+f}
\langle \epsilon (x , y) \rangle_{+|f} &=& {g \over 2 |z|} \cot (\vartheta \, g) \, , \nonumber \\
\langle \sigma (x , y) \rangle_{+|f} &=& \Biggl( {2g \over |z| \sin (\vartheta \, g)} \Biggr)^{1/8} \,  \Biggl(\sin {\vartheta g \over 2} \Biggr)^{1/2} \, ,
\end{eqnarray}
and
\begin{eqnarray} \label{epssig+-}
\langle \epsilon (x , y) \rangle_{+|-} &=& - {g \over 2|z|\sin (\vartheta \, g)} \, \Bigl(1-4 \sin^2 (\vartheta \, g)   \Bigr) \, , \nonumber \\
\langle \sigma (x , y) \rangle_{+|-} &=& - \Biggl( {2g \over |z|\sin (\vartheta \, g)} \Biggr)^{1/8} \, \cos ( \vartheta g)  \, ,
\end{eqnarray}
follow from Eq. (\ref{fIsing}). The generally valid result for the stress tensor in the unperturbed wedge \cite{dilaF},
\begin{eqnarray} \label{Tabwedge}
\langle T(z) \rangle_{a|b} = {1 \over z^2} \, \Biggl( g^2 t_{ab} + {\hat{c} \over 24} \, (1-g^2)
\Biggr) \, ,
\end{eqnarray}
follows from the transformation formula in footnote \cite{Ttrafo} and the Schwarzian derivative of the mapping (\ref{halfvswedge}) adressed below (\ref{opWEDGE}). For the Ising model $\hat{c} =1/2$ and $t_{+f} = 1/16$, $t_{+-} =1/2$, cf. the paragraph below Eq. (\ref{tressstrip}).   

From Eqs. (\ref{avdiffWEDGE}) one obtains
\begin{eqnarray} \label{opdiffWEDGE}
\phi(x, y)-\langle \phi(x,y) \rangle_{a|b} &\to& {\cal F}_{a|b}^{(\phi)} (x,y) \times  {\cal Y} \, , \quad \phi= \sigma, \, \epsilon \, , \nonumber \\
T(z) - \langle T(z)\rangle_{a|b} &\to&  {\cal F}_{a|b}^{(T)} (z) \times {\cal Y} \, .
\end{eqnarray}

The derivation of the operator expansions (\ref{opWEDGE}) and (\ref{opdiffWEDGE}) for the wedge 
geometry from the corresponding expansions (\ref{avUHP}) and (\ref{avdiffUHP}), respectively, in the upper half plane is possible since the conformal mapping for the perturbed and unperturbed averages on the left hand sides of (\ref{avWEDGE}) or of (\ref{avdiffWEDGE}) is the {\it same}. This is due to the simple nature of the perturbations, i.e. the switches, which do not change the shape of the infinite wedge system. This is different in the cases of a semi-infinite strip and a rectangle that we consider below \cite{swvsshape}. 

\subsection{``Zero lines'' of $\langle \phi(x,y) \rangle$  originating from an $a|b$ corner and the COE} \label{CEOzeroline}

The $a|b$ corners are origins of ``zero lines''. These are contour lines along which $\langle \phi \rangle$ vanishes, cf. Secs. \ref{seminfstrip} and \ref{recmix}. It follows from  (\ref{epssig+-}) and (\ref{epssig+f}) that two zero lines of $\langle \epsilon \rangle$ originate at a $+|-$ corner, with tangent vectors $\exp[i \pi /(6g)]$ and $\exp[5i \pi /(6g)]$), and one of $\langle \sigma \rangle$ with tangent vector $\exp[i \pi /(2g)]$. A   $+|f$ corner emanates one zero line of $\langle \epsilon \rangle$ with tangent vector $\exp[i \pi /(2g)]$ and one of  $\langle \sigma \rangle$ with tangent vector in the $f$ boundary of the corner.   

The COE (\ref{opdiffWEDGE}) predicts the $|z|$-dependence of $\langle \phi(x,y) \rangle$ for arbitrary fixed $\vartheta$ in leading and next-to-leading order. For the special values of $\vartheta$ where $\exp(i \vartheta)$ equals one of the above-mentioned tangent vectors and the leading contribution $\langle \phi(x,y) \rangle_{a|b}$ vanishes, a positive and negative sign of $ {\cal F}_{a|b}^{(\phi)} (x,y) \times \langle {\cal Y} \rangle$ implies that the zero line bends away from its tangent direction towards the side where $\langle \phi(x,y) \rangle_{a|b}$ is negative and positive, respectively. 

The COE (\ref{opdiffWEDGE}) provides even quantitative results for this bending away, since a zero line $z=z_0$ of $\langle \phi(x,y) \rangle$ near an $a|b$ corner is determined by the vanishing of $\langle \phi(x,y) \rangle_{a|b} + {\cal F}_{a|b}^{(\phi)} (x,y) \times \langle {\cal Y} \rangle$ for $z=z_0$. Parametrizing the line by the dependence of ${\rm arg} z_0 \equiv \vartheta_0$ on $|z_0|$, one finds for the Ising model with the explicit expressions given in Eqs. (\ref{calF+f})-(\ref{epssig+-}) and a corner with opening angle of 90 degrees ($g=2$) the following results for small $|z_0|$:
	
For $\phi = \epsilon$ and $a|b = +|f$
\begin{eqnarray} \label{zero_f+}
{\rm arg} z_0 \, = \, {\pi \over 4} -4 |z_0|^2 \times \langle {\cal Y} \rangle \, ,
\end{eqnarray}
for $\phi = \epsilon$ and $a|b = +|-$, where there are two lines $z=z_{0+}$ and $z=z_{0-}$, 
\begin{eqnarray} \label{zero_-+}
{\rm arg} \bigl(z_{0+} \, ,  \, z_{0-} \bigr) \, = \, \Bigl({\pi \over 12} \, , \, {5 \pi \over 12} \Bigr) -{1 \over 4} \, \bigl( |z_{0+}|^2 \, ,  \, |z_{0-}|^2 \bigr) \times \langle {\cal Y} \rangle \, ,
\end{eqnarray}
and for $\phi = \sigma$ and $a|b = +|-$
\begin{eqnarray} \label{zero_+-sig}
{\rm arg} z_0 \, = \, {\pi \over 4} -{1 \over 2} |z_0|^2 \times \langle {\cal Y} \rangle \, .
\end{eqnarray}

For a given geometry the values of $\langle {\cal Y} \rangle$ in Eqs. (\ref{zero_-+}) and (\ref{zero_+-sig}) are the same so that the zero line $z_0$ of $\sigma$ can be compared directly with the two zero lines $z_{0+}, \, z_{0-}$ of $\epsilon$. In particular, the signs of the bending away (clockwise or counterclockwise) from their asymptotic tangents are the same for the three lines.

The expressions in (\ref{zero_f+})-(\ref{zero_+-sig}) are consistent with expansions of exact results. For the semi-infinite strips on using the values of $\langle {\cal Y} \rangle$ given in Eqs. (\ref{semistripsdiffcornop}), see Eqs. (\ref{zero_-+}), (\ref{zerof+fprime}), and the remark below Eq. (\ref{sigsistripABC}). The expression in (\ref{zero_f+}) also agrees with the behavior (\ref{exeps+f+fxy}) of an $f+|f+$ rectangle of arbitrary aspect ratio due to the form of $\langle {\cal Y} \rangle$ given in Eq. (\ref{rectf+f+}).

Corresponding results for zero lines in the half plane originating from a switching point, can be obtained from the BOE (\ref{avdiffUHP}).

\subsection{Corner of a semi-infinite strip} \label{subsecsemistrip}

The operator expansions (\ref{opWEDGE}) and (\ref{opdiffWEDGE}) derived for an $a|a$ or $a|b$ wedge not only apply to perturbations arising from distant {\it switches} but also from deviations from the wedges {\it shape} at large distance from the apex. A useful example is the corner with apex at $z=0$ of the semi-infinite strip introduced in Sec. \ref{seminfstrip}. It can be used to confirm that the operator expansions (\ref{opWEDGE}) and (\ref{opdiffWEDGE}) again apply with the above prefactors ${\cal F}_a^{(\phi)} , \, {\cal F}_a^{(T)}$ and 
${\cal F}_{a|b}^{(\phi)} , \, {\cal F}_{a|b}^{(T)}$ taken for an enclosed angle of 90 degrees, i.e., for $g=2$.  

Since the perturbed and unperturbed density profiles are mapped with {\it different} conformal transformations (\ref{semiinfstrip}) and (\ref{halfvswedge}) onto the upper half $H$ plane, the BOE's (\ref{avUHP}) and (\ref{avdiffUHP}) in the upper half $H$ plane cannot be invoked. One possibility is to
evaluate the perturbed and unperturbed averages separately before investigating their difference for the situation ${\cal W} \gg |z|$ in which the shape perturbation is far away. This requires knowing the corresponding profiles in the upper half plane, which one does for many cases in the Ising model. Using the product representation in (\ref{generalstrip}) even allows to derive the COE for the semi-infinite strip from the BOE in the half plane without recourse to results for a particular model, see the discussion in Sec. \ref{generalargue}.

Adopting throughout this subsection the notation $ABC$ for the three sides of the strip introduced below Eq. (\ref{generalstrip}), the above notation $(a, \,b)$ for the boundary classes of the (vertical, horizontal) sides of a wedge of 90 degrees, with apex at $z=0$, is changed to
\begin{eqnarray} \label{notechange}
(a,b) \to (B,C) \, .
\end{eqnarray}

\subsubsection{Corner-operator averages} \label{Yforstrip}

For equal and different side classes $B=C$ and $B \neq C$ of the corner sides, the averages $\langle {\cal T} \rangle$ and $\langle {\cal Y} \rangle$ of the corresponding operators follow from the definitions (\ref{avWEDGE}) and (\ref{avdiffWEDGE}) using the stress tensor expressions in (\ref{tressstripcorner}) and (\ref{tressstripcornerprime}) with the results 
\begin{eqnarray} \label{semistripscornop}
\langle {\cal T} \rangle = \langle {\cal T} \rangle_{AB|B} = {1 \over 16} \Bigl[ - {\hat{c} \over 30} + t_{AB}  \Bigr] \Bigl ( {\pi \over {\cal W}} \Bigr)^4 \, ,
\end{eqnarray}
and
\begin{eqnarray} \label{semistripsdiffcornop}
\langle {\cal Y} \rangle = \langle {\cal Y} \rangle_{AB|C} = {1 \over 4} \Bigl[ t_{AC} - t_{AB} - {1 \over 3} t_{BC} \Bigr] \Bigl ( {\pi \over {\cal W}} \Bigr)^2 \, , \quad B \neq C \, .
\end{eqnarray}

\subsubsection{$B|B$ corner in a $BB|B$ strip} \label{unistrip}

In this case of uniform boundary conditions $B$, the averages $\langle \phi (x,y) \rangle$ of $\phi = \sigma$ or $\epsilon$ in the strip are given in Eq. (\ref{uniformstrip}) for arbitrary $z$. For $|z| \ll {\cal W}$ the quantity $\Psi$ in (\ref{uniformstrip}) becomes 
\begin{eqnarray} \label{semistripphiprime}
\Psi \to {2 \over|z| \sin (2 \vartheta)} \Biggl[ 1+{1 \over 120} \Biggl( {\pi |z| \over {\cal W}} \Biggr)^4 \sin ^2 (2 \vartheta) \Biggr] \, ,
\end{eqnarray}
yielding 
\begin{eqnarray} \label{semistripphiprimeprime}
\langle \phi(x,y) \rangle  \to {\cal A}_B^{(\phi)} \Biggl( {2 \over|z| \sin (2 \vartheta)} \Biggr)^{x_{\phi}}\times \Biggl[ 1+{x_{\phi} \over 120} \Biggl( {\pi |z| \over {\cal W}} \Biggr)^4 \sin ^2 (2 \vartheta) \Biggr] \, ,
\end{eqnarray}
and 
\begin{eqnarray} \label{semistripphiprime3}
\langle \phi(x,y) \rangle - \langle \phi(x,y) \rangle_{B|B} \to {\cal A}_B^{(\phi)} {2^{x_{\phi}} x_{\phi} \over 120} |z|^{4-x_{\phi}} \sin^{2-x_{\phi}} (2 \vartheta) \times \Biggl( {\pi \over {\cal W}} \Biggr)^4 \, ,
\end{eqnarray}
since $\langle \phi(x,y) \rangle_{B|B}$ equals the limit of the rhs of (\ref{semistripphiprimeprime}) for ${\cal W} \to \infty$. The rhs of (\ref{semistripphiprime3}) indeed  equals the product of 
\begin{eqnarray} \label{semistripphiprime4}
{\cal F}_B^{(\phi)} =   \mu_B^{(\phi)} 2^{x_{\phi}} |z|^{4-x_{\phi}} \sin^{2-x_{\phi}} (2 \vartheta) \equiv - {\cal A}_B^{(\phi)} (4 / \hat{c}) \, 2^{x_{\phi}} x_{\phi} |z|^{4-x_{\phi}} \sin^{2-x_{\phi}} (2 \vartheta) \, ,
\end{eqnarray}
that follows from (\ref{avWEDGE}) for $g=2$, $a \to B$, and of the operator average $\langle {\cal T} \rangle$ given in (\ref{semistripscornop}) for $A=B$, as predicted by the operator expansion (\ref{opWEDGE}).

\subsubsection{$B|B$ corner in $AB|B$ strips}  \label{ABB}

\paragraph{$+|+$ corner in an $f+|+$ strip} \label{stripf++} 
To confirm the COE in a strip with $A \neq B=C$, consider the Ising model with $\hat{c}=1/2$ and the special case $A=f$ and $B=+$. Since the switching point $z=i{\cal W}$ between $A$ and $B$ is mapped onto $H=-1$, using Eq.(4.1) in Ref. \cite{BX} for the {\it energy density} the corresponding half-plane profile reads
\begin{eqnarray} \label{epscaa}
\langle \epsilon(G,J) \rangle  = - {1 \over 2J} \, \cos \Theta \, , \quad \Theta = {\rm arg}(H+1) \, ,
\end{eqnarray}
i.e., $\Theta$ is the angle that the vector from the switching point $-1$ to point $H$ forms with the real axis. Inserting $H(z)$ from (\ref{semiinfstrip}) and expanding for $z \to 0$ yields
\begin{eqnarray} \label{epscaaprime}
\cos \Theta \to 1-{1 \over 32} |\tilde{z}|^4 \sin^2 (2 \vartheta) \, .
\end{eqnarray}
Due to the product representation (\ref{generalstrip}), the expansion of the $\langle \epsilon(x,y) \rangle$ follows on multiplying (\ref{semistripphiprimeprime}) for $\phi=\epsilon$, $x_{\phi}= x_{\epsilon} = 1$, and ${\cal A}_B^{(\phi)} = {\cal A}_+^{(\epsilon)} =-1/2$ with (\ref{epscaaprime}). This leads to an expression of the form (\ref{semistripphiprimeprime}), where the content $1+ 1/120  \, |\tilde{z}|^4 \sin^2 (2 \vartheta)$ of the square bracket is replaced by $1+\bigl( 1/120 -1/32 \bigr) \, |\tilde{z}|^4 \sin^2 (2 \vartheta)$. This replacement $1/120 \to 1/120 -1/32$ relating the $\langle \epsilon(x,y) \rangle$'s when moving from $BB|B = ++|+$ to $f+|+$ is consistent with the corresponding replacement  $\hat{c} / 30 \equiv 1/60 \to (\hat{c} / 30) -t_{AB} \equiv 1/60 -1/16$ relating the $\langle {\cal T} \rangle$'s, see the corresponding expression (\ref{semistripscornop}). Thus the validity of the COE (\ref{opWEDGE}) for $\langle \epsilon(x,y) \rangle$ in the case $f+|+$ follows from that in $++|+$.

A corresponding check for the {\it order parameter} runs along the same lines. Here the half-plane profile reads 
\begin{eqnarray} \label{sigcaa}
\langle \sigma(G,J) \rangle  = \Biggl({2 \over J}\Biggr)^{1/8} \, \Biggl[ {1+\cos \Theta \over 2 } \Biggr]^{1/4} \, ,
\end{eqnarray}
leading via (\ref{generalstrip}), (\ref{semistripphiprime}) with $x_{\phi}=x_{\sigma}=1/8$, ${\cal A}_B^{(\phi)} = {\cal A}_+^{(\sigma)} = 2^{1/8}$, and via (\ref{epscaaprime}) to
\begin{eqnarray} \label{sigcaaprime}
\langle \sigma(x,y) \rangle \to \Biggl({4 \over |z| \sin (2 \vartheta)}\Biggr)^{1/8} \, \Biggl[1+{1\over 8} \Bigl({1 \over 120} - {1 \over 32}  \Bigr) \Bigl( {\pi |z| \over {\cal W}} \Bigr)^4 \sin^2 (2 \vartheta)  \Biggr] \, 
\end{eqnarray}
for $f+|+$ boundary conditions. Again the difference from the uniform boundary case in (\ref{semistripphiprimeprime}) is consistent with the difference in the  $\langle {\cal T} \rangle$'s and confirms the COE for $\sigma$ in $f+|+$.

It is interesting to compare the effect on $\langle \phi \rangle$ of the distant perturbation of the half line $A$ in the present strip geometry with that in the simple wedge geometry described below Eq. (\ref{halfvswedge}) with $a,c,\zeta$ replaced by $B,A,{\cal W}$. Here it helps to consider their ratio given by the ratio $(\pi /2)^4 \, \bigl\{ 1-[\hat{c}/(30 t_{AB})] \bigr\}$ of the averages $\langle {\cal T \rangle}$ of the corresponding boundary operators, given in Eq. (\ref{semistripscornop}) and below (\ref{identapex}), since the prefactors ${\cal F}_B^{(\phi)}$ drop out. Due to the half line with boundary condition $A$ being closer to the $B|B$ corner in the strip geometry than in the wedge geometry one expects this ratio to be larger than 1. Indeed, in the Ising model it equals 5.88 and 4.46 for $AB=+-$ or $-+$ and for $+f$ or $f+$, respectively.    

\paragraph{$f|f$ corner in a $+f|f$ strip} \label{strip+ff}

Here we consider the case of $\sigma$ near an $f|f$ corner which preserves the Ising ($+ \leftrightarrow -$) symmetry so that corresponding corner operators  must be odd under this symmetry and ${\cal T}$ does not qualify. In accordance with Ref. \cite{fsigma} we denote the leading corner operator by ${\cal S}$ and confirm that it has scaling dimension 1. Taking from Eq. (4.1) of Ref. \cite{BX} the corresponding half plane profile
\begin{eqnarray} \label{sig+ff}
\langle \sigma(G,J) \rangle  = \Biggl({2 \over J}\Biggr)^{1/8} \, \Biggl[ {1-\cos \Theta \over 2 } \Biggr]^{1/4} \, 
\end{eqnarray}
and using Eq. (\ref{epscaaprime}) yields 
\begin{eqnarray} \label{sigmafstrip}
\langle \sigma (x,y) \rangle  \to {\pi \over 2} \, 2^{-1/4} \, |z|^{1-(1/8)} (\sin 2 \vartheta )^{3/8} \times {1 \over {\cal W}} \, .
\end{eqnarray}
This is indeed consistent with the last expression displayed in Ref. \cite{fsigma} when putting $g=2$ and $\zeta = {\cal W}$. The half line with boundary condition $+$ is closer to the corner in the present strip geometry than in the wedge geometry of Ref. \cite{fsigma}. This is reflected by the present prefactor being larger by $\pi /2$.  

\subsubsection{$B|C$ corner in $AB|C$ strip} \label{generalargue}

Here arbitrary classes $ABC$ with $B \neq C$ are considered, i.e., there is a switch at the corner $z=0$ and, correspondingly at $H=1$. Since the leading deviation of $\langle \phi \rangle$ from the unperturbed wedge that we wish to confirm is of order $(|z|/{\cal W})^2$, in the product representation (\ref{generalstrip}) we disregard in $\Psi$ of (\ref{semistripphiprime}) the term $\propto (|z|/{\cal W})^4$ and consider
\begin{eqnarray} \label{productABC}
\langle \phi(x,y) \rangle_{AB|C} \to \Biggl( {2 \over|z| \sin (2 \vartheta)} \Biggr)^{x_{\phi}} \times J^{x_{\phi}} \, \langle \phi(G,J) \rangle_{AB|C} \, .
\end{eqnarray}
Beginning with the BOE for the corresponding problem in the upper half $H$ plane, one confirms the COE (\ref{avdiffWEDGE}) for the semi-infinite strip as follows: First, expand $\langle \phi(G,J) \rangle_{AB|C}$ in (\ref{productABC}) about the $B|C$ switch at $H=1$ on using the BOE which resembles the BOE about $H=0$ in Eq. (\ref{avdiffUHP}) and yields
\begin{eqnarray} \label{avproductABC}
&&\langle \phi(G,J) \rangle_{AB|C} \to \langle \phi(G,J) \rangle_{B|C} + F_{B|C}^{(\phi)} (G-1,J) \times \langle \Upsilon(H=1) \rangle_{AB|C} =   \\
&&= J^{-x_{\phi}} {\cal B}_{B|C}^{(\phi)} (\theta_1) + {1 \over 2t_{BC}} \, J^{-x_{\phi}} |H-1| \times (\sin\theta_1) \, {d \over d \theta_1} \, {\cal B}_{B|C}^{(\phi)} (\theta_1) \times  \langle \Upsilon(H=1) \rangle_{AB|C} \nonumber
\end{eqnarray}
where (\ref{scalephiH}) and (\ref{scaleFH}) has been used in the last step. Here $\theta_1 \equiv {\rm arg}(H-1)$ is the angle that $H-1 = |H-1| \exp i \theta_1$ forms with the positive real axis and, via the mapping (\ref{semiinfstrip}) is, for large ${\cal W}/|z|$, given by
\begin{eqnarray} \label{theta1}
\theta_1 \equiv {\rm arg}(\cosh \tilde{z} -1) \to {\rm arg} {\tilde{z}^2 \over 2} + {\rm arg} \Bigl(1+ {\tilde{z}^2 \over 12} \Bigr) \to 2 \vartheta + \Biggl({ \pi |z| \over {\cal W} } \Biggr)^2  \times {\sin 2 \vartheta \over 12} \, .
\end{eqnarray}
The last factor in (\ref{avproductABC}) follows from Eq. (3.11) in Ref. \cite{BE21} and reads
\begin{eqnarray} \label{ups1}
\langle \Upsilon(H=1) \rangle_{AB|C} = (t_{AC} -t_{AB} -t_{BC}) /2\, ,
\end{eqnarray}
since $\zeta_1$ and $\zeta_2$ there are to be identified with 1 and $-1$, respectively. Since $|H-1| \to |\tilde{z}^2|^2 /2$, expanding (\ref{avproductABC}) up to order $(|z|/{\cal W})^2$ yields
\begin{eqnarray} \label{avproductABCprime}
&& J^{x_{\phi}} \times \langle \phi(G,J) \rangle_{AB|C} \to \\
&&\quad = {\cal B}_{B|C}^{(\phi)} (2 \vartheta) + {|z|^2 \over 2 t_{BC}} \, (\sin 2 \vartheta) \, {d \over d 2 \vartheta} \, {\cal B}_{B|C}^{(\phi)} (2 \vartheta) \times {1 \over 4} \bigl[t_{AC}-t_{AB}- {1 \over 3} t_{BC}  \bigr] \, \Bigl({ \pi \over {\cal W} } \Bigr)^2  \nonumber \, .
\end{eqnarray}
Inserting this in (\ref{productABC}) and taking the form of $\langle {\cal Y} \rangle$ in Eq. (\ref{semistripsdiffcornop}) into account leads to the expected expansion of $\langle \phi(x,y) \rangle$ shown in Eqs. (\ref{avdiffWEDGE}) and (\ref{scalephi}) for the present 90 degree corner where $g = 2$. This derivation is quite general, and uses no properties specific to the Ising model.

Besides this general argument it is instructive to confirm the COE by straightforward calculation for the two following $BB|C$ examples.

\subsubsection{$B|C$ corner in $BB|C$ strip} 

In this case of $A=B \neq C$, Eq. (\ref{semistripsdiffcornop}) tells us that
\begin{eqnarray} \label{Ysemiprimeprime}
\left\langle {\cal Y} \right\rangle =  {1 \over 6}  t_{BC} \, \Biggl({\pi \over {\cal W}}\Biggr)^2 \, .
\end{eqnarray}
Now consider two realizations.

\paragraph{$+|f$ corner in a $++|f$ strip} \label{strip++f}

For $\phi = \epsilon$ 
\begin{eqnarray} \label{epssemi}
\left\langle \epsilon(G,J) \right\rangle = {1 \over 2 J} \cos \theta \, , \quad \theta \equiv {\rm arg} (H-1) \, , 
\end{eqnarray}
so that
\begin{eqnarray} \label{epssemiprime}
\left\langle \epsilon(x,y) \right\rangle = {1 \over 2} \Psi \cos \theta  \, ,
\end{eqnarray}
with $\Psi$ from (\ref{uniformstrip}) which for small $|z|$ reduces to (\ref{semistripphiprime}). The transformation (\ref{semiinfstrip}) implies the expansion
\begin{eqnarray} \label{expandtheta}
\cos \theta \to \cos (2 \vartheta) - {1 \over 12} |\tilde{z}|^2 \, \sin^2 (2 \vartheta)
\end{eqnarray}
which should be compared with (\ref{epscaaprime}). On inserting this in (\ref{epssemiprime}) and expanding for small $|z|$ to leading and next-to-leading order, only the leading term of $\Psi$ in  (\ref{semistripphiprime}) contributes, and one finds
\begin{eqnarray} \label{epssemiexp}
\langle \epsilon (x,y) \rangle &\to& \langle \epsilon (x,y) \rangle_{+f} - {1 \over 12} |z| \sin (2 \vartheta) \times \Biggl( {\pi \over {\cal W}} \Biggr)^2 \, , \nonumber \\
\langle \epsilon (x,y) \rangle_{+f} &=& {1 \over |z|} \cot (2 \vartheta ) \, .
\end{eqnarray}
The first of these equations, together with $\langle {\cal Y} \rangle = (1/96)(\pi / {\cal W})^2$ from Eq. (\ref{Ysemiprimeprime}) and with the form of ${\cal F}_{+|f}^{(\epsilon)}$ given in the first equation (\ref{calF+f}), confirms the COE (\ref{opdiffWEDGE}). 

Besides deriving the form of the zero line $z=z_0$ of $\langle \epsilon \rangle$ from Eq. (\ref{epssemiexp}), a more direct way, like in Eqs. (\ref{zero-+-}), (\ref{zero_-+}) and (\ref{zerof+f}), (\ref{zerof+fprime}), is from its image $H=H_0 = 1+iJ$. By expanding $z_0 = ({\cal W} / \pi) \, {\rm arccosh}(1+iJ)$ to orders $J^{1/2}$ and $J^{3/2}$ one finds
\begin{eqnarray} \label{eps++fzero}
{\rm arg}z_0 \to {\pi \over 4} - {1 \over 24} \Bigl( {|z_0| \pi \over {\cal W}} \Bigr)^2 \, .
\end{eqnarray}

\paragraph{$+|-$ corner in $++|-$ strip}

Here the profile of the energy density in the half plane reads
\begin{eqnarray} \label{epssemi++-}
\left\langle \epsilon(G,J) \right\rangle = - {1 \over 2 J}  \Bigl(1-4\sin^2 \theta  \Bigr) 
\end{eqnarray}
with $\theta$ from (\ref{epssemi}), which implies 
\begin{eqnarray} \label{epssemi++-prime}
\left\langle \epsilon(x,y) \right\rangle = - {1 \over 2} \Psi \Bigl(1-4\sin^2 \theta  \Bigr) \, .
\end{eqnarray}
Since in conformity with (\ref{expandtheta}) 
\begin{eqnarray} \label{epssemi++-primeprime}
\sin^2 \theta  \to \sin^2 (2 \vartheta) \Biggl[ 1+ {1 \over 6} \Biggl({ \pi |z| \over {\cal W} } \Biggr)^2 \cos (2 \vartheta) \Biggr]
\end{eqnarray}
for $|z| \ll {\cal W}$ in leading and next-to-leading order, Eq. (\ref{epssemi++-prime}) is consistent with the operator expansion (\ref{opdiffWEDGE}) when the expressions of ${\cal F}_{+|-}^{(\epsilon)}$ in (\ref{calF+-}), of $\langle \epsilon \rangle_{+|-}$ in (\ref{epssig+-}), and the expression $\langle {\cal Y} \rangle = (1/12) (\pi / {\cal W})^2$ from (\ref{Ysemiprimeprime}) are taken into account. 

Deriving the two zero lines $z=z_{0+}$ and $z=z_{0-}$ either from the COE (\ref{opdiffWEDGE}) via (\ref{zero_-+}) or, more directly, from their images $H=H_{0+} = 1+J \exp(i \pi /6)$ and $1+J \exp(5 i \pi /6)$, yields
\begin{eqnarray} \label{zero_-+strip}
{\rm arg} \bigl( z_{0+} \, ,  \, z_{0-} \bigr) \, = \, \Bigl({\pi \over 12} \, , \, {5 \pi \over 12} \Bigr) -{1 \over 48} \, \bigl( |z_{0+}|^2 \, ,  \, |z_{0-}|^2 \bigr) \, \Bigl( {\pi \over {\cal W}} \Bigr)^2 \, .
\end{eqnarray}

\subsection{Squares: compact expressions and corner-perturbations prevented \\ by symmetry} \label{absence}

Consider squares with side length ${\cal L}$, with centers located at the origin, and with corners on the {\it axes} of the ($x_{\rm D}, y_{\rm D}$) coordinate system in the $z_{\rm D}=x_{\rm D}+iy_{\rm D}$ plane. By the conformal transformation in Appendix \ref{SquareDcircH} they are mapped to the upper half $H=G+iJ=|H|i \exp(i \psi)$ plane with their (NE, NW, SW, SE) sides mapped onto the intervals ($-\infty < G< -1, \, -1<G<0, \, 0<G<1, \, 1<G< \infty$) on the real axis $J=0$ and the horizontal diagonal, that extends along $y_{\rm D}=0$ between the left and right corners $-{\cal L}/\sqrt{2} < x_{\rm D} < {\cal L}/\sqrt{2}$, to the imaginary axis $G=0$ of the half plane. To simplify the following, the variable $Z_{\rm D} \equiv X_{\rm D}+iY_{\rm D} = z_{\rm D} \, {\bf K}\times \sqrt{2} /{\cal L}$ of Eq.  (\ref{ZDtozD}) is used, which at the left and right corners takes the values $Z_{\rm D}= \mp {\bf K}$. Here ${\bf K} \equiv {\bf K}(1/\sqrt{2}) = 1.85407$ with ${\bf K}(q)$ the complete elliptic integral.

Besides discussing how symmetry about the diagonal affects the behavior near the corners of the square, this Subsection presents compact expressions for the profiles all along the diagonal.

\subsubsection{Order parameter- and energy-densities in a square with a uniform boundary}

For a square in which all four edges have the {\it same} boundary universality class $a$, the densities of the primary operators $\phi= \sigma$ or $\epsilon$ in the corresponding half plane and circular disk are given by
\begin{eqnarray} \label{uniplane}
\langle \phi (G,J) \rangle = {\cal A}_{a}^{(\phi)} \, J^{-x_{\phi}}
\end{eqnarray}
and, due to the Moebius mapping $H(w_{\rm D}) = i(1+w_{\rm D})/(1-w_{\rm D})$, 
\begin{eqnarray} \label{unisqu}
\langle \phi (u_{\rm D}, v_{\rm D}) \rangle = {\cal A}_{a}^{(\phi)} \, \Biggl( {2 \over 1-u_{\rm D}^2 - v_{\rm D}^2 } \Biggr)^{x_{\phi}} \, ,
\end{eqnarray}
respectively. For the densities in the square, Eqs. (\ref{unitcircsqua}), (\ref{ZDtozD}) then yield
\begin{eqnarray} \label{uniSQU}
&& \qquad \langle \phi (x_{\rm D}, y_{\rm D} \rangle = {\cal A}_{a}^{(\phi)} \; \Bigl( 2 \, {\cal Q}(Z_{\rm D}) \, {\bf K} / {\cal L} \Bigr)^{x_{\phi}} \, , \nonumber \\
&& {\cal Q}(Z_{\rm D}) \equiv {|{\rm cn} Z_{\rm D}| \over |{\rm dn} Z_{\rm D}|^2 - (1/2) |{\rm sn} Z_{\rm D}|^2} = {\cal D}(Z_{\rm D}+ {\bf K}) = {\cal D}(Z_{\rm D}- {\bf K}) \, , 
\end{eqnarray}
where ${\rm sn}$, ${\rm cn}$, and ${\rm dn}$ are Jacobi functions and where 
\begin{eqnarray} \label{uniSQUprime}
\quad {\cal D}({\cal Z}) = \sqrt{2} \,  {|{\rm sn}{\cal Z} \, {\rm dn}{\cal Z}| \over 1- |{\rm cn}{\cal Z}|^2} \to 2^{3/2} \, {|{\cal Z}| \over {\cal Z}^2 + \bar{{\cal Z}}^2} \, \Bigl[1+{1 \over 160} \bigl( {\cal Z}^2 + \bar{{\cal Z}}^2 \bigr)^2 \, \Bigr] + O(|{\cal Z}|)^4 .
\end{eqnarray}
On the horizontal {\it diagonal} \cite{CSquareMidline} of the square, where $Z_{\rm D}=X_{\rm D}$ is real, Eq. (\ref{uniSQU}) reduces to the compact expression
\begin{eqnarray} \label{unireal}
{\cal Q}={1 \over {\rm cn}(X_{\rm D})} \; . 
\end{eqnarray}

The last form in Eq. (\ref{uniSQUprime}) serves to describe the neighborhood of the left and right corners. For $z_{\rm D}=-({\cal L}/\sqrt{2})+\widehat{z_{\rm D}} $ a small deviation $\widehat{z_{\rm D}} = |\widehat{z_{\rm D}}| \exp (i \varphi)$ apart from the {\it left} corner, the modulus of ${\cal Z}=Z_{\rm D}+ {\bf K}=\widehat{z_{\rm D}} \, {\bf K} \sqrt{2} / {\cal L}$ is small and (\ref{uniSQU}), (\ref{uniSQUprime}) yield the expansion  
\begin{eqnarray} \label{exuniSQU}
\qquad \langle \phi (x_{\rm D}, y_{\rm D} \rangle = {\cal A}_{a}^{(\phi)} \, \Biggl({2 \over |\widehat{z_{\rm D}}| \cos (2 \varphi)}  \Biggr)^{x_{\phi}} \times \Biggl[1+{x_{\phi} \over 10} \Biggl( |\widehat{z_{\rm D}}| \,{ {\bf K} \over {\cal L} }\Biggr)^4 \cos^2 (2 \varphi) \Biggr] \, .
\end{eqnarray}
In order to check the COE in Eq. (\ref{avWEDGE}), one finds, by a rotation $\widehat{z_{\rm D}} = z \exp(-i \pi /4)$ about the left corner, the required profile 
\begin{eqnarray} \label{exuniSQUxy}
\langle \phi (x, y \rangle =\Bigl[ \langle \phi (x_{\rm D}, y_{\rm D}) \rangle \Bigr]_{|\widehat{z_{\rm D}}| \to |z|, \, \varphi \to \vartheta - \pi /4} \, ,
\end{eqnarray}
since $z=|z|\exp(i \vartheta)$. Here $0< \vartheta < \pi /2$ and $-\pi /4 < \varphi < \pi /4$. Thus $\cos (2 \varphi)\to \sin (2 \vartheta)$, and the validity of the COE then follows on using the forms of ${\cal F}_a^{(\phi)}$ arising from Eq. (\ref{avWEDGE}), with $\mu_a^{(\phi)}$ given below (\ref{uniformBOE}) and of $\langle {\cal T} \rangle$ given in (\ref{calTsquare}) as well as of $\langle \phi \rangle_{a|a}$ given below (\ref{opWEDGE}).  

\subsubsection{Squares with mixed boundaries} \label{SQmixed}

There are special cases of $a|b$ corners in which the shape and boundary conditions of the system generate a profile $\langle \phi \rangle$ with a symmetry
which is incompatible with the form of the prefactors ${\cal F}_{a|b}^{(\phi)}$ in the expression (\ref{opdiffWEDGE}). In these cases $\langle {\cal Y} \rangle$ must vanish, and the perturbation is of higher order than predicted by (\ref{opdiffWEDGE}). For illustration consider squares with mirror symmetry about the horizontal diagonal $y_{\rm D}=0$. Due to the mapping in Appendix \ref{SquareDcircH}, this corresponds to a mirror symmetry about the imaginary axis $G=0$ in the upper half $H$ plane. Along the diagonal simple compact expressions for the profiles are obtained and, as in Eqs. (\ref{exuniSQU}), (\ref{exuniSQUxy}), their behavior near the left corners of the square is compared with the COE in Sec. \ref{diffbc}.

(i) Consider boundary conditions $(+,+,f,f)$ on the $({\rm NE,NW,SW,SE})$ sides of the square. In the corresponding $H$ plane this implies a single switch from $+$ to $f$ at the origin and the profiles $\langle \phi (G,J) \rangle$ can be taken from Eqs. (4.1) in Ref. \cite{BX}.  

The energy density in the square is antisymmetric about the horizontal diagonal of the square, i.e. it changes sign on replacing $\varphi \to - \varphi$, since the corresponding form ${\cal A}_{+}^{(\epsilon)} J^{-1} \sin \psi$ with ${\cal A}_{+}^{(\epsilon)} = -(1/2)$ in the upper half $H=|H|i \exp (i \psi)$ plane is antisymmetric about the imaginary axis $\psi =0$. Due to Eq. (\ref{exuniSQUxy}), this is compatible with $\langle \epsilon (x,y) \rangle_{+|f}= - |z| \tan (2 \varphi)$ but incompatible with ${\cal F}_{+|f}^{(\epsilon)} (x,y) = -8 |z| \cos (2 \varphi)$ from Eq. (\ref{calF+f}) so that $\langle {\cal Y} \rangle$ must vanish. 

Now consider the density of the order parameter in the square and its counterpart in the upper $H$ plane, ${\cal A}_{+}^{(\sigma)}J^{-1/8} \bigl[ \sin (\psi /2) + \cos (\psi /2) \bigr]^{1/2} \times 2^{-1/4}$ where ${\cal A}_{+}^{(\sigma)} = 2^{1/8}$. Obviously both have no symmetry properties of the type discussed above for the energy density. Still, the vanishing of $\langle {\cal Y} \rangle$ implies a dependence of $\langle \sigma \rangle$ on the distance $|\widehat{z_{\rm D}} \equiv z_{\rm D}+{\cal L}/\sqrt{2}|$ from the left corner of the square leading, in next-to-leading order, to a power law with an exponent $\lambda$ larger than the exponent $2-(1/8)$ predicted by ${\cal F}_{+|f}^{(\sigma)} \propto |z|^{2-(1/8)} \sin^{3/8} \vartheta \cos^{15/8} \vartheta$ from (\ref{calF+f}). That this is the case and that the exponent equals $\lambda = 4-(1/8)$ is implied by the result
\begin{eqnarray} \label{++ffSQU}
\langle \sigma (x_{\rm D}, y_{\rm D}=0) \rangle = {\cal A}_{+}^{(\sigma)} \; \Biggl( {2 \, {\bf K} / {\cal L} \over  {\rm cn} (X_{\rm D}) }\Biggr)^{1/8} \times 2^{-1/4} \, 
\end{eqnarray}
for the order parameter profile all along the horizontal diagonal of the square. Eq. (\ref{++ffSQU}) follows from observing that along the imaginary axis $\psi =0$ of the upper half $H$ plane, the profile of the order parameter given at the  beginning of this paragraph and the corresponding profile (\ref{uniplane}) for the uniform boundary are equal apart from an overall factor $2^{-1/4}$.  

(ii) For boundary conditions $(-,+,-,+)$ on the $({\rm NE,NW,SW,SE})$ sides of the square, the energy density and the order parameter are obviously symmetric and antisymmetric, respectively, about the horizontal diagonal of the square and about the imaginary axis of the half plane, i.e., about $\varphi =0$ and $\psi=0$. This is incompatible with the forms ${\cal F}_{+|-}^{(\epsilon)}(x,y) \propto - \sin (4 \varphi)$ and ${\cal F}_{+|-}^{(\sigma)}(x,y) \propto  \cos^{15/8} (2 \varphi)$, which are antisymmetric and symmetric, respectively, for $\varphi \to - \varphi$, i.e., on exchanging the values of $x$ and $y$. Likewise, in the corresponding half plane it is incompatible with the forms considered below (\ref{calF+-}) of $F_{+-}^{(\epsilon)}$ and $F_{+-}^{(\sigma)}$, which are antisymmetric and symmetric, respectively, for $\psi \to - \psi$ in our notation $H=|H|i \exp (i \psi)$. Thus for the  boundary conditions considered here the average $\langle {\cal Y} \rangle$ of the corner operator ${\cal Y}$ and that of the boundary operator $\Upsilon$ in the half plane introduced in (\ref{avdiffUHP}) must vanish, and the perturbations must be of higher order. The vanishing of $\langle {\cal Y} \rangle$ is confirmed by taking the square limit $q=q'=1/\sqrt{2}$ in the expression (\ref{rect-+-+}) below for the rectangle with corresponding boundary conditions.

For later comparison note the simple results,
\begin{eqnarray} \label{epshordia}
\langle \epsilon (G=0,J) \rangle = {3 \over 2J}-{8 \over3}{\zeta^2 J \over (\zeta^2 +J^2)^2} \, ,
\end{eqnarray}
following from Eqs. (16) in Ref. \cite{TWBG2}, and 
\begin{eqnarray} \label{epshordiaprime}
\langle \epsilon (x_{\rm D}, y_{\rm D}=0) \rangle = {{\bf K} \over {\cal L}} \Biggl( {3 \over {\rm cn}(X_{\rm D})}- {4 \over 3} \, {\rm cn}^3 (X_{\rm D}) \Biggr)  \, , \quad X_{\rm D} \equiv x_{\rm D} {{\bf K} \over {\cal L}/\sqrt{2}}\, ,
\end{eqnarray}
for the energy densities on the imaginary axis of the upper half plane, with switching points $-\zeta, \, 0, \, \zeta$ instead of $-1,0,1$, and on the horizontal diagonal of the square \cite{insufficient}, respectively. Note that the second term on the rhs of (\ref{epshordia}) vanishes for $\zeta =0$ and for $\zeta = \infty$, in which case the $-+-+$ boundary of the half plane reduces to a $-+$ and a $+-$ boundary, respectively. 

(iii) Unlike the two preceding cases, for boundary conditions $(f,+,f,+)$ there is no symmetry or antisymmetry \cite{fweak}, ruling out the leading perturbation of the $+|f$ wedge due to finite ${\cal L}$, and $\langle \Upsilon \rangle$ and $\langle {\cal Y} \rangle$ are nonvanishing. The counterparts of the relations (\ref{epshordia}) and (\ref{epshordiaprime}) read
\begin{eqnarray} \label{epshordia+f}
\langle \epsilon (G=0,J) \rangle = -(\sqrt{2}-1) {\zeta \over \zeta^2 +J^2} 
\end{eqnarray}
and
\begin{eqnarray} \label{epshordiaprime+f}
\langle \epsilon (x_{\rm D}, y_{\rm D}=0) \rangle = -(\sqrt{2}-1) {{\bf K}\over {\cal L}} \, {\rm cn}(X_{\rm D}) \to -(\sqrt{2}-1) \Biggl({{\bf K}\over {\cal L}}\Biggr)^2 \, \widehat{x_{\rm D}} \, 
\end{eqnarray}
where $\widehat{x_{\rm D}}=x_{\rm D}+{\cal L}/\sqrt{2}$ is the distance from the left corner. 
The expression (\ref{epshordia+f}) follows from Eq. (2.41) in Ref. \cite{BE21}. In agreement with the duality arguments given in \cite{fweak}, the expression  is nonvanishing, reflecting the lack of antisymmetry about the imaginary axis, except for $\zeta =0$ and $\zeta = \infty$ where the boundary condition $f+f+$ reduces to the single switch cases $+f$ and $f+$, respectively, for which the antisymmetry $\langle \epsilon (-G,J) \rangle =-\langle \epsilon (G,J) \rangle$ applies. The last expression in (\ref{epshordiaprime+f}) gives the behavior near the left corner. It is in agreement with the COE prediction since the energy density along the diagonal $\vartheta=\pi /4$ vanishes in the unperturbed $+|f$ wedge where ${\cal L}=\infty$, compare (\ref{epssig+f}), and since the leading behavior for large ${\cal L}$ is determined by ${\cal F}_{+|f}^{(\epsilon)}(\vartheta = \pi /4)= -8|z|= -8 \widehat{x_{\rm D}}$ and $\langle {\cal Y} \rangle = (1/8) (\sqrt{2}-1)({\bf K} / {\cal L})^2 $, compare (\ref{calF+f}) and (\ref{rectf+f+}) below in the square limit $q'=1/\sqrt{2}$. 

Now turn to the order parameter. Obviously it does not display a symmetry wrt the diagonal, since even the unperturbed $\langle \sigma \rangle_{+|f}$ does not, see Eq. (\ref{epssig+f}). The relations  corresponding to (\ref{epshordia+f}), (\ref{epshordiaprime+f}) read
\begin{eqnarray} \label{ophordia+f}
\langle \sigma (G=0,J) \rangle &=& \Biggl( {2 \over J} \Biggr)^{1/8} \, 2^{-1/2} \, \Biggl[ \sqrt{\bigl( 1+(\zeta /R) \bigr) \, \bigl( 1+(J /R) \bigr)} -  \\
&-& (\sqrt{2}-1)^2 \, \sqrt{\bigl( 1-(\zeta /R) \bigr) \, \bigl( 1-(J /R) \bigr)} \Biggr]^{1/2} \, , \quad R \equiv \sqrt{\zeta^2 + J^2} \nonumber
\end{eqnarray}
and
\begin{eqnarray} \label{ophordiaprime+f}
\langle \sigma (x_{\rm D}, y_{\rm D}=0) \rangle &=& \Biggl({{\bf K}\over {\cal L} \, {\rm cn}(X_{\rm D})}\Biggr)^{1/8} \, 2^{1/4} \, \bigl( \sqrt{2} -1 \bigr)^{1/2} \, \Bigl[ {\rm dn}(X_{\rm D}) +1 \Bigr]^{1/2} \, \to \nonumber \\
&\to& \widehat{x_{\rm D}}^{-1/8} + \widehat{x_{\rm D}}^{2-(1/8)} (\sqrt{2}-1) {\bf K}^2 /(4 {\cal L}^2)\, .
\end{eqnarray}
Eq. (\ref{ophordia+f}) reproduces, in the two single-switch limits $\zeta =0$ and $\zeta = \infty$, the result $\langle \sigma (G=0,J) \rangle = (2/J)^{1/8} \times 2^{-1/4}$ known from Eq. (4.1) in Ref. \cite{BX}. The last line in (\ref{ophordiaprime+f}) is consistent with the COE, since on using (\ref{exuniSQUxy})  its first term is reproduced by $\langle \sigma (x, y=0) \rangle_{+|f}= |z|^{-1/8}$ from (\ref{epssig+f}) for $\vartheta = \pi /4$ and its second term by the product of ${\cal F}_{+|f}^{(\sigma)}= 2 |z|^{2-(1/8)}$ from (\ref{calF+f}) for $\vartheta = \pi /4$ and $\langle {\cal Y} \rangle$ for the present square which is given in the text below (\ref{epshordiaprime+f}).

\subsection{Corner of a rectangle} \label{CofRect}

Finally consider the corner at $z=0$ of a rectangle extending over the domain $0<y<{\cal W}$, $0<x< {\cal H}$. The arrangement of boundary conditions $A,B,C$, and $D$ along the top, left, bottom, and right boundaries of the rectangle, introduced in-between Eqs. (\ref{uniformmid}) and (\ref{stiched}) above, we  denote now by $AB|CD$. Here the results for the average of the corner operator at $z=0$ are presented. The derivation of these results as well as confirming the validity of the corresponding COE is deferred to Appendix \ref{RecCorn}.    

\subsubsection{Rectangle with uniform boundary condition $a$} \label{rectangleuniform}
  
For uniform boundary conditions $aa|aa$ it is shown in Appendix \ref{RecCorn} that 
\begin{eqnarray} \label{unitaurect}
&& \qquad \langle T(z) \rangle - \langle T(z) \rangle_{a|a} \to 4z^2 \, \langle {\cal T} \rangle \, , \nonumber \\
&& \langle {\cal T} \rangle = - {\hat{c} \over 60} \, \bigl(1+ q^4 + q'^4 \bigr) \big/ \Lambda^4 \, .
\end{eqnarray}
The quantities $q, q'$ and the length $\Lambda$ depend on the aspect ratio and the size of the rectangle, as explained in Eqs. (\ref{alternative}) and (\ref{Lambdaprime}) or (\ref{aspq}) and (\ref{Lgeo}). From their form the expected ${\cal W} \leftrightarrow {\cal H}$ invariance of $\langle {\cal T} \rangle$ can be read off immediately.

It is easy to see that the results for $\langle {\cal T} \rangle$ given in (\ref{semistripscornop}) for the semi-infinite strip and in (\ref{calTsquare}) for the square are special cases of (\ref{unitaurect}). For the semi-infinite strip, ${\cal H} \to \infty$, so that by (\ref{aspq}) $q \to 1$,  $q' \to 0$, and the bracket in (\ref{unitaurect}) is equal to 2. Since by (\ref{Lgeo}) $1/\Lambda^4 \to \bigl({\bf K}(0) / {\cal W} \bigr)^4 \equiv \bigl( \pi / (2 {\cal W}) \bigr)^4$, the rhs of (\ref{unitaurect}) reduces to $\langle {\cal T} \rangle$ given in (\ref{semistripscornop}) when $A=B$. For the square with ${\cal W}={\cal H} \equiv {\cal L}$ the bracket in (\ref{unitaurect}) equals $3/2$ since $q=q' = 1/ \sqrt{2}$, and (\ref{unitaurect}) reduces to $\langle {\cal T} \rangle$ given in (\ref{calTsquare}).     

\subsubsection{Rectangle with $aa|bb$ boundary conditions} \label{aabbRectangle}

For an $aa|bb$ rectangle the top and left edges have boundary condition $a$ while the bottom and right edges have boundary condition $b$. In Appendix \ref{RecCorn} it is shown that in this case 
\begin{eqnarray} \label{rectdiffYcornaabb}
\langle T(z) \rangle - \langle T(z) \rangle_{a|b} \to  {8 \over 3} \, t_{ab} \, (q^2 - q'^2) /\Lambda^2 = 4 \, \langle {\cal Y} \rangle \, .
\end{eqnarray}
In the limit of the semi-infinite strip ${\cal H} \to \infty$ where $q \to 1$,  $q' \to 0$, Eq. (\ref{rectdiffYcornaabb}) reduces to Eq. (\ref{Ysemiprimeprime}). In the limit of the square ${\cal H} = {\cal W}$, $q=q'$, and $\langle {\cal Y} \rangle$ {\it vanishes}, in agreement with the symmetry-argument given above (\ref{++ffSQU}) in part (i) of Sec. \ref{SQmixed}. On exchanging the values of ${\cal W}$ and ${\cal H}$, the average $\langle {\cal Y} \rangle$ in (\ref{rectdiffYcornaabb}) retains its magnitude but changes sign. 

\subsubsection{Rectangle with $-+|-+$ boundary conditions}

For a rectangle with boundary condition $+$ on the two vertical edges and $-$ on the two horizontal edges, the stress tensor for $z \to 0$ reads
\begin{eqnarray} \label{rect-+-+}
&&\langle T(z) \rangle - \langle T(z) \rangle_{a|b} \to - {4 \over \Lambda^2} \,  \Big(q^2 -q'^2 \Big) \, \times \nonumber \\
&&\qquad \qquad \times \Biggl[1-{(q^2 -q'^2)^2 \over 9} \Biggr] \Bigg/ \Biggl[1+{(q^2 -q'^2)^2 \over 3} \Biggr] = \, 4 \, \langle {\cal Y} \rangle  \, ,
\end{eqnarray}
as shown in Appendix \ref{RecCorn} . The average $\langle {\cal Y} \rangle$ vanishes for the square according to the symmetry argument given above Eq. (\ref{epshordia}) in part (ii) of Sec. \ref{SQmixed}, and it changes its sign on exchanging ${\cal W}$ and ${\cal H}$. Again the result for $q' \to 0$ has been checked against the semi-infinite strip. 

Note that (\ref{rect-+-+}) and  (\ref{rectdiffYcornaabb}) have {\it opposite} signs. This is understood most easily in the limit $q' \to 0$, i.e. ${\cal H} \to \infty$, of the semi-infinite strip where ${\cal F}_{a|b}^{(\phi)}$ describes how the infinite wedge with apex at $z=0$ is perturbed by the upper horizontal edge. In the case of Eq. (\ref{rectdiffYcornaabb}) the upper horizontal edge with boundary condition $a$ {\it enhances} the $a$-effect of the vertical edge and reduces the $b$-effect of the lower horizontal edge. In contrast in the case of Eq. (\ref{rect-+-+}) the upper horizontal $-$ edge {\it reduces} the $+$ effect of the vertical $+$ edge and enhances the lower horizontal $-$ edge. For example, compare the $aa|bb = ++|--$ case with the $-+|-+$ case. In the corresponding expansion $\langle \sigma \rangle \to \langle \sigma \rangle_{+|-} + {\cal F}_{+|-}^{(\sigma)} \times \langle {\cal Y} \rangle$ of the order parameter profile, $\langle \sigma \rangle_{+|-}$ is negative (positive) near $\vartheta = 0$ ($\vartheta = \pi /2$) while ${\cal F}_{+|-}^{(\sigma)}$ is always positive, see Eq. (\ref{calF+-}). Thus for $aa|bb = ++|--$ and $-+|-+$ in the semi-infinite strip limit, $\langle {\cal Y} \rangle$ must be positive and negative, respectively, to generate the enhancement/reduction effect described above. In the other limit $q' \to 1$, i.e. ${\cal W} \to \infty$, the semi-infinite strip extends in vertical rather than horizontal direction, and the enhancement vs. reduction is reversed.

\subsubsection{Rectangle with $f+|f+$ boundary conditions} \label{recf+f+}

For vertical edges $+$, as before, but boundary condition $f$ on the two horizontal edges, one obtains
\begin{eqnarray} \label{rectf+f+}
\langle T(z) \rangle - \langle T(z) \rangle_{a|b} \to  \Bigl(q' - {1+q'^2 \over 3} \Bigr) \Big/ \Lambda^2 = 4 \, \langle {\cal Y} \rangle \, ,
\end{eqnarray}
as shown in Appendix \ref{RecCorn}.  While the enhancement vs. reduction effects on the two edges meeting at $z=0$ have, in the two limits $q' \to 0$ and $q' \to 1$, the same signs as in the case of Eq. (\ref{rect-+-+}), their  vanishing, i.e. the vanishing of $\langle {\cal Y} \rangle$, does not happen for the square where $q'=1/\sqrt{2}$ but instead for $q'=(3-\sqrt{5})/2 \approx 0.382$, corresponding to the aspect ratio ${\cal H}/{\cal W}=1.470511$ between the horizontal $f$ edges and the vertical + edges. This is in agreement with the property that ``$+$ dominates $f$'' that we found in Sec. \ref{competitionf+}.
The consequence of the vanishing on the zero line of the energy density entering the rectangle's corner is discussed in point (3) of this section. The role of nonvanishing $\langle {\cal Y} \rangle$ for the COE in the square is discussed in part (iii) of Sec. \ref{SQmixed}.

\section{SUMMARY AND CONCLUDING REMARKS}

Boundary critical phenomena have been investigated both theoretically and with simulation for various geometries. On the theoretical side studies began with the simple  geometry of a half space where the boundary is an infinite plane. Simulation studies involve finite systems for which the boundary has a more complicated form. For lattice simulations in two spatial dimensions a paradigmatic geometry is a rectangular domain. 

In this paper a field theoretical study of critical density profiles in rectangular domains is presented. We do not consider pseudo-rectangles with periodic boundaries in one direction but rather rectangles with four genuine boundary sides. This offers the opportunity to study the interesting boundary effects coming from the corners. The main emphasis is on mixed boundaries with equal boundary conditions on opposing sides but different for the horizontal and vertical sides. Since the main interest is in the Ising model, the combinations $-+$ and $f+$ are considered.

The density profiles $\langle \phi \rangle$ of the order parameter $\phi = \sigma$ and the energy $\phi = \epsilon$ in the rectangle at criticality are evaluated by means of conformal mappings from their known counterparts \cite{BX,TWBG1,TWBG2,BE21} in the upper half plane. The results crucially depend on the aspect ratio ${\cal H}/{\cal W}$ where ${\cal H}$ and ${\cal W}$ are the lengths of the horizontal and vertical sides of the rectangle. In the limits ${\cal H} \gg {\cal W}$ and ${\cal H} \ll {\cal W}$ the effects coming from the two shorter sides decouple, and the behavior degenerates to that of two semi-infinite strips. The latter are interesting in their own right and are discussed in Sec. \ref{seminfstrip}. Results for ${\cal H}/{\cal W}$ of order 1 are presented in Sec. \ref{recmix}. 

\begin{figure}
	\begin{center}
		\includegraphics*[height=0.50\textheight]{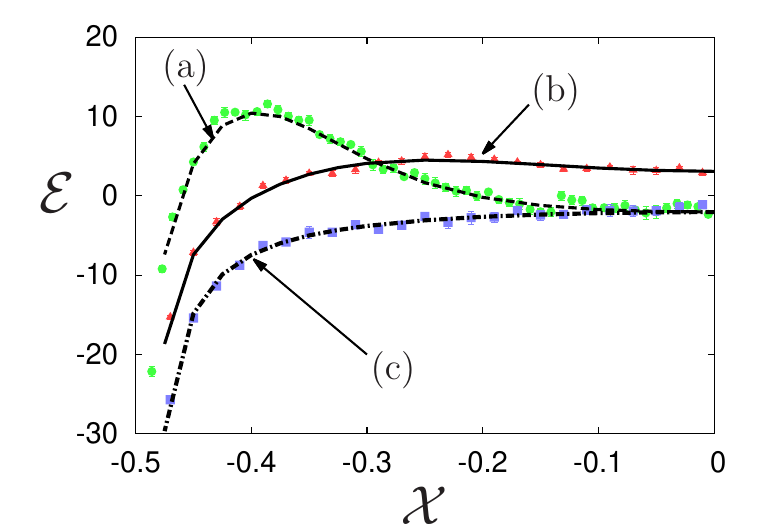}\\*
	\end{center}
	\caption{Dependence of the energy density along the horizontal midlines of the three rectangles (a), (b), (c) of FIG. \ref{Henkel} with boundary conditions $-$ and $+$ on the horizontal and vertical sides, respectively. Shown is the scale-invariant form ${\cal E} \equiv ({\cal H} \times {\cal W})^{1/2} \, \langle \epsilon \rangle$ introduced in Eq. (\ref{scinvhoreps}). Moving from ${\cal X}=-0.5$ to ${\cal X}=0$ means moving along the horizontal midlines from their left ends at the left vertical boundary sides to the centers of the corresponding rectangles. Analytic results (line of dashes, full line, dash-dotted line) are compared with those from simulations (circles, triangles, squares). While the former are obtained via Eq. (\ref{scinvhorepsprime}), the latter are  taken from the Monte Carlo simulations in Ref. \cite{OAV} for Ising models on a square lattice with 220 $\times$ 100 spins for rectangle (a), with 100 $\times$ 220 spins for rectangle (c), and with 100 $\times$ 100 spins for the square (b). 
	The simulation method used in Ref. \cite{OAV} is explained in Appendix \ref{lattmc} and is a correspondingly adapted version of the Monte Carlo simulation described in the paper in Ref. \cite{engydens}. Good agreement is found between the analytic and simulation results which are normalized without any adjustible parameter according to Ref. \cite{primaryfields} and Appendix \ref{lattmc}, respectively. For the center of the square, for example, the simulation result of ${\cal E} \approx 3.14$ agrees well with the analytic expression ${\cal E}=(5/3){\bf K}(1/\sqrt{2}) =3.090$.    }\label{-+completehorizontalOleg}
\end{figure}

The mixed boundary conditions and variable aspect ratio lead to a rich behavior of the density profiles. To illustrate this, now consider two examples. They involve the energy density $\langle \epsilon \rangle$ as defined in \cite{primaryfields}, which when positive and negative signifies stronger and weaker local disorder, respectively, than in the infinite bulk. Thus, in the half plane with uniform boundary condition $f$ and $+$ or $-$, $\langle \epsilon \rangle$ is positive and negative, respectively. In the midline of a $+-$ wedge the competition between the two differently ordered sides leads to strong disorder with $\langle \epsilon \rangle >0$ while approaching one of the two sides of the wedge away from the tip leads to $\langle \epsilon \rangle <0$ \cite{BXwedge}. These phenomena help to understand the interesting behavior of $\langle \epsilon \rangle$ in the rectangle $-+-+$ with horizontal $-$ and vertical $+$ sides that is discussed in detail in Sec. \ref{+-+-}. In the center of the corresponding square (${\cal H}/{\cal W}=1$), $\langle \epsilon \rangle$ is positive due to the strong disordering effects of the four $+-$ corners. On the other hand for ${\cal H}/{\cal W} \gg 1$ or $\ll 1$ the ordering $-$ or $+$ sides are much closer to the center than the corners, and $\langle \epsilon \rangle$ is negative in the center, cf. FIG. \ref{Henkel}. Thus at some intermediate value of the aspect ratio $\langle \epsilon \rangle$ in the center must vanish. This happens for ${\cal H}/{\cal W} =1.5172$ and its inverse.  An even more remarkable consequence of this competition is the appearence of two symmetric maxima when $\langle \epsilon \rangle$ moves along midlines of the rectangle, see FIG. \ref{-+completehorizontalOleg} (a). The maxima appear for the square and, for all aspect ratios, when moving along the longer midline. In particular a maximum appears along the midline of the corresponding semi-infinite strip, as discussed in Sec. \ref{semi-+-}. Along the shorter midline of the rectangle the maxima only appear when the aspect ratio is sufficiently close to 1. We note that unlike $\langle \epsilon \rangle$, the stress tensor density $\langle T \rangle$ does vanish at the center of the $-+-+$ square, due to the combination of the 90 degree symmetry of the square's shape and the $+-$ symmetry of the stress tensor, see Eq. (\ref{Tcenter-+}).

Another interesting competition between disorder and order arises from the combination of disordering and ordering sides in a rectangle with horizontal $f$ and vertical $+$ sides that are discussed in Sec. \ref{+f+f}. Consider, e.g., the diagonal of the corresponding square. On approaching the $+f$ corners, $\langle \epsilon \rangle \to 0$, see Eq. (\ref{epshordiaprime+f}), but in the center of the square the competition is not balanced and $\langle \epsilon \rangle$ is negative, cf. curve (iii) in FIG. \ref{f+completehorizontalOleg}. A vanishing $\langle \epsilon \rangle$ in the center of the rectangle requires the length ${\cal H}$ of the  $f$ side to be longer by a factor 1.279 than the length ${\cal W}$ of the vertical $+$ side \cite{fweak}. Likewise the vanishing of the stress tensor in the center requires ${\cal H}/{\cal W}=2.413$, see Eq. (\ref{Tcenterf+}) and Appendix \ref{Janus}. In this sense $+$ dominates over $f$ \cite{instable}. 

Results of simulations on a square lattice \cite{OAV} for the two above-mentioned examples compare surprisingly well with the analytic predictions, without using any adjustible parameter, an apparent confirmation of universality. See FIG. \ref{-+completehorizontalOleg} and FIG. \ref{f+completehorizontalOleg} as well as Appendix \ref{lattmc}. 

In the detailed discussion of the density profiles and their dependence on the aspect ratio in Sec. \ref{recmix}, two points are investigated in particular:

(i) The behavior at the center and along the entire horizontal and vertical midlines of the rectangle. As discussed in the two above examples, the simple results for $\langle \epsilon \rangle$, $\langle \sigma \rangle$, $\langle T \rangle$ at the center given in Eqs. (\ref{epsaddreccent}), (\ref{midpointsig+-+-}), (\ref{Tcenter-+}) and (\ref{ep+f+f+00}), (\ref{sig+f+f+center}), (\ref{Tcenterf+}) help to understand important features of the aspect-ratio dependence in the $-+-+$ and $f+f+$ rectangles, respectively. Near the ends where the midlines touch the centers of the boundary sides, the boundary operator expansion (BOE) discussed in Appendix \ref{BOEmid} connects the aspect-ratio dependence of $\langle \epsilon \rangle$ and $\langle \sigma \rangle$. See, in particular, the discussion in the long paragraph following Eq. (\ref{vertBOE}). For the behavior along the entire midlines, see FIGs \ref{-+completehorizontalOleg} and \ref{f+completehorizontalOleg}.

(ii) To understand the behavior in the entire rectangle, it is useful to discuss the structure of the ``zero-lines'' along which the profiles vanish, separating regions in the rectangle of positive and negative $\langle \epsilon \rangle$ or $\langle \sigma \rangle$, i.e., regions where the values of the profiles are larger and smaller, respectively, than in the bulk. The lines originate and end in the corners with mixed boundary conditions, see Sec. \ref{CEOzeroline} and FIGs. \ref{Henkel} and \ref{Ted}. 

Note that the density profiles can be used to evaluate the free energy of interaction with a small embedded particle \cite{SMED}. An example involving the density of the stress tensor is presented in Appendix \ref{Janus}.

Section \ref{coordaxes} of the paper is devoted to a systematic study of the profiles $\langle \phi \rangle$ near a {\it corner}. We concentrate on the SW corner of the rectangle and place its tip or apex at the origin of the $z=x+iy$ plane, as described in Appendix \ref{firstquadrant}. To leading order in the small distance $|z|$, the profile equals the one inside the wedge formed by the two corner sides when their length is infinite, i.e., it is independent of the size and aspect ratio of the rectangle. The dependence on the latter arises from the deviations of the rectangle from the infinite wedge which appear from a point $(x,y)$ close to the tip as distant perturbations of the infinite wedge. To evaluate their effect (and that of other distant perturbations) on $\langle \phi \rangle$, the operator $\phi(x,y)$ can be replaced by a series of operators $O$ located right in the tip and multiplied with amplitudes that carry the $(x,y)$-dependence. While both the operators $O$ and their amplitudes are independent of the perturbations, the latter enter the averages $\langle O \rangle$. For a given opening angle of the corner, the operators $O$ depend on the boundary conditions of its two sides only, while their amplitudes depend in addition on whether one considers $\phi = \epsilon$ or $\phi = \sigma$. For an opening angle of 180 degrees the present ``Corner Operator Expansions'' (COE)   reduce to the boundary operator expansions (BOE) for a flat boundary like the ones discussed in Ref. \cite{BE21}. For the opening angle of 90 degrees, like in the SW corner of the rectangles, the leading operator has a scaling dimension twice as large as the dimension of the corresponding BOE operator. So the dimension of the operator ${\cal Y}$ for two different sides of the corner is $(1/{\rm length})^2$. For equal sides it is $(1/{\rm length})^4$ except for the COE of $\sigma$ in an $ff$ corner \cite{fsigma} where in the Ising model it is  $(1/{\rm length})$, and the two corresponding operators are ${\cal T}$ and ${\cal S}$, respectively. For the SW corner of the rectangles the COE is confirmed to lowest order by explicitly evaluating the averages of ${\cal T}$ and ${\cal Y}$ for arbitrary aspect ratio, see Sec. \ref{CofRect} and Appendix \ref{RecCorn}. As for  ${\cal S}$, the corresponding COE given in Ref. \cite{fsigma} is confirmed for the  $f|f$ wedge with an $f+$ switch and for the $+f|f$ semi-infinite strip in Ref. \cite{fsigma} and in paragraph \ref{strip+ff}, respectively.

As indicated above the aspect-ratio dependence of the profiles near the centers of the rectangle's {\it sides} is evaluated. Together with the BOE's in Appendix \ref{BOEmid} the results follow from the expressions for the stress tensors at these centers given in Eqs. (\ref{Smidendsprime}) ff.

Both the BOE's and the COE's are based on general principles of conformally invariant field theories in two spatial dimensions and are not limited to the Ising model. 

\vspace{1cm}

{\bf ACKNOWLEDGMENTS}

\vspace{0.3cm}

I thank T. W. Burkhardt for useful discussions and O. A. Vasilyev for allowing me to show in FIGs.  \ref{-+completehorizontalOleg} and \ref{f+completehorizontalOleg} his unpublished Monte Carlo results.  

\appendix

\section{BOUNDARY OPERATOR EXPANSIONS \, (BOE)} \label{BOEmid}

Boundary operator expansions (BOE's) are a useful tool for investigating the density profiles $\langle \phi \rangle$, $\phi = \epsilon$ or $\sigma$, near an {\it internal} point of the horizontal and vertical boundary sides of our semi-infinite strips and rectangles. Here ``near'' means that $\phi$ is located much closer to the boundary than to the corners where the boundaries intersect. For the {\it horizontal} boundary of universality class $a$ of the {\it upper} half $H=G+iJ$ plane $J>0$, the expansion reads
\begin{eqnarray} \label{horizBOE}
{\phi(G,J) \over {\cal A}_a^{(\phi)} J^{-x_{\phi}}} \to 1 - J^2 \, {4 x_{\phi} \over \hat{c}} \, T(H=G) \, , \quad J \to 0 \, ,
\end{eqnarray}
as discussed in detail in the paragraph containing Eq. (\ref{uniformBOE}) and in Sec. III A of Ref. \cite{BE21}. The corresponding expansion near the {\it vertical} boundary of the {\it right} half $H=G+iJ$ plane $G>0$ follows by a rotation of 90 degrees, in which $T$ acquires a minus sign in front \cite{horizvertminusT}, so that
\begin{eqnarray} \label{vertBOE}
{\phi(G,J) \over {\cal A}_a^{(\phi)} G^{-x_{\phi}}} \to 1+ G^2 \, {4 x_{\phi} \over \hat{c}} \, T(H=iJ) \, , \quad G \to 0 \, . 
\end{eqnarray}
Note that at the two boundaries, $T$ and $-T$ are the components of the Cartesian stress tensor perpendicular to the boundary \cite{CardBoundaryCritPhen}. Due to their local character these expansions apply also to the horizontal and vertical boundary sides of the semi-infinite strips and rectangles. 

For the density profiles along the midlines of rectangles, the BOE's provide information on the effect of the aspect ratio near the midline ends at the centers of the boundary sides. This arises from the dependence on the aspect-ratio of the average of $T$ at the centers, which can even change sign at some value $({\cal H}/{\cal W})_0$ of the aspect ratio. Thus, the asymptotic behavior of the profiles near the boundary, which is independent of the aspect ratio, can be enhanced or weakened in next order for aspect ratio smaller or larger than $({\cal H}/{\cal W})_0$, and this change appears for $\langle \epsilon \rangle$ and $\langle \sigma \rangle$ at the same value of the aspect ratio. Near the left vertical boundary of a horizontal midline, in particular, the equivalent of Eq. (\ref{vertBOE}) in the rectangle tells us that the asymptotic behavior of decreasing magnitude of $\langle \phi \rangle$ with increasing distance from the boundary is supported and suppressed, respectively, for positive and negative value of the average of the boundary operator $T$. In a rectangle with horizontal $-$ and vertical + boundary conditions, for example, we show near Eq. (\ref{q01+-}) that the average of $T$  is positive and negative for small and large ${\cal H}/{\cal W}$, respectively. The ensuing enhancement and suppression in the profiles $\langle \epsilon \rangle$ and $\langle \sigma \rangle$ is as expected intuitively.  

It is instructive to check the BOE's by calculating the stress tensor and the profile independently. For the latter it is convenient to split off its behavior for {\it uniform} boundary conditions, as in Eq. (\ref{generalstrip}), and investigate the near-boundary behavior for the two factors separately. 

For example, consider the density profile along the midline in the {\it semi-infinite strip} with $ABC$ boundary conditions of Sec. \ref{seminfstrip} close to its end point $z=i {\cal W}/2$ at the vertical $B$ boundary. Here the corresponding BOE is Eq. (\ref{vertBOE}) with $G$, $J$, and $T(H=iJ)$ replaced by $x$, ${\cal W}/2$, and $T(z=i{\cal W}/2)$, respectively, and ${\cal A}_a^{(\phi)}$ replaced  by ${\cal A}_B^{(\phi)}$. On multiplying and dividing by the profile $\langle \phi (x,{\cal W}/2) \rangle_{B}$ of the semi-infinite strip with {\it uniform} boundaries $B$, one writes the average $LHS$ of the left hand side as a product $LHS =I \times II$ of the two factors
\begin{eqnarray} \label{product}
I = {\langle \phi(x,{\cal W}/2) \rangle_B \over {\cal A}_B^{(\phi)} x^{-x_{\phi}}} \, ; \quad II = {\langle \phi(x,{\cal W}/2) \rangle \over \langle \phi(x,{\cal W}/2) \rangle_B } \equiv { \langle \phi ( G=0, J ) \rangle \over {\cal A}_B^{(\phi)} J^{-x_{\phi}} } \, , \; J \equiv J(x, y= {\cal W}/2) \, .
\end{eqnarray}
In the last step of $II$ the rescaling factor in the transformation (\ref{semiinfstrip}) from the strip to the upper half plane drops out. Introducing the product facilitates checking the BOE by expanding, instead of the product, each of the two factors separately for small $x$ up to order $x^2$. For $I$ this involves checking the BOE for the simpler case of a {\it uniform} strip,
\begin{eqnarray} \label{productprime}
I \to 1 + x^2 \, {4 x_{\phi} \over \hat{c}} \, \langle T(i {\cal W}/2) \rangle_B \, ,
\end{eqnarray}
the validity of which follows readily from Eqs. (\ref{uniformstrip}) and (\ref{tressstripprime}). For $II$ checking the BOE requires showing that  
\begin{eqnarray} \label{productprimeprime}
II \to 1 + x^2 \, {4 x_{\phi} \over \hat{c}} \, \bigl[\langle T(i {\cal W}/2) \rangle -  \langle T(i {\cal W}/2) \rangle_B  \bigr]\, 
\end{eqnarray}
in which case one can use the BOE in (\ref{horizBOE}) in the corresponding {\it half plane} with $ABC$ boundary conditions described below Eq. (\ref{generalstrip}). These imply $H=0$ and $\langle T(H=0) \rangle = 2(t_{AB} + t_{BC}) - t_{AC}$ on using Eq. (1.3) in Ref. \cite{BE21}. With $J=J(x, y= {\cal W}/2) \to \pi x / {\cal W}$ the result
\begin{eqnarray} \label{productprimeprimeprime}
II \to 1-J^2 {4 x_{\phi} \over \hat{c}} \langle T(H=0) \rangle \to 1- \Biggl({\pi x \over {\cal W}}  \Biggr)^2 \,  {4 x_{\phi} \over \hat{c}} \bigl[2(t_{AB} + t_{BC}) - t_{AC}\bigr]
\end{eqnarray}
of the BOE in the half plane agrees with the rhs of (\ref{productprimeprime}) when the expression (\ref{tressstripprime}) for $\langle T(i {\cal W}/2)$ is taken into account. These arguments show that the BOE at the boundary point $z=i {\cal W}/2$ of the semi-infinite strip is consistent with the BOE at $H=0$ in the upper half plane with the corresponding $ABC$ boundary conditions and are quite general, not limited to the Ising model. Of course, the half-plane profiles of the Ising model given in Eqs. (\ref{epssistripABC}) and (\ref{sigsistripABC}) obey the BOE (\ref{horizBOE}) in the $ABC$ half plane, as one verifies by expanding them along the imaginary axis $G=0$ for small $J$. 

For the {\it rectangle} near the end of the left midline $x_{\rm M}=- {\cal H}/2$ with vertical $B$ sides, one has to check correspondingly whether
\begin{eqnarray} \label{productrect}
&&I \equiv {\langle \phi(x_{\rm M} , y_{\rm M} =0) \rangle_B \over {\cal A}_B^{(\phi)} \widehat{x_{\rm M}}^{-x_{\phi}} } \to  1 + \widehat{x_{\rm M}}^2 \, {4 x_{\phi} \over \hat{c}} \, \langle T(z_{\rm M}=- {\cal H}/2) \rangle_{\rm uniform} \, , \nonumber \\
&&II \equiv {\langle \phi(x_{\rm M} , y_{\rm M} =0) \rangle \over \langle \phi(x_{\rm M} , y_{\rm M} =0) \rangle_B } \equiv { \langle \phi ( G=0, J ) \rangle \over {\cal A}_B^{(\phi)} J^{-x_{\phi}} } \to \nonumber \\
&& \qquad \to 1 + \widehat{x_{\rm M}}^2 \, {4 x_{\phi} \over \hat{c}} \, \bigl[\langle T(z_{\rm M}=- {\cal H}/2) \rangle - \langle T(z_{\rm M}=- {\cal H}/2) \rangle_{\rm uniform} \bigr]
\end{eqnarray}
for $x_{\rm M} \equiv - {\cal H}/2 + \widehat{x_{\rm M}}$ with $\widehat{x_{\rm M}}$ small, so that in turn $J \equiv J(x_{\rm M} , y_{\rm M} =0)$ in $II$ becomes small and equal to $(1+u)/2 \to q' \widehat{x_{\rm M}}/ \Lambda$ in leading order, see (\ref{Moebprime}), (\ref{unitcircrect}), and (\ref{ZMtozM}). To verify $II$, consider the example $\langle \phi \rangle = \langle \epsilon \rangle$ in a rectangle with horizontal $-$ and vertical $B=+$ sides. Here $II$ is given by the curly bracket in (\ref{epsrec+-+-+}), which for small $J$ tends to $1-4 J^2 \Theta / P $. Using the basic relation $S=q'^2$ in the expression for $\Theta$ and $P$ then yields $II \to 1-32 (\widehat{x_{\rm M}} / \Lambda)^2 (1+q^2)q^2 /(1-q^2 q'^2)$, which confirms the second relation in (\ref{productrect}) on comparing with the stress tensor expressions (\ref{TzMleft+-}) and (\ref{Smidendsprime}). Concerning $I$, one verifies that the horizontal midline behavior of the profile for uniform boundary condition in (\ref{uniprimaryprim}) with $a=B$ obeys the first relation in (\ref{productrect}) with the corresponding stress tensor average given in (\ref{Smidendsprime}).

\section{VARIOUS WAYS TO MAP THE INTERIOR OF RECTANGLES \\ AND SQUARES TO THE UPPER HALF $H$ PLANE}

For a rectangle of side lengths ${\cal H}$ and ${\cal W}$ in the horizontal and vertical directions, respectively, it is convenient to characterize the aspect ratio ${\cal H}/{\cal W}$ via
\begin{eqnarray} \label{aspq}
{{\cal H} \over {\cal W}}={{\bf K}(q) \over {\bf K}(q')} \, , \quad q'\equiv \sqrt{1-q^2 }
\end{eqnarray}
in terms of a parameter $q$, where ${\bf K}$ is the complete elliptic integral of the first kind, and to  introduce the dilatation factor 
\begin{eqnarray} \label{Lgeo}
{1 \over {\Lambda}}=\Biggl( {{\bf K}(q) \over {\cal H}} \, {{\bf K}(q') \over {\cal W}} \Biggr)^{1/2} \equiv {\bf K}(q)/{\cal H} \equiv {\bf K}(q')/{\cal W} \, 
\end{eqnarray}
between the rectangle and a rectangle with  horizontal and vertical side lengths ${\bf K}(q)$ and ${\bf K}(q')$.  Exchanging the values of the horizontal and vertical lengths ${\cal H}$ and ${\cal W}$  exchanges the values of $q$ and $q'$ while ${\Lambda}$ remains unchanged.

\subsection{Rectangle $0<x<{\cal H}, \, 0<y<{\cal W}$ to half plane} \label{firstquadrant}

The conformal transformation
\begin{eqnarray} \label{uphalfrect}
H(Z)={\rm sn}^2 (Z,q) \, , \quad dH/dZ = 2 \, {\rm sn}(Z,q) {\rm cn}(Z,q) {\rm dn}(Z,q)
\end{eqnarray}
with
\begin{eqnarray} \label{Ztoz}
Z=z/{\Lambda}
\end{eqnarray}
maps the rectangle in the $z$ plane with corners at $z=0 \, , \, {\cal H} \, , \, {\cal H} +i {\cal W}\, , \, i {\cal W}$ and the center at $z=({\cal H} +i {\cal W})/2$ to the upper half $H$ plane with the image points of the corners at $H= 0, 1, 1/q^2, \infty$ on its boundary and with the image of the center at $H=1+i(q'/q)$. It encompasses as special cases the square for ${\cal H}={\cal W}$ where $q=1/\sqrt{2}$ and the semi-infinite strip ${\cal H} \to \infty$ with ${\cal W}$ fixed \cite{nonunique} where $q=1$.

Since (\ref{uphalfrect}) is easily expanded about the corner $z=0$, this mapping is most convenient for verifying the corner operator expansion (COE).  

In the following we often suppress the arguments $(Z,q)$ and $q$ of the Jacobian elliptic functions ${\rm sn}, {\rm cn}, {\rm dn}$ of Eq. (\ref{uphalfrect}) and use the abbreviated notation ${\bf K}(q) \equiv {\bf K}$ and ${\bf K}(q') \equiv {\bf K}'$ for the complete elliptic integrals in (\ref{aspq}) and (\ref{Lgeo}).

For the discussion of the stress tensor in Appendix \ref{unifRecCorn} note the Schwarzian derivative
\begin{eqnarray} \label{ShatS}
S(z) = \check{S}(Z) /\Lambda^2
\end{eqnarray}
of $H(Z(z))$, where

\begin{eqnarray} \label{stressrecta}
&&\check{S}(Z) \equiv {H''' H' - (3/2) H''^2 \over H'^2} = \nonumber \\
&&=- {1 \over 2 \, {\rm sn}^2} \times {3(1+q^4 {\rm sn}^8)-4(1+q^2)(1+q^2 {\rm sn}^4){\rm sn}^2 +2(2+q^2 +2 q^4) {\rm sn}^4 \over 1-(1+q^2) {\rm sn}^2 + q^2 {\rm sn}^4 } \nonumber \\
&&\to -{3 \over 2 \, Z^2} - {8 \over 5} \, (1-q^2+q^4) \, Z^2 + O(Z^4) \, .
\end{eqnarray}
Here the primes denote derivatives wrt $Z$, and for the Jacobian functions ${\rm sn} \equiv {\rm sn}(Z,q)$ relations such as 8.145, 8.154, and 8.158 in Gradsteyn+Ryshik have been used. Note the absence of a term $\propto Z^0$ in the expansion for small $Z$.   

\subsection{Rectangle $-{\cal H}/2<x_{\rm M}<{\cal H}/2, \, -{\cal W}/2<y_{\rm M}<{\cal W}/2$ to unit circle to half plane} \label{RectMcircH}

Here the rectangle is mapped to the half plane in two steps. 

(i) The conformal transformation
\begin{eqnarray} \label{unitcircrect}
w(Z_{\rm M}) = {{\rm sn}(Z_{\rm M},q) \, {\rm dn}(Z_{\rm M},q) \over {\rm cn}(Z_{\rm M},q)} \, , \, \quad {d w \over d Z_{\rm M}} = {{\rm dn}^2 (Z_{\rm M},q) \over {\rm cn}^2 (Z_{\rm M},q)}-q^2 \, {\rm sn}^2 (Z_{\rm M},q) \, ,
\end{eqnarray}
with
\begin{eqnarray} \label{ZMtozM}
Z_{\rm M}=z_{\rm M} / {\Lambda}\, ,
\end{eqnarray}
maps the rectangle in the $z_{\rm M}$ plane with corners at $z_{\rm M}=({\cal H} +i {\cal W})/2, \, (-{\cal H} +i {\cal W})/2 \, , \, -({\cal H} +i {\cal W})/2 \, , \, ({\cal H} -i {\cal W})/2$, i.e., at $Z_{\rm M}= ({\bf K}(q)+i{\bf K}(q'))/2 , \, (-{\bf K}(q)+i{\bf K}(q'))/2 , \, -({\bf K}(q)+i{\bf K}(q'))/2 , \, ({\bf K}(q)-i{\bf K}(q'))/2$ and the center at $z_{\rm M}=Z_{\rm M}=0$ onto the interior of the circular unit-disk in the entire $w_{\rm M}$ plane with the image points of the corners at $w_{\rm M}= q+iq' \, , \, -q+iq' \, , \, -(q+iq') \, , \, q-iq'$ on its boundary and with the image of the center at $w_{\rm M}=0$. In the useful notation
\begin{eqnarray} \label{qqprime}
q=\cos \alpha \, , \quad  q'=\sin \alpha 
\end{eqnarray}
introduced in Eq. (\ref{alternative}), the images of the corner are at $w_{\rm M}= \exp i \alpha \, , \, -\exp (- i \alpha) \, , \, -\exp i \alpha \, , \, \exp (- i \alpha)$ \cite{symmetry}. The subscript M on $z$ and $Z$ should remind us that the {\it midlines} of the rectangle and their images in the circular disk are located on the coordinate axes. The mapping in Eq. (\ref{unitcircrect}) is the inverse of the transformation
\begin{eqnarray} \label{invcircrect}
Z_{\rm M}(w)=\int\limits_{0}^{w} \, {d\omega \over \sqrt{1-2\omega^2 \cos(2 \alpha)+\omega^4}} \, , \quad \cos(2 \alpha) = q^2 - q'^2 \, .
\end{eqnarray}
For $w \equiv u$ on the real axis and for $w \equiv i v$ on the imaginary axis, Eq. (\ref{invcircrect}) can be written as
\begin{eqnarray} \label{invcircrect'}
Z_{\rm M}(u) \equiv X_{\rm M}(u)={1 \over 2} \, F\Biggl({2 u \over 1+u^2} \, , q \Biggr)
\end{eqnarray}
and 
\begin{eqnarray} \label{invcircrect''}
Z_{\rm M}(i v) /i \equiv Y_{\rm M}(v)={1 \over 2} \, F\Biggl({2 v \over 1+v^2} \, , q' \Biggr) \, ,
\end{eqnarray}
respectively, compare 3.138.5 in  Gradsteyn and Ryshik. Here $F(t,Q)=\int\limits_{0}^{t} ds [(1-s^2)(1-Q^2 s^2)]^{-1/2}$ is the elliptic integral of the first kind \cite{ZMF}. Eqs. (\ref{invcircrect'}) and (\ref{invcircrect''}) imply that $Z_{\rm M}(w=1)=(1/2)F(1,q) =(1/2) {\bf K}(q)$, consistent with $z_{\rm M}=(1/2){\cal H}$ and $Z_{\rm M}(w=i)=(i/2)F(1,q') =(i/2) {\bf K}(q')$ consistent with $z_{\rm M}=(1/2){\cal H}$ and $z_{\rm M}=(i/2){\cal W}$, respectively, see (\ref{ZMtozM}).

(ii) Finally the Moebius transformation
\begin{eqnarray} \label{Moebprime}
H(w)=i {1+w \over 1-w} \, , \quad {dH \over d w} = {2i \over (1-w)^2} \, 
\end{eqnarray}
maps the unit disk to the upper half $H$ plane with the corner images at $H=-\cot (\alpha/2) \, , \, -\tan (\alpha/2) \, , \, \tan (\alpha/2) \, , \, \cot (\alpha/2)$, cf. Eq. (\ref{GK}),  and the centers of the rectangle and unit disk to $H=i$. Note that $\tan (\alpha/2) = q' /(1+q)$ and $\cot (\alpha/2) = q' /(1-q)$. For the special case ${\cal W}={\cal H}$ of the square, $q=q'=1/\sqrt{2}$, $\alpha = \pi /4$, and $\tan(\alpha /2) = \tan (\pi /8) = 1/(\sqrt{2} +1)$.  

This rectangle is related to the one in \ref{firstquadrant} by $z_{\rm M}=z-({\cal H}+i{\cal W})/2$. However, the simple detour via the circular disk provides useful insight, since the disk shares more symmetries with the rectangle than the upper half plane: Mirror imaging a point $w$ about the $u$ or $v$ axis translates to mirror imaging the corresponding point $z_{\rm M}$ about the $x_{\rm M}$ or $y_{\rm M}$ axis. In particular, the center and midlines of the rectangle on the real and imaginary axes are mapped to the center of the disk and the real and imaginary axes inside the disk. Accordingly, the configurational symmetry of the corner images on the circumference of the disk resembles that of the corners of the rectangle.

The horizontal midlines $Z_{\rm M} =X_{\rm M}$ of the rectangle  and $w_{\rm M}=u_{\rm M}$ of the circular disk are mapped onto the imaginary axis $H=iJ,\, J>0$ in the half plane, while the vertical midlines are mapped onto the upper unit circle $|H|=i \exp (i \psi) , \,- \pi /2 <\psi < \pi /2$ in the half plane. The corresponding explicit forms
\begin{eqnarray} \label{invhalfrect}
Z_{\rm M}(u(J)) \equiv X_{\rm M}(u(J))={1 \over 2} \, F\Biggl({J^2-1 \over J^2+1} \, , q \Biggr) \, , \quad q \equiv \cos \alpha 
\end{eqnarray}
and 
\begin{eqnarray} \label{invhalfrectprime}
Z_{\rm M}(i v(\psi)) /i \equiv Y_{\rm M}(v(\psi))={1 \over 2} \, F\Bigl( \sin \psi \, , q' \Bigr) \, , \quad q' \equiv \sin \alpha 
\end{eqnarray}
of the inverse mapping follow from Eq. (\ref{invcircrect'}) with $u=(J-1)/(J+1)$ and from (\ref{invcircrect''}) with $v=\tan(\psi /2)$, respectively.

Finally recall the transformation formula (\ref{stripfromplane}) of the densities of primary operators $\phi = \epsilon$ or $\sigma$ under a conformal mapping, which in the case considered here reads  
\begin{eqnarray} \label{primarytrafo}
\langle \phi (x_{\rm M}, y_{\rm M})\rangle = \Bigg| {dw \over d z_{\rm M}} \Bigg|^{x_{\phi}} \times \langle \phi (u, v)\rangle = \Bigg| {dw \over d z_{\rm M}} \, {dH \over d w}\Bigg|^{x_{\phi}} \times \langle \phi (G, J)\rangle \, .
\end{eqnarray}
Together with the relations (\ref{unitcircrect}) and (\ref{Moebprime}), this allows to evaluate $\langle \phi (x_{\rm M}, y_{\rm M})\rangle$ once $\langle \phi (G, J)\rangle$ is known. In the simplest case of a {\it uniform} boundary condition $a$,
\begin{eqnarray} \label{uniprimaryprim}
\langle \phi (G,J \rangle_a = {\cal A}_a^{(\phi)} \times J^{-x_{\phi}} \, &,& \quad \langle \phi (u,v) \rangle_a = {\cal A}_a^{(\phi)} \times \Biggl( {2 \over 1-u^2 -v^2} \Biggr)^{x_{\phi}} \, , \nonumber \\
\langle \phi (x_{\rm M},y_{\rm M} \rangle_a &=& {\cal A}_a^{(\phi)} \times\Biggl({2 \over \Lambda} \, {|{\rm dn}^2 -q^2 {\rm sn}^2 {\rm cn}^2| \over |{\rm cn}^2| - |{\rm sn}^2 {\rm dn}^2|}  \Biggr)^{x_{\phi}}
\end{eqnarray}
where ${\rm sn} = {\rm sn} \bigl( (x_{\rm M}+i y_{\rm M})/\Lambda,q \bigr)$ etc. One checks that for the square with $q^2 =1/2$ and ${\cal H}={\cal W}\equiv {\cal L}$ the last expression in Eq. (\ref{uniprimaryprim}) is consistent with the result (\ref{uniSQU}) on taking the 45 degree rotation into account \cite{consistsquare}.  

For {\it arbitrary mixed} boundary conditions the center $x_{\rm M}= y_{\rm M}=0$ of the rectangle maps onto $u=v=0$ and onto $G=0, J=1$, so that 
\begin{eqnarray} \label{midpoint}
\langle \phi (x_{\rm M}=0,y_{\rm M}=0) \rangle = {\Lambda}^{-x_{\phi}} \langle \phi (u=0,v=0) \rangle = (2/{\Lambda})^{x_{\phi}} \langle \phi (G=0,J=1) \rangle  \, , 
\end{eqnarray}
with the length $\Lambda$ from (\ref{Lgeo}), since in this case $|d w/dZ_{\rm M}|=1$ and $|dH/dw|=2$.

The Schwarzian derivative $S_{\rm M}(z_{\rm M})$ of the transformation $H(w(Z_{\rm M}(z_{\rm M})))$ reads 
\begin{eqnarray} \label{SMhatSM}
S_{\rm M}(z_{\rm M}) = \check{S}_{\rm M}(Z_{\rm M}) / \Lambda^2 
\end{eqnarray}
where    
\begin{eqnarray} \label{stressrectprim}
\check{S}_{\rm M}(Z_{\rm M}) &=& {w_{\rm M}''' w_{\rm M}' - (3/2) (w_{\rm M}'')^2 \over (w_{\rm M}')^2} =   \\
&=&-2 \, {(2q^2-1)(1+q^4 {\rm sn}^8)+4q^2(q^2 -2)(1+q^2  {\rm sn}^4) {\rm sn}^2 -2q^2 (2q^4 -5) {\rm sn}^4 \over 1+q^4  {\rm sn}^8 -4q^2 (1+q^2  {\rm sn}^4) {\rm sn}^2 + 2q^2 (2q^2 +1) {\rm sn}^4} \nonumber
\end{eqnarray}
with the primes denoting derivatives wrt $Z_{\rm M}$ and $ {\rm sn} \equiv  {\rm sn}(Z_{\rm M},q)$. Here one uses that the Schwarzian of the Moebius transformation $H(w)$ vanishes. Note that $\check{S}_{\rm M}(- Z_{\rm M})= \check{S}_{\rm M}(Z_{\rm M})$, $\check{S}_{\rm M}(0)= 2(1-2q^2)=2(q'^2 - q^2)=-2\cos (2 \alpha)$, and the relation
\begin{eqnarray} \label{SMhatSMprime}
\check{S}\biggl(Z={{\bf K}(q)+i {\bf K}(q') \over 2} +Z_{\rm M} \biggr)  = \check{S}_{\rm M}(Z_{\rm M})  
\end{eqnarray}
between (\ref{stressrecta}) and (\ref{stressrectprim}).

The Schwarzian derivative in Eqs. (\ref{SMhatSM}), (\ref{stressrectprim}) determines the stress tensor for {\it uniform} boundary conditions, cf. Ref. \cite{Ttrafo}, which we now evaluate at the centerpoints  $z_{\rm M}=\pm {\cal H}/2$ and $z_{\rm M}=\pm i{\cal W}/2$ of the vertical and horizontal sides of the rectangle, which correspond to the points $Z_{\rm M}=\pm {\bf K}(q)/2$ and $Z_{\rm M}=\pm i{\bf K}(q')/2$. In the more complete notation $\check{S}_{\rm M}(Z_{\rm M}, q)$ for the rhs of Eq. (\ref{stressrectprim}) this expression yields 
\begin{eqnarray} \label{Smidends}
\check{S}_{\rm M}(\pm {\bf K}(q)/2, q) &=& 2(1+q^2) \, , \\
\check{S}_{\rm M}(\pm i{\bf K}(q')/2, q) &=& -2(2-q^2) \equiv -2 (1+q'^2) = -\Bigl[ \check{S}_{\rm M}(\pm {\bf K}(q)/2, q) \Bigr]_{q \leftrightarrow q'} \nonumber
\end{eqnarray}
where the relations ${\rm sn}^2 (\pm {\bf K}(q)/2, q)=1/(1+q')$ and  ${\rm sn}^2 (\pm i{\bf K}(q')/2, q)=-1/q$ has been used. With the help of (\ref{SMhatSM}) and (\ref{Lgeo}), one arrives at
\begin{eqnarray} \label{Smidendsprime}
{\cal W}^2 \, \langle T(z_{\rm M} =\pm {\cal H} /2)  \rangle_{\rm uniform} &=& 2 (1+q^2) \, {\bf K}^2(q') \, \hat{c}/12 \\
{\cal H}^2 \, \langle T(z_{\rm M} =\pm i{\cal W} /2)  \rangle_{\rm uniform} &=& -2 (1+q'^2) \, {\bf K}^2(q) \, \hat{c}/12 = - \Bigl[ {\cal W}^2 \, \langle T(z_{\rm M} =\pm {\cal H} /2)  \rangle_{\rm uniform} \Bigr]_{q \leftrightarrow q'} \nonumber \, .
\end{eqnarray}
The expected limiting behavior for $q \to 1$ and $q \to 0$ is easily checked. For example, the rhs of the first equation in (\ref{Smidendsprime}) tends to $\pi^2 \hat{c}/12$ and $({\cal W} / {\cal H} )^2 \times \pi^2 \hat{c}/24$ for $q \to 1$ and $q \to 0$, respectively. This is consistent with the result (\ref{tressstripprime}) in the {\it semi-infinite} strip extending in horizontal direction and with the result ${\cal H}^2 \langle T(z_{\rm M}) \rangle=\pi^2 \hat{c}/24$ in an {\it infinite} strip extending in vertical direction, respectively, when both objects have a uniform boundary condition.

For the  {\it mixed} boundary conditions considered in Sec. \ref{recmix}, one uses the transformation formula (compare Ref. \cite{Ttrafo})
\begin{eqnarray} \label{THandzM}
\langle T(z_{\rm M})\rangle&=& (dH/dz_{\rm M})^2 \, \langle T(H) \rangle + \langle T(z_{\rm M}) \rangle_{\rm uniform} 
\end{eqnarray} 
for the centerpoints $z_{\rm M} = \bigl[ -{\cal H} /2 \, , \, - i{\cal W} /2 \bigr]$ of the W and S sides of the rectangle, where according to (\ref{unitcircrect}) and (\ref{sidecenter}),
\begin{eqnarray} \label{THandzMmid} 
(dH/dz_{\rm M})^2 = \Lambda^{-2} \bigl[ -q'^2 \, , 4q^2  \bigr] \, , \quad \langle T(H) \rangle = \bigl[ \langle T(H=0) \, , \, \langle T(H=1) \rangle \rangle \bigr] \, .
\end{eqnarray} 
For the two rectangles in the Ising model ($\hat{c}=1/2$) with mixed boundary conditions considered in Secs. \ref{+-+-} and \ref{+f+f}, the stress tensors $\langle T(H) \rangle$ follow from Eqs. (2.53) and (2.49) in Ref. \cite{BE21} on replacing $z$ by $H$ and, as described in our Eq. (\ref{GK}), on putting the switching points $\zeta_1, \, \zeta_2, \, \zeta_3, \, \zeta_4$ equal to $-t^{-1}, \, -t, \, t, \,  t^{-1}$. For the rectangle with horizontal $-$ and vertical $+$ boundaries in Sec. \ref{+-+-}, one finds 
\begin{eqnarray} \label{TH1+-} 
\langle T(H=0) \rangle= {1+q^2 \over q'^2} \, {4q^2 \over 1-q^2 q'^2} \, , \quad \langle T(H=1) \rangle= {(1+q'^2) q'^2 \over (1-q^2 q'^2) q^2} \, ,
\end{eqnarray} 
which together with (\ref{Smidendsprime})-(\ref{THandzMmid}) yields
\begin{eqnarray} \label{TzMleft+-}
{\cal W}^2 \, \langle T(z_{\rm M}=-{\cal H}/2)\rangle&=& (1+q^2) \, {\bf K}^2(q') \, \Bigl( {1 \over 12} - {4q^2 \over 1-q^2 q'^2} \Bigr) \, , \nonumber \\
{\cal H}^2 \, \langle T(z_{\rm M}=-i{\cal W}/2)\rangle &=&-\Bigl[ {\cal W}^2 \, \langle T(z_{\rm M}=-{\cal H}/2)\rangle \Bigr]_{q \leftrightarrow q'} \, .
\end{eqnarray} 
The first line in (\ref{TzMleft+-}) implies that $\langle T(z_{\rm M}=-{\cal H}/2)\rangle$ changes sign on varying the aspect ratio. It is positive and negative for $q^2 < q_0^2$ and $q^2 < q_0^2$, where
\begin{eqnarray} \label{q01+-} 
q_0^2 = \bigl[49-\sqrt{49^2 -4}\bigr]/2 = 0.0204 \, ,
\end{eqnarray} 
corresponding to ${\cal H}/{\cal W}$ smaller and larger than $({\cal H}/{\cal W})_0 = 0.4721$, respectively.   
 
The relation in the second line of (\ref{TzMleft+-}) has the same form as in Eq. (\ref{Smidendsprime}) as expected, since the change $q \leftrightarrow q'$ together with the stress-tensor preserving change $+ \leftrightarrow -$ leads merely to a rotation of our rectangle by 90 degrees.

Now consider the rectangle with horizontal $f$ and vertical $+$ boundaries, i.e. $ABCD=f+f+$. Except  for the limit of the semi-infinite (and infinite) strip, where Eq. (\ref{tressstripaba}) holds, $T$ is not invariant under $+ \leftrightarrow f$ and a relation corresponding to the second equation in (\ref{TzMleft+-}) does not apply. Here one finds
\begin{eqnarray} \label{TH1+f} 
\langle T(H=0) \rangle= {1-q' \over q'^2} \, , \quad \langle T(H=1) \rangle= {q' \over 4 q^2}
\end{eqnarray} 
which yields
\begin{eqnarray} \label{TzMleft+f}
{\cal W}^2 \, \langle T(z_{\rm M}=-{\cal H}/2)\rangle&=& {\bf K}^2(q') \,  \Biggl( {1+q^2 \over 12} - 1+q' \Biggr) \, , \nonumber \\
{\cal H}^2 \, \langle T(z_{\rm M}=-i{\cal W}/2)\rangle &=& - {\bf K}^2(q) \, \Biggl( {1+q'^2 \over 12} -q' \Biggr) \, .
\end{eqnarray} 
Indeed it is only in the two limits $q=1$ and $q=0$, where $\bigl[-1+q']_{q \leftrightarrow q'} \equiv -1+q$ equals $-q'$, that the relation in (\ref{TzMleft+-}) applies to the expressions in (\ref{TzMleft+f}).

As in the discussion of Eq.  (\ref{Smidendsprime}), for $q \to 1$ the two expressions for $\langle T(z_{\rm M}=-{\cal H}/2)\rangle$ in (\ref{TzMleft+-}) and (\ref{TzMleft+f}) must reduce to the semi-infinite strip results $(\pi /{\cal W})^2 [(\hat{c}/12)-4 t_{AB}]$ in (\ref{tressstripprime}) for $ABC=-+-$ and $ABC=f+f$, respectively, which for $\hat{c}=1/2$ are given by $(\pi /{\cal W})^2 [(1/24)-2]$ and $(\pi /{\cal W})^2 [(1/24)-(1/4)]$. For $q \to 0$ these two expressions must be independent of the distant horizontal $-$ or $f$ boundaries and lead to the result $\langle T(z_{\rm M}=-{\cal H}/2)\rangle = (\pi /{\cal H})^2 /48$ for a vertical infinite strip with uniform boundary condition. All this is easily checked.

In the special case  $q=q'=1/\sqrt{2}$, in which the rectangle becomes a {\it square} with ${\cal H}={\cal W} \equiv {\cal L}$, the above expressions yield 
\begin{eqnarray} \label{TzMleft+-square} 
{\cal L}^2 \, \langle T(z_{\rm M}=-{\cal L}/2)\rangle&=& - {\bf K}^2(1/\sqrt{2}) \times 31/8  \  = - 13.321 \, ,  \nonumber \\
{\cal L}^2 \, \langle T(z_{\rm M}=-i{\cal L}/2)\rangle &=& 13.321
\end{eqnarray} 
for horizontal $-$ and vertical $+$ boundaries and
\begin{eqnarray} \label{TzMleft+fsquare}
{\cal L}^2 \, \langle T(z_{\rm M}=-{\cal L}/2)\rangle&=& {\bf K}^2(1/\sqrt{2}) \,  \Biggl( - {7 \over 8} + {1 \over \sqrt{2}} \Biggr)  = - 0.577 \, ,  \nonumber \\
{\cal L}^2 \, \langle T(z_{\rm M}=-i{\cal L}/2)\rangle &=& - {\bf K}^2(1/\sqrt{2}) \, \Biggl( {1 \over 8} - {1 \over \sqrt{2}} \Biggr) = 2.001 \, ,
\end{eqnarray} 
for horizontal $f$ and vertical $+$ boundaries. The effect of the $-$ boundaries on the stress tensor in the + boundary and vice versa are considerably stronger than the effect of the $f$ boundaries on the stress tensor in the + boundary and of the + boundaries on the stress tensor in the $f$ boundary. These findings are quite plausible. That the influence of the $+$ boundaries on the stress tensor in the $f$ boundary is stronger than the effect of the $f$ boundaries on the stress tensor in the + boundary is consistent with the property that + dominates $f$ that we found in Sec. \ref{competitionf+}. That the two effects be of equal strength, i.e. that $[(1+q^2)/12] -1 +q'$ equals $[(1+q'^2)/12] -q'$ and thus $\langle T(z_{\rm M}=-{\cal H}/2)\rangle = - \langle T(z_{\rm M}=-i{\cal W}/2)\rangle$, requires  $q'=6-[\sqrt{122}/2] =0.4773$, which is smaller than $1/\sqrt{2}$, so that the length ${\cal H}$ of the $f$ side must be longer by a factor ${\bf K}(q)/{\bf K}(q') = 1.313$ than the length ${\cal W}$ of the + side.

Another quantity of interest is the stress tensor in the {\it center} of the rectangle. Proceeding like above one finds the model independent result
\begin{eqnarray} \label{Tcenteruni}
\Lambda^2 \langle T(z_{\rm M}=0) \rangle_{\rm uniform} = 2(q'^2 -q^2) \hat{c} /12
\end{eqnarray}
for uniform boundary conditions. For mixed boundary conditions in the Ising model the result is
\begin{eqnarray} \label{Tcenter-+}
\Lambda^2 \langle T(z_{\rm M}=0) \rangle = {q'^2 -q^2 \over 12} + 16 q^2 q'^2 {q'^2 -q^2 \over (q'^2 -q^2)^2 +3}
\end{eqnarray}
for $ABCD=-+-+$ and 
\begin{eqnarray} \label{Tcenterf+}
\Lambda^2 \langle T(z_{\rm M}=0) \rangle = {q'^2 -q^2 \over 12} + q' (1-q')
\end{eqnarray}
for $ABCD=f+f+$. See Appendix \ref{Janus} for an application. 

\subsection{Square ${\cal L} \times {\cal L}$ with corners at $z_{\rm D}=\pm {\cal L}/\sqrt{2}$ and $\pm i{\cal L}/\sqrt{2}$} \label{SquareDcircH}

The conformal transformation
\begin{eqnarray} \label{unitcircsqua}
w_{\rm D}(Z_{\rm D}) = {{\rm sn}(Z_{\rm D},q) \over \sqrt{2} \, {\rm dn}(Z_{\rm D},q)} \, , \quad {dw_{\rm D} \over dZ_{\rm D}} ={{\rm cn}(Z_{\rm D},q) \over \sqrt{2} \, {\rm dn}^2 (Z_{\rm D},q)} , \quad q={1 \over \sqrt{2} }
\end{eqnarray}
with
\begin{eqnarray} \label{ZDtozD}
Z_{\rm D}=z_{\rm D} \, \sqrt{2} \, {\bf K}(q)/{\cal L} 
\end{eqnarray}
maps the ${\cal L}  \times {\cal L} $ square in the $z_{\rm D}$ plane with corners at $z_{\rm D}=(1,i,-1,-i)\times {\cal L} /\sqrt{2}$ and the center at $z_{\rm D}=0$ onto the interior of the unit circle in the  $w_{\rm D}$ plane, with corner images at $w_{\rm D}= 1,i,-1,-i$ on its boundary and its center at $w_{\rm D}=0$. The subscript D on $z$ and $w$ is a reminder that the {\it diagonals} of the square and their images in the circular disk lie on the coordinate axes. In the following we often suppress the modulus $q=1/\sqrt{2}$ in ${\bf K}(1/\sqrt{2}) \equiv 1.85407$ and the Jacobian functions.
 
The final mapping to the upper half $H$ plane is the same Moebius mapping $H(w_{\rm D})=i(1+w_{\rm D})/(1-w_{\rm D})$ as in Eq. (\ref{Moebprime}), so that the corner images are at $H= - \infty, -1, 0,1$.

The mapping considered here is most convenient for evaluating the behavior of the profiles along the horizontal diagonal of the square from its counterpart along the imaginary axis of the upper half $H$ plane.

The Schwarzian derivative $S_{\rm D}(z_{\rm D})$ of the transformation $H(w_{\rm D}(Z_{\rm D}(z_{\rm D})))$ reads 
\begin{eqnarray} \label{SDhatSD}
S_{\rm D}(z_{\rm D}) = \check{S}_{\rm D}(Z_{\rm D}) \, 2 ({\bf K}/{\cal L})^2  \, ,
\end{eqnarray}
where  
\begin{eqnarray} \label{stresssquarepri}
\check{S}_{\rm D}(Z_{\rm D}) = {w_{\rm D}''' w_{\rm D}' - (3/2) (w_{\rm D}'')^2 \over (w_{\rm D}')^2}  = -{3 \over 8} \; {{\rm sn}^2 Z_{\rm D} \; \Bigl( 1+ {\rm cn}^2 Z_{\rm D} \Bigr)^2 \over {\rm cn}^2 Z_{\rm D} \; {\rm dn}^2 Z_{\rm D}} \ = \nonumber \\ 
={\cal S} (Z_{\rm D} + {\bf K}) = {\cal S} (Z_{\rm D} - {\bf K}) \, , \quad {\cal S} ( {\cal Z}) \equiv -{3\over 2}\, {{\rm cn}^2({\cal Z}) \over {\rm sn}^2({\cal Z}) \, {\rm dn}^2({\cal Z})} 
\end{eqnarray}
with the primes denoting derivatives wrt $Z_{\rm D}$. Here the vanishing of the Schwarzian of the Moebius transformation $H(w_{\rm D})$ was used, and the result follows from (\ref{unitcircsqua}) via derivatives and functional relations of the Jacobian elliptic functions as given, e.g., in 8.158,, 8.154, and 8.151.2 in Gradsteyn and Ryshik. Related useful properties are $\check{S}_{\rm D}(Z_{\rm D}+ 2 {\bf K}) = \check{S}_{\rm D}(Z_{\rm D})$ and
\begin{eqnarray} \label{ninetyrotate}
\check{S}_{\rm D}( Z_{\rm D}) = - \check{S}_{\rm D}(i Z_{\rm D}) = \check{S}_{\rm D}(- Z_{\rm D}) \, ,
\end{eqnarray}
reflecting the invariance of the stress tensor $\langle T(z_{\rm D}) \rangle = (\hat{c} /12) S_{\rm D}(z_{\rm D})$ in a square with uniform boundary condition when rotated by 90 and 180 degrees, compare Ref. \cite{Ttrafo}. The expansions of the Schwarzian derivative about the left and right corners of the square where $|Z_{\rm D} + {\bf K}|$ and $|Z_{\rm D} - {\bf K}|$ are small, respectively, follow from the expansion
\begin{eqnarray} \label{expcorn}
{\cal S}({\cal Z}) \to  -{3 \over 2} {1 \over {\cal Z}^2} + {3 \over 10} {\cal Z}^2 + O({\cal Z}^4) \, . 
\end{eqnarray}

The form of the stress tensor $\langle T(z_{\rm D}) \rangle$ near the left and right corner of the square with uniform boundary conditions follows from Eqs. (\ref{ZDtozD})-(\ref{stresssquarepri}) and (\ref{expcorn}) and is given by
\begin{eqnarray} \label{expcornprime}
\langle T(z_{\rm D}) \rangle \to {\hat{c} \over 12} \Biggl[ -{3 \over 2} \Bigl(z_{\rm D} \pm {{\cal L} \over \sqrt{2}}  \Bigr)^{-2} + {3 \over 10} \Bigl(z_{\rm D} \pm {{\cal L} \over \sqrt{2}}  \Bigr)^{2} \times \Bigl( {2 {\bf K}^2 \over {\cal L}^2} \Bigr)^2 \Biggr] \, , \quad z_{\rm D} \to \mp  {{\cal L} \over \sqrt{2}}  
\end{eqnarray}
to leading and next-to-leading order. By a rotation $z= \exp(i \pi /4) \times (z_{\rm D}+{\cal L}/ \sqrt{2})$ or $z= \exp(-3i \pi /4) \times (z_{\rm D}-{\cal L}/ \sqrt{2})$ the square we are considering turns into  the square in the $z$ plane with corners at $z=0, \, {\cal L}, \, (1+i) {\cal L}, \, i {\cal L}$, i.e., into the rectangle of Appendix \ref{firstquadrant} with ${\cal H}={\cal W}={\cal L}$ and (\ref{expcornprime}) yields the behavior 
\begin{eqnarray} \label{expcornprimeprime}
\langle T(z) \rangle \to {\hat{c} \over 12} \Biggl[ -{3 \over 2} z^{-2} - {3 \over 10} z^{2} \times \Bigl( {2 {\bf K}^2 \over {\cal L}^2} \Bigr)^2 \Biggr] \, , \quad |z| \ll {\cal L}   
\end{eqnarray}
of the stress tensor near its lower left corner. With the definition (\ref{avWEDGE}) this implies 
\begin{eqnarray} \label{calTsquare}
\langle {\cal T} \rangle = - {\hat{c} \over 40}  \Bigl( { {\bf K} \over {\cal L}} \Bigr)^4 \,    
\end{eqnarray}
for the average of the corner operator for the ${\cal L} \times {\cal L}$ square, consistent with the expression (\ref{unitaurect}) for the rectangle.  This should be compared with the corresponding average in Eq. (\ref{semistripscornop}) for the semi-infinite strip of width ${\cal W}$ with a uniform boundary. For ${\cal L}={\cal W}$ the average for the square is larger by a factor $12 ({\bf K}/\pi)^4 \approx 1.45$ than for the semi-infinite strip, indicating that the square more strongly perturbes the infinite wedge near its apex than the semi-infinite strip.

Dropping the restriction to the neighborhood of corners and allowing {\it arbitrary} positions inside the square, we note that the quantity $\check{S}_{\rm D}$ is related to $\check{S}_{\rm M}$ in (\ref{stressrectprim}) and $\check{S}$ in (\ref{stressrecta}) in the limit $q^2 =1/2$ of the square by
\begin{eqnarray} \label{SDtoSM}
\check{S}_{\rm D}(Z_{\rm D}) = {i \over 2} \check{S}_{\rm M} \biggl( Z_{\rm M}={1+i \over 2} Z_{\rm D} \biggr) \, \equiv \, {i \over 2} \check{S}_{\rm M} \biggl( Z_{\rm M}= - {1+i \over 2} Z_{\rm D} \biggr) =  
\nonumber \\
={i \over 2} \check{S} \biggl( Z={1+i \over 2} ({\bf K}+Z_{\rm D}) \biggr) \equiv  {i \over 2} \check{S} \biggl( Z={1+i \over 2} ({\bf K}- Z_{\rm D}) \biggr) \, .
\end{eqnarray}
The reason is (i) that the square in the $z_{\rm D}$ plane is related by a rotation of 45 degrees to the square in the $z_{\rm M}$ plane of Appendix \ref{RectMcircH} and (ii) the relation between $\check{S}_{\rm M}$ and $\check{S}$ given in Eq. (\ref{SMhatSM}). In the relations $(Z,Z_{\rm M},Z_{\rm D})=\bigl( {\bf K} / {\cal L} \bigr) \times (z,z_{\rm M},\sqrt{2} z_{\rm D})$  between the arguments of the Jacobian functions and the side length ${\cal L}$ of the square following from (\ref{Ztoz}), (\ref{ZMtozM}), (\ref{ZDtozD}), note in the last case the additional factor $\sqrt{2}$ and a corrresponding additional factor 2 in the relations $(S,S_{\rm M},S_{\rm D}) = ({\bf K} / {\cal L})^2 \times (\check{S},\check{S}_{\rm M}, 2 \check{S}_{\rm D})$ following from Eqs. (\ref{ShatS}), (\ref{SMhatSM}), (\ref{SDhatSD}). 

\section{DERIVATION OF AVERAGES OF CORNER OPERATORS AND CONFIRMATION OF THE COE FOR RECTANGLES}   \label{RecCorn}

To derive the averages of corner operators for rectangles presented in Section  \ref{CofRect} from known stress tensor averages in the upper half $H$ plane, we use the conformal transformation in Appendix \ref{firstquadrant}. Here the combination $AB|CD$ of boundary conditions for the rectangle introduced at the beginning of Section \ref{CofRect} is transformed to the combination $B,C,D$, and $A$ for the intervals $-\infty < G<0$, $0<G<1$, $1<G<1/q^2$, and $1/q^2 <G< +\infty$, respectively, onto the real axis of the upper $H=G+iJ$ plane.

In the following it is convenient to rewrite the usual transformation law of the stress tensor given in Ref. \cite{Ttrafo} in the form 
\begin{eqnarray} \label{stressre}
\langle T(z) \rangle \,  \Lambda^2 =  \langle T(Z) \rangle = \Biggl( {d H \over dZ} \Biggr)^2 \; \langle T(H) \rangle + {\hat{c} \over 12} \; \check{S}(Z) \, ,
\end{eqnarray}
where $\check{S}(Z)$ is the Schwarzian derivative of $H(Z)$ given in Eq. (\ref{stressrecta}).

\subsection{Rectangle with uniform boundary condition} \label{unifRecCorn}

For rectangles with a {\it uniform} boundary condition $\langle T(H) \rangle =0$, and 
\begin{eqnarray} \label{stressrecuni}
\langle T(Z) \rangle \to -{\hat{c} \over 8 \, Z^2} - {\hat{c} \over 15} \, (1+q^4+q'^4) \, Z^2 + O(Z^4) \, 
\end{eqnarray}
where $q'=\sqrt{1-q^2}$. Eqs. (\ref{stressre}) and (\ref{stressrecuni}) together with the form of $\langle T(z) \rangle_{a|a}$ given below Eq. (\ref{opWEDGE}) yield the result (\ref{unitaurect}). 

Checking the COE (\ref{avWEDGE}) requires evaluating the leading and next-to-leading contributions to  the profile $\langle \phi (x,y) \rangle$ near the corner $z=0$ of the rectangle, which follow from 
\begin{eqnarray} \label{phiunirect}
\langle \phi (x,y) \rangle=\Bigg| {1 \over \Lambda} {dH \over dZ} {1 \over J} \Bigg|^{x_{\phi}} \, {\cal A}_{a}^{(\phi)} \equiv \Bigg| {1 \over \Lambda} {\rm cn \, dn \over sn} \times {\rm sn^2 \over Im \, sn^2} \Bigg|^{x_{\phi}} \, 2^{x_{\phi}} \, {\cal A}_{a}^{(\phi)} \, ,
\end{eqnarray}
on substituting the Jacobian functions ${\rm sn}\equiv {\rm sn}(Z,q)$ etc. and expanding them for small $Z$. Eq. (\ref{phiunirect}) is obtained by proceeding like in (\ref{stripfromplane}), (\ref{uniformstrip}) and using (\ref{uphalfrect}) and (\ref{Ztoz}). It is easy to see that in leading order the rhs of (\ref{phiunirect}) reproduces the profile $\langle \phi(x,y) \rangle_{a|a}$ given below Eq. (\ref{opWEDGE}), which is $\propto |Z|^{-x_{\phi}}$, and that the next term $\propto |Z|^{2-x_{\phi}}$ {\it vanishes}. This is consistent with the COE (\ref{avWEDGE}) which predicts the next-to-leading term to be $\propto |Z|^{4-x_{\phi}}$ and will prove useful below for calculating the profiles for mixed boundary conditions. Verifying the explicit form of the next-to-leading term is quite instructive, and we present some details. First one finds that
\begin{eqnarray} \label{phiunirectprime}
{\rm cn \, dn \over sn} \to Z^{-1} (1+AZ^2 + BZ^4) \, , \quad A=-(1+q^2)/3 \, , \quad B=-(1-16q^2 +q^4)/45 \, .
\end{eqnarray}
Assuming for simplicity that $Z=|Z| \exp (i \pi/4)$ so that $\vartheta =\pi /4, \, \sin 2\vartheta =1$ \cite{arbtheta}, yields 
\begin{eqnarray} \label{phiunirectprimeprime}
\Bigg|{\rm cn \, dn \over sn}\Bigg| \to {1 \over |Z|} \Biggl[1+ {|Z|^4 \over 90} \bigl( 7-22q^2 +7q^4 \bigr)  \Biggr] \, , \quad {\rm |sn^2| \over Im \, sn^2} \to 1+{(1+q^2)^2 \over 18} |Z|^4 \, ,
\end{eqnarray}
which on substituting in (\ref{phiunirect}) yields an expression which is consistent with the COE (\ref{avWEDGE}) with $\sin 2 \vartheta =1$ in ${\cal F}_a^{(\phi)}$ and with $\langle {\cal T} \rangle$ from (\ref{unitaurect}).

\subsection{Rectangle with $aa|bb$ boundary conditions} \label{aabbRectangleprime}

To check the prediction of the COE (\ref{opdiffWEDGE}) we consider the energy density $\phi=\epsilon$ in our $aa|bb$ rectangle for $a=+$ and $b=f$. The corresponding boundary conditions in the upper half $H$ plane are $+$ for $-\infty < G<0$, $f$ for $0<G< 1/q^2$, and $+$ for $1/q^2 <G< +\infty$, for which the profile reads
\begin{eqnarray} \label{eps+ff+}
\langle \epsilon (G,J) \rangle ={1 \over 2J} \, {1-|1-2q^2 H| \over \sqrt{\bigl[1-|1-2q^2 H|\bigr]^2 +4 \bigl(2q^2 J \bigr)^2}} \, .
\end{eqnarray}
Eq. (\ref{eps+ff+}) follows from the single switch expression in Eq. (4.1) in Ref. \cite{BX} for the energy density in a similar way as described in \cite{BXprof}. Transforming the profile to the rectangle geometry by means of (\ref{uphalfrect}) and expanding for small $|Z|$ in leading and next-to-leading order and using the expression (\ref{rectdiffYcornaabb}) for $\langle {\cal Y} \rangle$ one verifies the validity of the COE determined by the $+|f$ expressions in (\ref{epssig+f}) and (\ref{calF+f}) for the energy density.

The form (\ref{rectdiffYcornaabb}) of $\langle {\cal Y} \rangle$ follows on inserting
\begin{eqnarray} \label{stressaabb}
\Bigl( {d H \over dZ} \Bigr)^2 \; \langle T(H) \rangle = 4t_{ab} \, \Bigl({{\rm  cn \, dn} \over {\rm sn} } \Bigr)^2\times \Bigl[ 1+{(q \, {\rm sn})^2 \over 1-(q \, {\rm sn})^2} \Bigr]^2
\end{eqnarray}
in Eq. (\ref{stressre}) and on expanding the result up to order $Z^0$ with the help of (\ref{phiunirectprime}). To obtain Eq. (\ref{stressaabb}), one uses the form of $\langle T(H) \rangle$ for the two switches at $H=0$ and $H=1/q^2$ that follows from Eq. (1.2) in Ref. \cite{BE21}.

\subsection{Rectangle with boundary conditions $f+|f+$} \label{COE+f+frec}

\subsubsection{Deriving the average of the corner operator}

Here the average $\langle {\cal Y} \rangle$ of the $a|b$ corner operator ${\cal Y}$ is derived for the $f+|f+$ rectangle of Sec. \ref{recf+f+}. In the upper half $H$ plane this corresponds to boundary conditions $+$ for $-\infty < G < 0$, $f$ for $0<G<1$, $+$ for $1<G<1/q^2$, and $f$ for $1/q^2 < G< +\infty$. The corresponding stress tensor
\begin{eqnarray} \label{TH+f+f}
\langle T(H) \rangle &=& {1 \over 16}\Biggl({1 \over H}-{1 \over H-1}\Biggr)^2+ {1 \over 16}\Biggl({1 \over H-1/q^2}\Biggr)^2 + \nonumber \\
&& \qquad +{1\over 8} {1-q' \over 1+q'} \Biggl({1 \over H}-{1 \over H-1}\Biggr) {1 \over H-1/q^2}
\end{eqnarray}
follows from Eq. (2.49) in Ref. \cite{BE21} on setting $\zeta_1, \, \zeta_2, \, \zeta_3, \, \zeta_4 \equiv G_{\rm I}, \, G_{\rm II}, \, G_{\rm III}, \, G_{\rm IV}$ equal to $0,\, 1,\, 1/q^2,\, \infty$. Using Eqs. (\ref{uphalfrect}) and (\ref{stressre}) and expanding for small $Z$, one finds 
\begin{eqnarray} \label{stressref+f+}
\langle T(Z) \rangle - {\hat{c} \over 12} \; \check{S}(Z) \equiv (dH/dZ)^2 \langle T(H) \rangle \to {1 \over 4 Z^2} +q' - {1+q'^2 \over 3}+ O(Z^2) \, .
\end{eqnarray}
On using $\hat{c}=1/2$, the expansion (\ref{stressrecta}) of $S(Z)$, and the expression (\ref{Tabwedge}) for $\langle T(Z) \rangle_{+f}$, as well as the dilatation relations (\ref{Ztoz}) and (\ref{stressre}), the above relation (\ref{stressref+f+}) leads to the result (\ref{rectf+f+}) for  $\langle {\cal Y} \rangle$. 

\subsubsection{Confirming the COE for the energy density}

The corresponding energy and order parameter densities in the upper half plane follow from Eqs. (\ref{arbeps}) and (\ref{arbsig}) on inserting the expressions
\begin{eqnarray} \label{CS}
\bigl[C_{\rm II,I} \, , \,  S_{\rm II,I}\bigr] &=& \bigl[-(G-|H|^2) \, , \, J\bigr] \big/ \sqrt{(G-|H|^2)^2 +J^2} \, \nonumber \\
\bigl[C_{\rm IV,III} \, , \,  S_{\rm IV,III}\bigr] &=& \bigl[1-q^2 G \, , \, q^2 J\bigr] \big/ |1 -q^2 H| \, , \quad \Xi = (q'-1)/(q'+1) 
\end{eqnarray}
determined by (\ref{C1C3}) and (\ref{Xi}) for our switching points given below (\ref{TH+f+f}). For the energy density this yields
\begin{eqnarray} \label{eps+f+f}
\langle \epsilon (G,J) \rangle = - {1 \over 2J} \, {1 \over \sqrt{(|H|^2 -G)^2 +J^2}} \, {1 \over |1-q^2 H|} \, \Bigl[ (|H|^2 -G) (1-q^2 G) - J^2 (1-q')^2 \Bigr] \, .
\end{eqnarray}
For the expansion for small distance $|z|$ from the rectangle's corner one needs the expansion for small distance $|H|$ from the upper half plane's switch point $H=0$ for which (\ref{eps+f+f}) yields in leading and next-to-leading order 
\begin{eqnarray} \label{exeps+f+f}
\langle \epsilon (G,J) \rangle &\to&  {1 \over 2|H|} \, \Bigl[ {G \over J} - J \Bigl( 2q'-q'^2 \Bigr) \Bigr] \, \equiv \nonumber \\
&\equiv&  \langle \epsilon (G,J) \rangle_{+|f} + F_{+f}^{(\epsilon)} (G,J) \times \langle \Upsilon (H=0) \rangle \, .
\end{eqnarray}
This expansion is consistent with the boundary-operator expansion about the switching point $H=0$ \cite{BE21}, as we indicate in the last expression in (\ref{exeps+f+f}). Indeed, $-J/(2|H|)=F_{+f}^{(\epsilon)}(G,J)/8$, cf. Eqs. (3.28), (3.29) in Ref. \cite{BE21}, and the average $\langle \Upsilon (H=0) \rangle = (2q'-q'^2)/8$ of the boundary operator $\Upsilon$ follows from the stress tensor in (\ref{TH+f+f}) via $\langle T(H) \rangle - \langle T(H) \rangle_{+f} \to \langle \Upsilon (H=0) \rangle /H$ as $H \to 0$. 

For the energy density $\langle \epsilon (x,y) \rangle$ in the rectangle, the transformation (\ref{uphalfrect}), (\ref{Ztoz}) together with (\ref{exeps+f+f}) yields  
\begin{eqnarray} \label{exeps+f+fxy}
\langle \epsilon (x,y) \rangle \Lambda &=& 2|{\rm cn} \,{\rm dn}| \times |{\rm sn}| \, \langle \epsilon (G,J) \rangle \nonumber\\
&\to& {\cot (2 \vartheta) \over |Z|} -2|Z| \sin (2 \vartheta) \times \Bigl( q' - {1+q'^2 \over 3} \Bigr) \, .
\end{eqnarray}
Comparing with (\ref{epssig+f}), (\ref{calF+f}), and (\ref{rectf+f+}) confirms the COE (\ref{opdiffWEDGE}) for the boundary conditions considered here. 

\subsubsection{Confirming the COE for the order parameter}

Inserting the expansions
\begin{eqnarray} \label{1pmC}
\sqrt{1 \pm C_{\rm II,I}} &\to& \sqrt{|H|\mp G \over |H|} \, \Biggl(1 \pm {1 \over 2} {J^2 \over |H| \mp G}  \Biggr) \, , \nonumber \\
\sqrt{1 + C_{\rm IV,III}} &\to& 2^{1/2} + O(J^2) \, , \quad \sqrt{1 - C_{\rm IV,III}} \to 2^{-1/2} \, J q^2
\end{eqnarray}
to leading and next-to-leading order, which follow from (\ref{CS}), into Eq. (\ref{arbsig}) yields the expected BOE
\begin{eqnarray} \label{FsigUps}
\langle \sigma (G,J) \rangle &\to&  \langle \sigma (G,J) \rangle_{+|f} + F_{+f}^{(\sigma)} (G,J) \times \langle \Upsilon (H=0) \rangle \, 
\end{eqnarray}
close to the switching point $H=0$. Here we have used the form of $\langle \Upsilon (H=0) \rangle$ given below (\ref{exeps+f+f}) and the identities 
\begin{eqnarray} \label{FsigUpsprime}
&&\Biggl({2 \over J}  \Biggr)^{1/8} \, 2^{-1/4} \, \Biggl( {|H|-G \over |H|} \Biggr)^{1/4} = \Biggl({2 \over |H| \sin \varphi}  \Biggr)^{1/8} \, \bigl( \sin (\varphi /2) \bigr)^{1/2} = \langle \sigma (G,J) \rangle_{+|f}    \\
&&\Biggl({2 \over J}  \Biggr)^{1/8} \, 2^{-1/4} \, {2J(|H|+G)^{1/2} \over \bigl[ |H|(|H|-G) \bigr]^{1/4}} = 4 |H|^{7/8} \, \bigl( \sin (\varphi /2) \bigr)^{3/8} \, \bigl( \cos (\varphi /2) \bigr)^{15/8} = F_{+f}^{(\sigma)} (G,J) \nonumber
\end{eqnarray}
that follow from $H=G+iJ=|H| \exp (i \varphi)$ and from (\ref{epssig+f}) and (\ref{calF+f}) when $g=1$, $z \to H$, and $\vartheta \to \varphi$.

In order to evaluate the small $|z|$ behavior of $\langle \sigma (x,y) \rangle$ from (\ref{FsigUps}) by means of the transformation (\ref{uphalfrect}), we expand the Jacobian functions, obtaining 
\begin{eqnarray} \label{HtoZ}
&&\Lambda^{-1/8} \Bigg| {dH \over dZ} \Bigg|^{1/8} \langle \sigma (G,J) \rangle_{+|f} \to \langle \sigma (x,y) \rangle_{+|f} \, - \, \Bigl( {\cal F}_{+|f}^{(\sigma)}(x,y) /4 \Bigr)\Lambda^{-2} \, {2-q'^2 \over 6} \, , \nonumber \\
&&\Lambda^{-1/8} \Bigg| {dH \over dZ} \Bigg|^{1/8} F_{+|f}^{(\sigma)}(G,J) \times \langle \Upsilon (H=0) \rangle \to \Bigl( {\cal F}_{+|f}^{(\sigma)}(x,y) /4 \Bigr)\Lambda^{-2} \, {2q' -q'^2 \over 2}
\end{eqnarray} 
where ${\cal F}_{+|f}^{(\sigma)} (x,y) /4 =2^{1/8} |Z\Lambda|^{2-1/8} \sin^{3/8} \vartheta \cos^{15/8} \vartheta$ follows from (\ref{calF+f}) for $g=2$. The sum of the two expressions confirms the COE (\ref{opdiffWEDGE}) since
\begin{eqnarray} \label{sumHtoZ}
- {2-q'^2 \over 6} + {2q' -q'^2 \over 2} = q' - {q'^2 +1 \over 3} = 4 \langle {\cal Y} \rangle\Lambda^2 \, ,
\end{eqnarray} 
compare (\ref{rectf+f+}). 

\section{ISING LATTICE MODEL FOR MONTE CARLO SIMULATION} \label{lattmc}

Consider the usual Ising model on a square lattice with ferromagnetic nearest neighbor interaction for which the bulk Hamiltonian reads
\begin{eqnarray} \label{lattham}
\mathbf{H}=- \beta \sum \limits_{\{\mathbf{n}\}}  S_\mathbf{n} \left( S_\mathbf{n+e_1}
+ S_\mathbf{n+e_2}\right)
\end{eqnarray}
with spins $S_{\bf n}=\pm 1$ on the lattice sites ${\bf n}=(n_1 , n_2)$ where the two components $n_1$ and $n_2$ in horizontal and vertical directions are integers and  $\mathbf{e_1}=(1,0)$, $\mathbf{e_2}=(0,1)$ lead to the right and upper nearest neighbor sites of $\bf n$. The interaction constant $J$ is absorbed into the dimensionless inverse temperature $\beta$ so that at the critical point $\beta=\beta_c=J/\left(k_B T_c \right) = \ln(\sqrt{2}+1)/2 \simeq 0.44068\dots~$.

To express the primary operators $\phi({\bf r})$ in terms of the spin variables, place the spins $S_{\bf n}$ at the position ${\bf r}_{\bf n} =\lambda {\bf n}$ in the ${\bf r}$ plane which does not affect their interaction (\ref{lattham}). Here $\lambda$ is a microscopic length. For the energy density $\epsilon ({\bf r})$ this expression reads
\begin{eqnarray} \label{expresseps}
\epsilon ({\bf n} \lambda)= {\pi \over \lambda} E_{\bf n} \, , \quad  E_{\bf n} \equiv -S_{\bf n} S_{\bf n'} + \langle S_{\bf n} S_{\bf n'} \rangle_{\rm bulk} \, ,
\end{eqnarray}
where ${\bf n'}$ is a nearest neighbor of ${\bf n}$. The relation applies inside averages with the necessary condition that the distance of the position ${\bf n}\lambda$ from boundaries and other operators is much larger than $\lambda$. In this limit it does not matter which one of the 4 nearest neighbors ${\bf n'}$ is chosen. 

The prefactor $\pi /\lambda$ provides the required normalization \cite{primaryfields} in the bulk, $\langle \epsilon ({\bf n} \lambda) \epsilon ({\bf 0}) \rangle_{\rm bulk} = 1/(|{\bf n}| \lambda)^2 \equiv 1/({\rm distance})^{2x_{\epsilon} }$ with $x_{\epsilon} =1$. This follows from the asymptotic behavior $\langle E_{\bf n} E_{\bf 0} \rangle_{\rm bulk} \to 1/ (\pi |{\bf n}|)^2$ for $|{\bf n}| \gg 1$ on the Ising lattice derived, e.g.,  by R. Hecht, Phys. Rev. {\bf 158},  557 (1967), see his Eqs. (3.1), (3.3), (3.14b) and the remark below (3.13), and is confirmed in the simulations of Ref. \cite{OAV}. 

For the systems in Sec. III  inside a rectangular region in the ${\bf r}$ plane with horizontal and vertical sides of length ${\cal H}$ and $\cal W$, respectively, the corresponding lattice consists of ${\cal W}/ \lambda$ rows and ${\cal H}/ \lambda$ columns. A simple choice followed in Ref. \cite{OAV} to microscopically realize the boundary conditions is as follows: For the two vertical sides with boundary condition + one freezes the spins in the leftmost and rightmost column in the up direction $S=1$ while for the upper and lower horizontal sides with boundary condition $-$ or $f$ the spins in the top and bottom rows are frozen in the down direction $S=-1$ or can freely flip, respectively. In any case the corner spins of the lattice are allowed to flip freely. Unlike the bulk, in the bounded system $\langle E_{\rm n} \rangle$ and $\langle \epsilon({\bf r}_{\rm n} \rangle$ are nonvanishing. Since the scale-invariant expression introduced in Eq. (\ref{scinvhoreps}) can be expressed in terms of a product of quantities in the lattice model as $\bigl( {\cal H} {\cal W} \bigr)^{1/2} \, \langle \epsilon({\bf r}_{\rm n} \rangle = \pi \times \bigl( ({\cal H}/\lambda) ({\cal W}/\lambda) \bigr)^{1/2} \, \langle E_{\rm n} \rangle$, where $\lambda$ drops out, it can be evaluated by means of lattice Monte-Carlo simulations. We also note a corresponding relation ${\cal A}_a^{(\epsilon)} \equiv n_2 \lambda \langle \epsilon ({\bf n} \lambda) \rangle_a = \pi \times n_2 \langle E_{\bf n} \rangle_a $ for the amplitudes of the energy density profiles in the upper half plane with uniform boundary conditions $a=+,-,f$ given in Eq. (\ref{IsingA}).

The simulations in Ref. \cite{OAV} were performed at the critical point 
$\beta =\beta_c$. One Monte Carlo (MC) step consists of the Wolff cluster update followed by 
$\mathcal{H} / \lambda  \times \mathcal{W} / \lambda$  single spin Metropolis updates.
The bulk energy per bond $-  \left< S_\mathbf{n}S_\mathbf{n'}\right>_{\rm bulk}$
was computed for the square system of size $2000 \times 2000$ with periodic boundary conditions, 
the average taken over $10^6$ MC steps and over the bulk of the system.  
To calculate $\langle E_{\bf n} \rangle$ for the rectangular system with prescribed boundary conditions the necessary averages $- \left<S_\mathbf{n} S_\mathbf{n'} \right> $ were taken over $10^6$ MC steps for FIG. \ref{-+completehorizontalOleg} and over $10^8$ MC steps for FIG. \ref{f+completehorizontalOleg}.

\section{A SMALL JANUS PARTICLE IN THE CENTER \\ OF THE RECTANGLE} \label{Janus}
As an example for an embedded particle consider a circular ``Janus'' particle of radius $R$ with boundary conditions + and $-$ on its surface segments that extend counterclockwise from $R \exp ( i \alpha_{\rm J})$ to $- R \exp ( i \alpha_{\rm J})$ and from $R \exp (- i \alpha_{\rm J})$ to $ R \exp ( i \alpha_{\rm J})$, respectively. Compare Fig. 11 (a) in Ref. \cite{SMED} where $\alpha_{\rm J}$ was denoted by $\alpha$. The angle $\alpha_{\rm J}$ characterizes the orientation of the Janus particle via its two switching points and should not be confused with our variable $\alpha$ that characterizes the aspect ratio ${\cal H}/{\cal W}$ of the embedding rectangle. Placing the center of the Janus at the center of the rectangles considered in Sec. \ref{recmix} the free energy of interaction that depends on the orientation $\alpha_{\rm J}$ is given by 
\begin{eqnarray} \label{Jan1}
F_{\rm orientation} \to -16 R^2 \cos (2 \alpha_{\rm J}) \, \langle T(z_{\rm M} =0) \rangle \, 
\end{eqnarray}
if $R$ is much smaller than ${\cal H}$ and ${\cal W}$, see Eqs. (3.4) and (4.2) in Ref. \cite{SMED}. Using the results for the stress tensor given in (\ref{Tcenteruni})-(\ref{Tcenterf+}) yields
\begin{eqnarray} \label{Jan2}
F_{\rm orientation} \to {4 \over 3} {R^2 \over {\cal H} \, {\cal W}} \cos (2 \alpha_{\rm J}) \times \tau(\alpha), \quad  \tau(\alpha) \equiv {\bf K}(\cos \alpha)  {\bf K}(\sin \alpha) \times \nonumber \\
\times \cos (2 \alpha) \Biggl[1,\, 1+48 {(\sin 2 \alpha)^2 \over (\cos 2 \alpha)^2 +3}, \, 1-12 {\sin \alpha \, (1- \sin \alpha) \over \cos 2 \alpha}  \Biggr]
\end{eqnarray}
for the embedding rectangles with boundary conditions $[{\rm uniform}, \, - + - +,\, f+f+]$.

For the uniform and $-+-+$ rectangles the function $\tau(\alpha)$ is antisymmetric about $\alpha = \pi /4$, i.e., on exchanging the values of ${\cal H}$ and ${\cal W}$ it keeps its value but changes its sign. It is positive and negative for $0< \alpha< \pi /4$ and $\pi /4 < \alpha < \pi /2$, i.e., for $\infty > {\cal H}/{\cal W} > 1$ and $1 > {\cal H}/{\cal W} > 0$, respectively. While for the uniform boundary conditions  $\tau$ is a monotonic function of $\alpha$, for $-+-+$ it is not and shows maxima and minima. For the boundary condition $f+f+$, however, $\tau(\alpha)$ displays no antisymmetry. It is non-monotonic and changes from positive to negative at $\alpha= {\rm arcsin} [(6-\sqrt{26})/10] < \pi /4$, requiring the horizontal $f$ sides of length ${\cal H}$ to be longer by a factor of 2.4135 than the length ${\cal W}$ of the + sides. Thus on minimizing the free energy $F_{\rm orientation}$, the switch points of the Janus turn to the closer sides of the rectangle in all cases {\it except} for an $f+f+$ rectangle with aspect ratio in the interval $1<{\cal H}/{\cal W}<2.4135$. Here the switches turn to the + sides although they are further apart than the $f$ sides. This confirms the supremacy of + over $f$ found in Sec. \ref{+f+f} above, see paragraph \ref{+f+f} 1 (a).

While the functions $\tau$ are quite different for the three boundary conditions, their leading behavior for $\alpha \to 0$ (as well as that for $\alpha \to \pi /2$) is the same. This reflects the fact that for an infinite strip with ++ and $ff$ boundary conditions the value of the stress tensor is the same. As expected, the leading behavior of $F_{\rm orientation}$ in  Eq. (\ref{Jan2}) for $\alpha \to 0$  reproduces the form $F_{\rm orientation} \to \cos (2 \alpha_{\rm J}) \times (\pi R/{\cal W})^2 /3$ for an infinite horizontal strip with equal boundary conditions known from Ref. \cite{SMED}.

\newpage

\newpage

\newpage
\begin{figure}
	\begin{center}
		\includegraphics[width=0.27\textwidth]{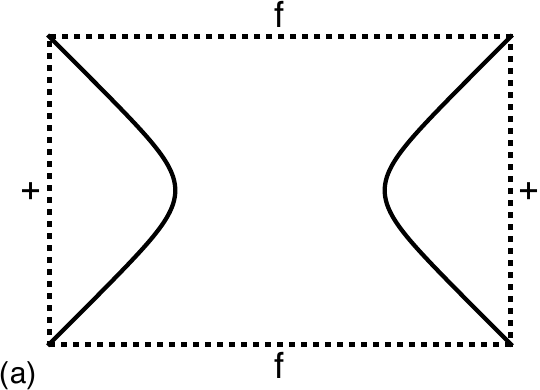}\qquad\qquad
		\includegraphics[width=0.27\textwidth]{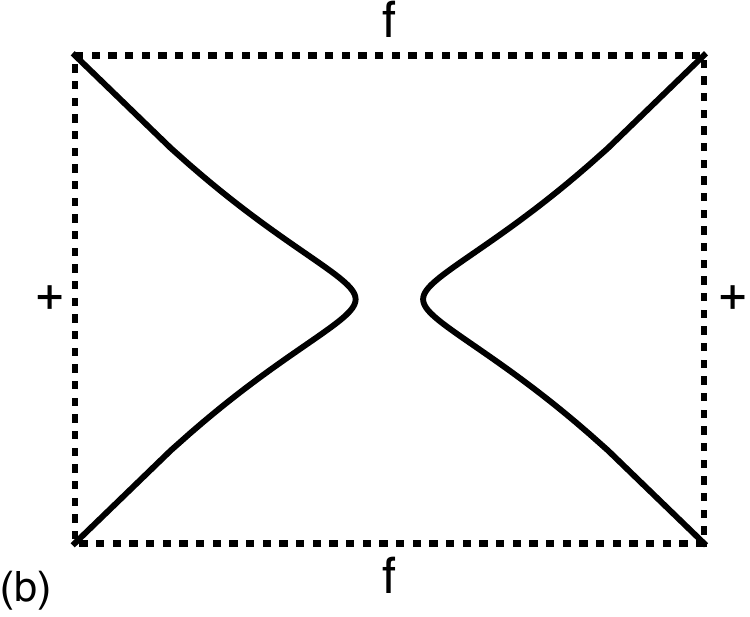}\\[1.0cm]
		\includegraphics[width=0.27\textwidth]{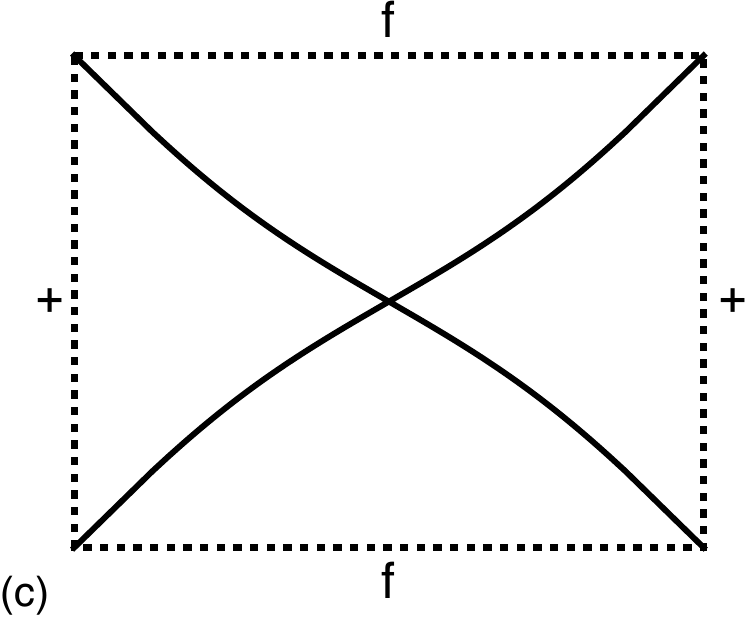}\\[1.0cm]
		\includegraphics[width=0.27\textwidth]{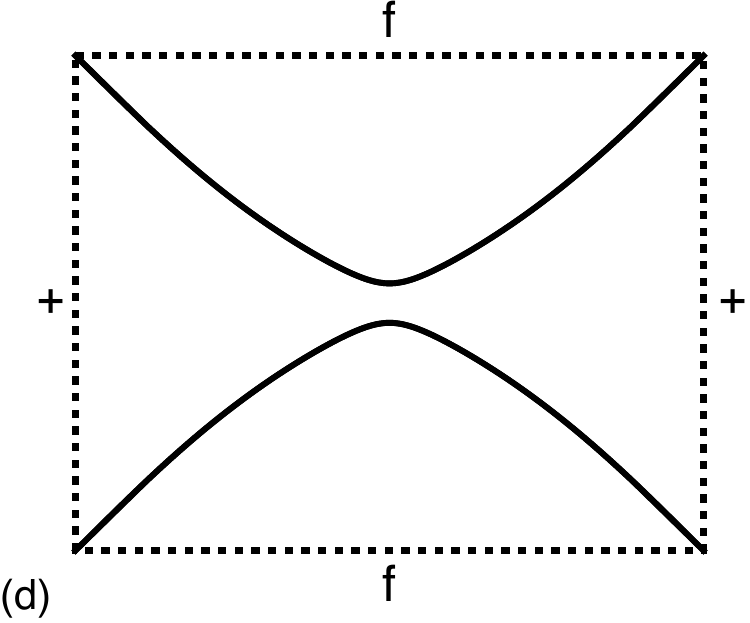}\qquad\qquad
		\includegraphics[width=0.27\textwidth]{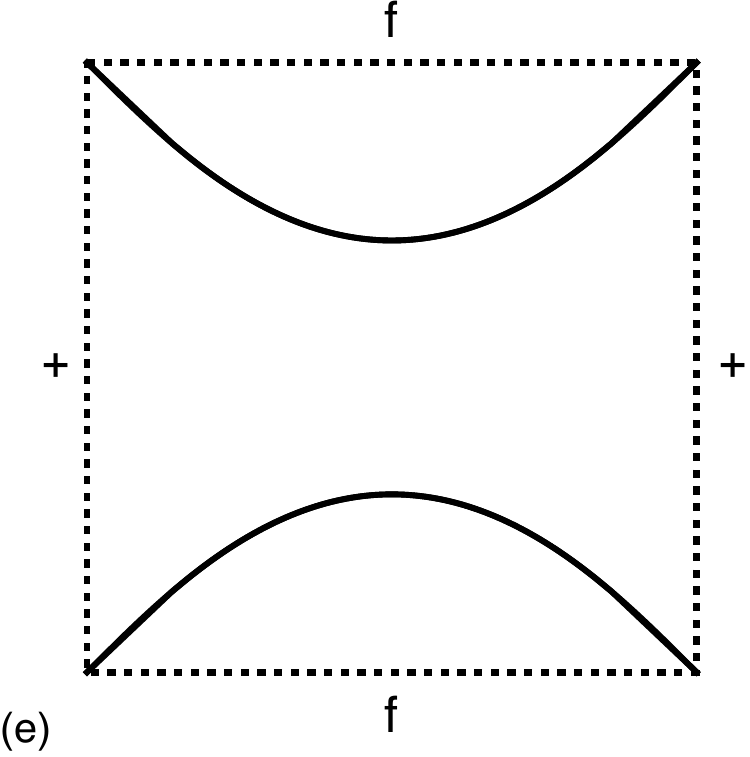}\\
	\end{center}
	\caption{Zero lines of $\langle \epsilon \rangle$ in the rectangle with boundary conditions $f$ and  $+$ on the horizontal and vertical sides. The lines separate regions of positive and negative $\langle \epsilon \rangle$ induced by the disordering $f$ and the ordering $+$ sides, respectively.  Results for the aspect ratios ${\cal H}/{\cal W}= 1.500, \, 1.289, \, 1.279, \, 1.269, \, 1.000$ are shown in (a), (b), (c), (d), (e). As described in points (1)-(3) of Sec. \ref{competitionf+}, the topology of the lines changes at ${\cal H}/{\cal W}=1.279$, i.e. at (c), for which the two zero lines form an intersection at the center where $\langle \epsilon \rangle$ vanishes. On the way from (a) over (c) to (e) the left and right regions of negative $\langle \epsilon \rangle$ shown in (a) and (b) approach each other and coagulate at (c) to form in (d) and (e) the central region of negative $\langle \epsilon \rangle$. At the same time the central region of positive $\langle \epsilon \rangle$ in (a) and (b) splits into an upper and lower region of positive $\langle \epsilon \rangle$ in (d) and (e). The opening angle of the left and right regions at the intersection point in (c) equals $\pi /3$, see Ref \cite{crossangle}. In between (a) and (b), at ${\cal H}/{\cal W}=1.471$, the turning away of the lines from their asymptotic corner tangents towards the $+$ boundaries as in (a) changes to turning away towards the $f$ boundaries as in (b)-(e).}\label{Ted}
\end{figure}

\newpage
\begin{figure}
	\begin{center}
		\includegraphics*[height=0.50\textheight]{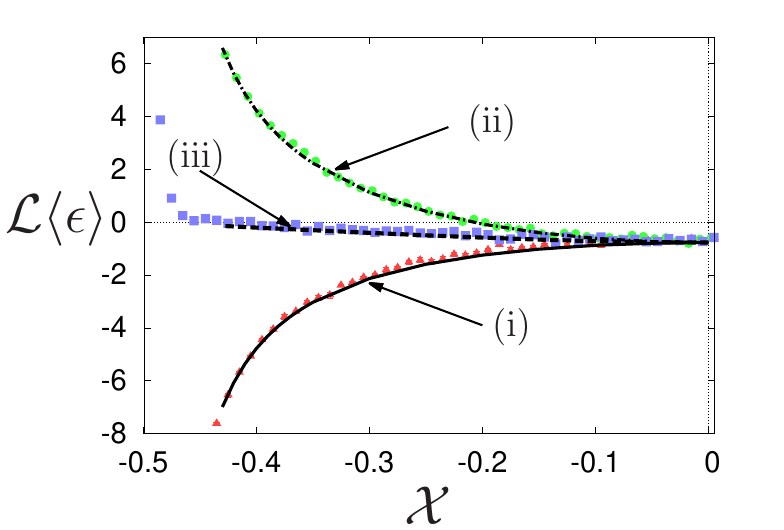}\\*
	\end{center}
	\caption{Energy density $\langle \epsilon (x_{\rm M},y_{\rm M}) \rangle$ in the square of FIG. \ref{Ted} (e) with side length ${\cal H}={\cal W} \equiv {\cal L}$, boundary conditions $f$ along the two horizontal sides, and $+$ along the vertical sides. The data denoted by (i), (ii), and (iii) represent the scale-free expressions ${\cal L} \times \langle \epsilon ({\cal L} {\cal X},0) \rangle$, ${\cal L} \times \langle \epsilon (0,{\cal L} {\cal X}) \rangle$, and ${\cal L} \times \langle \epsilon ({\cal L} {\cal X},{\cal L} {\cal X}) \rangle$ of $\langle \epsilon \rangle$ along the horizontal midline, the vertical midline, and along the diagonal of the square as ${\cal X}$ varies from $-1/2$ to $0$. The three expressions meet at ${\cal X}=0$, the center of the square, where $\langle \epsilon \rangle$ is negative. The field theory results for (i) and (ii) follow from Eqs. (\ref{epshoriu}) and (\ref{epsvertv}) on substituting them in the transformation (\ref{primarytrafo}) and using (\ref{unitcircrect}), (\ref{ZMtozM}). The result (iii) follows from Eq. (\ref{epshordiaprime+f}) which yields ${\cal L} \times \langle \epsilon ({\cal L} {\cal X},{\cal L} {\cal X}) \rangle = -(\sqrt{2} -1) {\bf K} {\rm cn}(2 {\cal X}  {\bf K})$ where the modulus $q$ of the Jacobi function ${\rm cn}$ and of the complete elliptic integral ${\bf K}$ equals $q=1/\sqrt{2}$ so that, e.g., ${\bf K} = {\bf K} (1/\sqrt{2}) = 1.854$. All the values in (i) and (iii) are negative since the horizontal midline and the diagonal are entirely within the central region of FIG. \ref{Ted} (e). The vertical midline, however, crosses the zero lines in FIG. \ref{Ted} (e) and the values of (ii) are negative and positive for $|{\cal X}|< 0.208$ and $|{\cal X}|> 0.208$ as predicted by Eq. (\ref{Y0Sq}). Like in FIG. \ref{-+completehorizontalOleg} there is good agreement between the results of conformal field theory (full line, dash-dotted line, line of dashes) and of the numerical simulations (triangles, circles, squares) obtained in Ref. \cite{OAV}  on a 200 $\times$ 200 lattice of Ising spins. For the center of the square, for example, the simulation result of $\approx -0.779$ agrees well with the analytic expression $-(\sqrt{2}-1){\bf K} =-0.768$ from Eq. (\ref{ep+f+f+00SQ}).}\label{f+completehorizontalOleg}
\end{figure}


\begin{thebibliography}{99}
\bibitem{unlike}In order to avoid confusion between the coordinates for the rectangle and half plane  geometries, in this paper we replace the notation $z=x+iy; \, y>0$ for the upper half plane used in Refs. \cite{BX}-\cite{BE21}  by the notation $H=G+iJ; \, J>0$.
\bibitem{primaryfields}The two primary operators $\phi = \epsilon$ and $\phi = \sigma$  are normalized by the conditions $$\langle \phi \rangle_{\rm bulk} =0 \, , \quad \langle \phi \phi \rangle_{\rm bulk} =1/({\rm mutual \, distance})^{2 x_{\phi}}$$ of their one- and two-point functions in the bulk. Thus $\langle \epsilon \rangle$ is the (normalized) deviation of the energy density in the system from the energy density in the bulk. As for the sign, $\langle \epsilon \rangle >0$ and $\langle \epsilon \rangle <0$ implies in the language of the Ising lattice model a short range order of spins that is weaker and stronger, respectively, than the order in the bulk. This is in agreement with the minus sign in front of the rhs of the short distance expansion $\langle \sigma \sigma \rangle - \langle \sigma \sigma \rangle_{\rm bulk}  \to - (1/2) ({\rm mutual \, distance})^{x_{\epsilon} - 2 x_{\sigma}} \times \langle \epsilon \rangle$, see, e.g., Eq. (2.14) in Ref. \cite{BE21}. For the Ising model the amplitudes ${\cal A}_a^{(\phi)}$ in (\ref{uniformstrip}) are given, e.g., in the paragraph below Eq. (\ref{uniformBOE}). The explicit expression of the primary operator $\epsilon$ in terms of the spins in the Ising lattice model is given in Eq. (\ref{expresseps}).
\bibitem{Ttrafo}Recall the general transformation formula 
%
\begin{eqnarray} \label{THandz}
%
\langle T(z) \rangle = \Biggl( {dH \over dz} \Biggr)^2 \, \langle T(H) \rangle  \, + \, {\hat{c} \over 12} \, S(H(z)) \, , \quad S(H(z)) = {H''' \over H'} -{3 \over 2} \Biggl( {H'' \over H'} \Biggr)^2   \nonumber
%
\end{eqnarray}
%
for stress-tensor averages in two geometries $H$ and $z$ that are related by a conformal mapping $H(z)$. Here $\hat{c}$ is the conformal charge of the critical system, and $S$ is the Schwarzian derivative of the mapping with the primes denoting derivatives wrt $z$.
\bibitem{domainwallprime}The zero-lines where $\langle \epsilon \rangle =0$ are the counterparts of the ``domain boundaries'' of Refs. \cite{TWBG2} and \cite{domainwall} where $\langle \sigma \rangle =0$.
\bibitem{tangent}To derive Eqs. (\ref{zero-+-prime}) one uses that ${\rm arccosh} (1+h) \to \sqrt{2h} \, [1- (h/12) ]$ for $h \to 0$.
%
The factor 1/2 between the asymptotic angles $\pi /12$, $5 \pi /12$ and the corresponding angles $\pi /6$, $5 \pi /6$ in the $H$ plane addressed in \cite{BXprof} is due to the corner generated at $z=0$.
\bibitem{confuse}The quantities $s=\sin \alpha$ and $S=\sin^2 \alpha$ introduced in Eq. (\ref{alternative}) should not be confused with the surface operator $s$ in the BOE of $\sigma$ near an $f$ boundary and with the Schwarzian derivative introduced in Refs. \cite{fsigma} and \cite{Ttrafo}, respectively.
\bibitem{brief}A brief account, limited to the tip of a needle with equal boundary conditions on both sides, has been given in Ref. \cite{BE21}.
\bibitem{fsigma}For the combination $(\phi,a)=(\sigma,f)$ the BOE has a form different from (\ref{uniformBOE}). The up-down spin symmetry of the Ising model is not broken by an $f$ boundary so that --- like $\sigma$ --- its corresponding boundary operator $s$ must be odd under an up-down change which does not apply to the stress tensor. The scaling dimension $x_s$ of $s$ equals 1/2 in the Ising model. From the arguments in Sec. \ref{equalbc} we conclude that for an $f|f$ corner and $\phi = \sigma$ the counterpart of the COE (\ref{opWEDGE}) reads 
%
\begin{eqnarray} \label{sigmafwedge}
%
\sigma(x,y) \to {\rm const} \, |z|^{g x_s -x_{\sigma}} (\sin\vartheta g)^{x_s -x_{\sigma}} \times {\cal S} \nonumber
%
\end{eqnarray}
%
where $\cal S$ is a corner operator of scaling dimension $g x_s$. For the example mentioned below Eq. (\ref{halfvswedge}) of an $f|f$ wedge with an $f+$ switch at $x= \zeta$ the exact expression of $\langle \sigma (x,y) \rangle$ is given by 
%
\begin{eqnarray} \label{sigmafwedgeprime}
%
\langle \sigma (x,y) \rangle = \Biggl( {g \over 2 |z| \sin \vartheta g} \Biggr)^{1/8} \Biggl[ {M-1+k \cos \vartheta g \over M} \Biggr]^{1/4} \, , \; M=\Bigl( 1-2k \cos \vartheta g +k^2 \Bigr)^{1/2} \, , \; k= \Bigg( {|z| \over \zeta} \Biggr)^g \nonumber
%
\end{eqnarray}
%
in the Ising model and by expanding the numerator inside the square bracket for small $k$ yields
%
\begin{eqnarray} \label{sigmafwedgeprimeprime}
%
\langle \sigma (x,y) \rangle  \to (g/8)^{1/8} |z|^{(g/2)-(1/8)} (\sin \vartheta g)^{3/8} \times \zeta^{-g/2} \nonumber
%
\end{eqnarray}
%
in agreement with the COE above. For $\langle \sigma \rangle$ close to an $f|f$ corner in other geometries considered in the main text --- like the semi-infinite strip and the square --- the above COE applies as well. Of course the geometry must contain a symmetry-breaking element such as a + boundary segment in order that the odd operator ${\cal S}$ has a nonvanishing average. An example is discussed in detail in paragraph \ref{strip+ff}.
\bibitem{swvsshape}While the validity of the relations (\ref{identapex}) and (\ref{identdiffapex}) is confined to the corners of wedges the sides of which are perturbed by switches but remain straight lines, the operator expansions (\ref{opWEDGE}) and (\ref{opdiffWEDGE}) apply to corners of a semi-infinite strip or a rectangle as well. The amplitudes ${\cal F}_a^{(T)}=g^2 z^{2 g -2}$ and  ${\cal F}_{a|b}^{(T)} = g^2 z^{g -2}$ given in Eqs. (\ref{avWEDGE}) and (\ref{avdiffWEDGE}) are independent of $a$ and $a|b$, respectively, and imply the normalizations of the operators ${\cal T}$ and ${\cal Y}$. This is in line with Eq. (3.7) in Ref. \cite{BE21}.
\bibitem{dilaF}$\langle T(z) \rangle_{a|b}$ in (\ref{Tabwedge}) determines the corner contribution to the change of the universal free energy $F$ under a dilatation, see Refs. \cite{CardPe,Imamura}. For our rectangles with sides $ABCD=baba$ where $g=2$ and the $t_{ab}$'s of its four corners are equal, this reads
$$ \bigl( {\cal H} \partial_{\cal H} + {\cal W} \partial_{\cal W}  \bigr) \, F({\cal H}, {\cal W}) = 8  t_{ab}-{\hat{c} \over 4}  $$ as noted and numerically confirmed for $ab=++$, $ff$, $+-$, $+f$ in Ref. \cite{Wu}. The above expression is independent of the size and aspect ratio of the rectangle and for the Ising model given by $(-1, -1, 31, 3)/8$.
\bibitem{BXwedge}For a $+-$ wedge with an opening angle of $\pi$, this order-disorder effect is contained in Eq. (4.1) of Ref. \cite{BX}.
\bibitem{instable}It would be interesting to know whether or not the domination of $+$ over $f$ is connected with the instability in the half plane of $f$  against a small + perturbation which leads to a renormalization flow from $f$ to $+$, compare, e.g., Ref. \cite{nocrossov}.
\bibitem{horizvertminusT}A simple way to understand the different signs of $T$ in Eqs. (\ref{horizBOE}) and (\ref{vertBOE}) is by joining together the positive horizontal and vertical boundaries to the boundary of a wedge extending over the first quadrant of the $H=G+iJ$ plane. For uniform boundary conditions $\langle \phi(G,J) \rangle$ remains invariant against mirror imaging about the diagonal $J=G$ of the wedge while the form $\langle T(G+iJ) \rangle \propto 1/(G+iJ)^2$ implies that $\langle T(i|H|) \rangle = - \langle T(|H|) \rangle$, i.e., when mirror imaging the location of the boundary operator $T$, its average remains invariant except for a minus sign which has to be corrected for in the BOE.
\bibitem{nonunique}In the limit ${\cal H} \to \infty$ with ${\cal W}$ fixed of the semi-infinite strip the corresponding mapping (\ref{uphalfrect}), (\ref{Ztoz}) with $q=1, \, q' =0$ takes the form $$H(Z)={\rm sn}^2 (Z,1)= \tanh^2(Z)=(\cosh \tilde{z} -1)/(\cosh \tilde{z} +1)$$ since, due to (\ref{aspq}) and (\ref{Lgeo}), $Z=\tilde{z}/2$ with $\tilde{z}=\pi z /{\cal W}$. Thus the mapping equals the mapping (\ref{semiinfstrip}) apart from a conformal transformation $H \leftrightarrow (H-1)/(H+1)$ that maps the upper half plane onto itself.
\bibitem {symmetry}The transformation $w(Z_{\rm M}) \equiv w(Z_{\rm M},q)$ in (\ref{unitcircrect}) obeys $$w(Z_{\rm M}^{*},q) =\bigl(w(Z_{\rm M},q) \bigr)^{*} \, , \quad w(-Z_{\rm M},q) = - w(Z_{\rm M},q) \, ,$$ preserving the mirror symmetries about the coordinate axes and $$ w( i Z_{\rm M},q')= i w(Z_{\rm M},q) $$ relating corresponding positions in a rectangle and its partner with the inverse aspect ratio and same size to those in the corresponding circular disks. Likewise, the local dilatation factor ${\cal R}(Z_{\rm M},q) \equiv |d w(Z_{\rm M},q) /d Z_{\rm M}|$, compare Eq. (\ref{primarytrafo}), obeys $$ {\cal R}(Z_{\rm M}^{*},q) =  {\cal R}(-Z_{\rm M},q) = {\cal R}(Z_{\rm M},q) = {\cal R}(iZ_{\rm M},q')\, . $$
\bibitem{ZMF}Using the new integration variable $\omega=s \sqrt{Q}$ in the integral representation of $F$ given below (\ref{invcircrect''}) and identifying $Q=\exp(\pm 2i \alpha)$ leads to the two alternative forms 
$$Z_{\rm M}(w)=e^{\pm i \alpha} F(we^{\mp i \alpha} , \, e^{\pm 2i \alpha})$$ 
of the transformation (\ref{invcircrect}) in terms of the elliptic function $F$.
\bibitem{consistsquare}The consistency of (\ref{uniprimaryprim}) with (\ref{uniSQU}) follows from the identity $\big|{\rm dn}^2-(1/2){\rm sn}^2 {\rm cn}^2 \big| \big/ \bigl[ |{\rm cn}^2| - |{\rm sn}^2 {\rm dn}^2| \bigr] = |{\rm cn}| \big/ \bigl[ |{\rm dn}^2| - (1/2) |{\rm sn}^2 | \bigr]$, where the arguments of the Jacobi functions on the lhs are $\bigl( Z_{\rm M}, \, q=1/\sqrt{2} \bigr)$ and those on the rhs are $\bigl( (1+i) Z_{\rm M}, \, q=1/\sqrt{2} \bigr)$.
\end{thebibliography}
\end{document}